\documentclass[12pt,twoside]{elsart}
\usepackage{epsfig,graphics}
\usepackage[T1]{fontenc}
\usepackage{ae,aecompl}
\usepackage{caption2}
\newcommand{\CP}{\mbox{\it CP}}
\newcommand{\ra}{\mbox{~$\rightarrow$}~}

\newcommand{\cd}{\mbox{$\cdot$}}
\newcommand{\gmu}{\mbox{$\mu$}}
\newcommand{\mum}{\mbox{$\mu^-$}}
\newcommand{\mup}{\mbox{$\mu^+$}}

\newcommand{\gpi}{\mbox{$\pi$}}
\newcommand{\pip}{\mbox{$\pi^+$}}
\newcommand{\pim}{\mbox{$\pi^-$}}
\newcommand{\pipm}{\mbox{$\pi^{\pm}$}}

\newcommand{\gK}{\mbox{$K$}}

\newcommand{\Km}{\mbox{$K^{-}$}}
\newcommand{\Kpm}{\mbox{$K^{\pm}$}}

\newcommand{\Ks}{\mbox{$K_S$}}

\newcommand{\gp}{\mbox{$p$}}
\newcommand{\agp}{\mbox{$\overline{p}$}}

\newcommand{\gL}{\mbox{$\Lambda$}}
\newcommand{\agL}{\mbox{$\overline{\Lambda}$}}

\newcommand{\gXi}{\mbox{$\Xi$}}

\newcommand{\Xipm}{\mbox{$\Xi^{\pm}$}}
\newcommand{\Xim}{\mbox{$\Xi^-$}}
\newcommand{\aXim}{\mbox{$\overline{\Xi}$$^+$}}

\newcommand{\gOm}{\mbox{$\Omega$}}

\newcommand{\Omm}{\mbox{$\Omega^-$}}
\newcommand{\Ompm}{\mbox{$\Omega^{\pm}$}}
\newcommand{\aOmm}{\mbox{$\overline{\Omega}$$^+$}}
\newcommand{\alXi}{\makebox{$\alpha_{\Xi}$}}

\newcommand{\aalXibig}{\makebox{$\overline{\alpha}_{\Xi}$}}

\newcommand{\alL}{\makebox{$\alpha_{\Lambda}$}}

\newcommand{\aalLbig}{\makebox{$\overline{\alpha}_{\Lambda}$}}

\newcommand{\AXiL}{\makebox{${A}_{\Xi\Lambda}$}}

\newcommand{\AXiLratiobigtxt}{\makebox{$(\alXi\alL - \aalXibig\aalLbig) /
                                        (\alXi\alL + \aalXibig\aalLbig)$}}

\def\kp3pi{K^+ \rightarrow \pi^+ \pi^+ \pi^-}
\def\km3pi{K^- \rightarrow \pi^- \pi^- \pi^+}
\def\kpm3pi{K^{\pm} \rightarrow \pi^{\pm} \pi^+ \pi^-}

\begin{document}

\begin{frontmatter}

\title{HyperCP: A high-rate spectrometer for the study of charged hyperon and 
       kaon decays \\[0.2in]
       {\large The Fermilab HyperCP Collaboration} \\[0.2cm]
       {\normalsize\rm  \today}
}

\author[iit]{R.A. Burnstein},
\author[iit,alekaddress]{A. Chakravorty},
\author[as,chanaddress]{A. Chan},
\author[as]{Y.C. Chen},
\author[ucb,choongaddress]{W.-S. Choong},
\author[usa]{K. Clark},
\author[uva]{E.C. Dukes\corauthref{cor}},
\author[uva,durandetaddress]{C. Durandet},
\author[ug]{J. Felix},
\author[lbl]{R. Fuzesy},
\author[lbl]{G. Gidal},
\author[lbl,guaddress]{P. Gu},
\author[umich]{H.R. Gustafson},
\author[as,hoaddress]{C. Ho},
\author[uva]{T. Holmstrom},
\author[uva]{M. Huang},
\author[fnal]{C. James},
\author[usa]{C.M. Jenkins},
\author[lbl,jonesaddress]{T.D. Jones},
\author[iit]{D.M. Kaplan},
\author[iit]{L.M. Lederman},
\author[ul,lerosaddress]{N. Leros},
\author[umich]{M.J. Longo},
\author[umich]{F. Lopez},
\author[uva]{L.C. Lu},
\author[iit]{W. Luebke},
\author[ucb,lbl]{K.-B. Luk\corauthref{cor}},
\author[uva]{K.S. Nelson},
\author[umich]{H.K. Park},
\author[ul]{J.-P. Perroud},
\author[iit,rajaramaddress]{D. Rajaram},
\author[iit]{H.A. Rubin},
\author[as]{P.K. Teng},
\author[lbl]{B. Turko},
\author[fnal]{J. Volk},
\author[iit]{C.G. White},
\author[iit,whiteaddress]{S.L. White},
\author[lbl]{P. Zyla}

\address[as]{Academia Sinica, Nankang, Taipei 11529, Taiwan, ROC}
\address[ucb]{University of California at Berkeley, Berkeley, CA 94720,
              USA}
\address[fnal]{Fermi National Accelerator Laboratory, Batavia, IL 60510,
               USA}
\address[ug]{Universidad de Guanajuato, 37000 Le\'{o}n, Mexico}
\address[iit]{Illinois Institute of Technology, Chicago, IL 60616, USA}
\address[ul]{Universit\'{e} de Lausanne, IPHE, CH-1015 Lausanne, Switzerland}
\address[lbl]{Lawrence Berkeley National Laboratory, Berkeley, CA 94720,
              USA}
\address[umich]{University of Michigan, Ann Arbor, MI 48109, USA}
\address[usa]{University of South Alabama, Mobile, AL 36688, USA}
\address[uva]{University of Virginia, Charlottesville, VA 22901, USA}

\corauth[cor]{Corresponding authors.  
Tel.: 1-434-982-5364, fax: 1-434-982-5375 (E.C. Dukes); 
Tel.: 1-510-486-7054, fax: 1-510-643-8497 (K.-B. Luk). \\
{\it E-mail addresses:} craigdukes@virginia.edu (E.C. Dukes),
k\_luk@lbl.gov (K.-B. Luk).}
\thanks[alekaddress]{Present address:  St.\ Xavier University, Chicago, IL
                     60655, USA.}
\thanks[chanaddress]{Present address: Physics Dept., Brookhaven National 
                     Laboratory, Upton, NY 11973, USA.}
\thanks[choongaddress]{Present address: Life Sciences Division, Lawrence 
                       Berkeley National Laboratory, Berkeley, CA 
                       94720, USA.}
\thanks[durandetaddress]{Present address: Paradise Valley Community College, 
                         18401 N. 32nd St., Phoenix, AZ 85032, USA.}
\thanks[guaddress]{Present address: KLA-Tencor, 160 Rio Robles, San Jose, 
                   CA 95134, USA.}
\thanks[hoaddress]{Present address: IS2 Research Inc., 20 Gurdwara Rd., 
                   \#3-6, Nepean, ON K2E8B, Canada.}
\thanks[jonesaddress]{Present address: Boeing Co., St.\ Louis, MO 63130, 
                      USA.}
\thanks[lerosaddress]{Present address: CERN, CH-1211 Gen\`{e}ve 23, Switzerland.}
\thanks[rajaramaddress]{Present address: University of Michigan, Ann Arbor, 
                        MI 48109, USA.}
\thanks[whiteaddress]{Present address: University of Alabama at Birmingham, 
                      Birmingham, AL 35233, USA.}

\clearpage

\begin{abstract}
The HyperCP experiment (Fermilab E871) was designed to search for rare 
phenomena in the decays of charged strange particles, in particular {\em CP} 
violation in \gXi\ and \gL\ hyperon decays with a sensitivity of $10^{-4}$.  
Intense charged secondary beams were produced by 800\,GeV/$c$ protons 
and momentum-selected by a magnetic channel. 
Decay products were detected in a large-acceptance, high-rate magnetic 
spectrometer using multiwire proportional chambers, trigger hodoscopes, 
a hadronic calorimeter, and a muon-detection system. 
Nearly identical acceptances and efficiencies for hyperons
and antihyperons decaying within an evacuated volume 
were achieved by reversing the polarities of the channel 
and spectrometer magnets. 
A high-rate data-acquisition system enabled 231 billion events to be
recorded in twelve months of data-taking. \\[0.2in]
\noindent
{\it PACS:}
07.05.Fb,
29.30.Aj,
29.40.Cs,
29.40.Mc,
29.40.Vj \\[0.2in]
\noindent
{\it Keywords:}
HyperCP;
Magnetic spectrometer;
Hadronic calorimeter;
Hyperon \CP\ violation;
Rare hyperon and kaon decays;
Fermilab
\end{abstract}

\end{frontmatter}

\clearpage

\section{Introduction}
\label{sec:intro}
Although expected to be ubiquitous in weak decays, {\em CP} violation 
is a small effect, and thus far it has been observed only in the 
decays of neutral $K$ and $B$ mesons. 
Searches for (and perhaps discoveries of) this phenomenon in the 
decays of other particles will provide new insights
and will help to determine whether the phase in the Cabibbo--Kobayashi--Maskawa 
quark mixing matrix~\cite{CKM} is the sole source of {\em CP} violation, 
or whether there are other sources. 
Many beyond-the-standard-model theories predict \CP-violating effects,
often with larger signals than found in
standard-model calculations.
Hyperon decays offer promising possibilities for \CP-violation
searches: hyperons are copiously produced and readily detected, 
and their decays are particularly sensitive to certain exotic sources of \CP\ 
violation~\cite{bSM_calc}.

The primary goal of the HyperCP experiment (Fermilab E871) 
was to search for \CP\ violation in \gXi\ and \gL\ hyperon decays with  
a sensitivity far beyond previous limits,
and at which some theories predict an effect.
The search method was to compare the asymmetries in proton and antiproton
decay distributions in \gL\ra\gp\pim\ and \agL\ra\agp\pip\ decays,
where the \gL\ and \agL\ were produced from unpolarized 
\Xim\ and \aXim\ decays.
The signature for \CP\ violation is a nonzero value of
$\AXiL = \AXiLratiobigtxt$, where \alXi\alL\ and \aalXibig\aalLbig\ are, 
respectively, the magnitudes of the asymmetries in the proton and antiproton 
angular distributions in the \gL\ and \agL\ rest frames.
To achieve the desired statistical sensitivity of $10^{-4}$ in \AXiL,
the HyperCP spectrometer was designed to operate at very high rates.
To keep biases below the statistical sensitivity an apparatus with 
minimal differences in acceptances and efficiencies for \Xim\ and \aXim\ 
decays was also essential.
HyperCP recorded 2.5 billion \Xim\ and \aXim\ decays;
the two best previous searches for \CP\ violation in hyperon decays
had datasets of 280\,000 \Xim\ and \aXim\ events
(E756~\cite{E756}) and 96\,000 \gL\ and \agL\ events (PS185~\cite{PS185}).

An experiment optimized for the physics just described inevitably
records large numbers of charged kaon and $\Omega$ decays as well.
Hence the spectrometer was designed to have good acceptance for
such decays.
Also, a simple muon-detection system was added to the spectrometer
to allow searches for rare and forbidden kaon and hyperon decays
with muonic final states at the level of $\mathcal{O}(10^{-7})$.

HyperCP was approved in June 1994. A new spectrometer was 
fabricated in two-years' time, with beam first delivered to the
experiment in November 1996.
Following a five-month running-in period, data were taken from 
April to August 1997 and, after minor upgrades, from June 1999 to January 2000.
We describe here the 1999 spectrometer and, where appropriate, point out the 
differences between the apparatus used in the two runs.
Further details of the HyperCP data acquisition system can be found 
in~\cite{HyperCPDAQ,DAQ97,IEEE}; forthcoming papers will provide thorough
descriptions of the HyperCP calorimeter and wire chambers.

\section{Overview of the Spectrometer}
\label{sec:overview}
The experiment was mounted at Fermilab in the Meson Center beamline.
The overall layout of the HyperCP apparatus is shown in 
Fig.~\ref{fig:site_layout}; Table~\ref{tab:material} gives the
$z$ positions of the spectrometer elements.
The target area was in enclosure MC6 of the 
Meson Detector Building and was surrounded by thick shielding.
The HyperCP spectrometer was in enclosure MC7, which
extended beyond the building.
Because this enclosure was rather narrow and could not be enlarged,
it constrained certain aspects of the spectrometer design.
Cables from the spectrometer passed over a bridge to the
Electronics Hall in enclosure MP7, where the data were digitized
by the front-end elements of the data-acquisition (DAQ) system. 
Fiber-optic cables carried the digitized data from the Electronics Hall 
to the Control Room, which was three connected Porta-Kamp 
trailers~\cite{Porta-Kamp} alongside enclosure MC8. 
Radiation levels precluded beam-on occupancy of the Detector Hall, 
but the Electronics Hall could be occupied even when beam was being 
delivered to the experiment; this was an important consideration for 
efficient trigger and DAQ-system commissioning and maintenance.

Elevation and plan views of the HyperCP spectrometer are shown in
Fig.~\ref{fig:HyperCP}.  
Intense charged secondary beams with mean momenta of about 160\,GeV/$c$ 
were produced by 800\,GeV/$c$ protons striking $2{\times}2$\,mm$^2$ targets,
with the secondaries charge- and momentum-selected by means of a curved 
channel embedded in a dipole magnet (Hyperon Magnet).
The central orbit of the secondary beam emerged from the channel at a 19.51\,mrad angle above the horizontal.
The mean momentum of the secondary beam was chosen to be
fairly low, corresponding to a small value of $x_F$ (${\approx}0.2$), 
in order to minimize the differences between the \Xim\ and \aXim\ yields, 
but to be still high enough that the acceptance was relatively good.
Typically the incident proton beam was collinear with the entrance
axis of the channel, but about 10\% of the data were taken
at nonzero targeting angles in order to produce polarized hyperons.
The secondary-beam intensity was usually ${\approx}$13\,MHz,
dominantly pions in the negative-polarity beam and with roughly
equal numbers of pions and protons in the positive-polarity beam.
It was dumped in a massive concrete and iron structure between 
enclosures MC7 and MC8.
To switch between the positive and negative secondary-beam polarities, 
the polarities of both the Hyperon and Analyzing Magnets were reversed,
and target lengths were changed in order to keep the intensity of the
secondary beam the same. 
This resulted in nearly identical acceptances and
efficiencies for hyperons and antihyperons, 
which was essential for minimizing biases.

A high-rate magnetic spectrometer of nine multiwire proportional 
chambers (MWPCs) and two dipole magnets (Analyzing Magnets)
situated back-to-back followed a 13\,m long evacuated pipe 
(Vacuum Decay Region).
A simple trigger, used to select candidate \gXi\ra\gL\gpi\ra\gp\gpi\gpi\
decays as well as other charged hyperon and kaon decays,
required the coincidence of at least one charged particle each on the left
and right sides of the secondary beam.
Left and right are defined with respect to an observer 
looking downstream along the beam line, and are also referred to 
respectively as the ``same-sign'' and ``opposite-sign'' sides, 
or pion and proton sides, since both pions from the \gXi\ decay 
had the same charge as the secondary beam, while the proton had the 
opposite charge and hence was deflected in the opposite direction
by the field of the Analyzing Magnets. 
The trigger employed two hodoscopes (Same-Sign and Opposite-Sign Hodoscopes),
situated on either side of the secondary beam,
and located sufficiently far downstream of the 
Analyzing Magnets that the proton and pions from the \gXi\ decays 
had all separated from the secondary beam. 
To suppress backgrounds from interactions of the secondary beam with
material in the spectrometer,
the trigger also required a minimum energy deposit in a
hadronic calorimeter (Hadronic Calorimeter). 
Care was taken to keep the material in
the spectrometer at a minimum, with helium-filled bags placed between 
the wire chambers and the hodoscopes, and in the apertures of
the Analyzing Magnets.  
The total material in the spectrometer, including helium bags,
from the collimator exit to the Opposite-Sign Hodoscope, was 
about 2.3\% of an interaction length and 4.7\% of a radiation 
length.\footnote{The spectrometer-material thickness in interaction lengths, 
calculated using the correct interaction cross sections and
fractions of protons and pions in the secondary beam
(approximately two-thirds pions and one-third protons in positive running
and all pions in negative running)
was 1.6\% (1.4\%) for positive (negative) secondary-beam polarity, 
and 0.6\% (0.5\%) for the material upstream of the Analyzing Magnets.}

A muon-detection system consisted of two stations on either side
of the charged secondary beam.
Each station had three layers of proportional-tube planes interspersed 
with iron shielding.
Muon triggers were formed with signals from two planes of
scintillation hodoscopes located behind the iron absorbers in each station.
Just to the rear of the Muon Stations were two hodoscopes (collectively
called the Beam Hodoscope) that were centered on the secondary beam 
and used to measure its intensity.

A fast front-end latch system coupled with a small event size
and a high-rate data-acquisition system enabled up to 100\,000 events 
per spill second to be recorded on magnetic tape.
A total of 231 billion events were written during the 1997 and 1999 runs.

\section{Beam and Beamline}
\label{sec:beam}
After acceleration to 800\,GeV/$c$ in the Fermilab Tevatron, 
primary protons were extracted at a rate of about 
$2.5{\times}10^{11}$\,s$^{-1}$ and sent to the Switchyard. 
A small fraction, typically fewer than $1{\times}10^{10}$\,protons/s, 
were split from the main proton beam and sent to the Meson Center (MC)
beamline, whose elements are shown in Fig.~\ref{fig:beamline}. 
Upon leaving the Switchyard, the Meson Center beam rose gradually and was
deflected back into the horizontal by magnets in enclosure M01. 
The beam was transported to enclosure MC6 (Fig.~\ref{fig:site_layout}), 
where the secondary beam was created. 

The primary proton beam was bunched by the 53\,MHz RF cavities 
of the Tevatron into ``RF buckets,''
each about 1\,ns wide and separated by 18.9\,ns. 
Typically about $1{\times}10^{11}$ 800\,GeV/$c$ protons were delivered 
in a 40-second spill, producing a secondary beam of ${\approx}$13\,MHz
intensity (at a $0^{\circ}$ targeting angle).
Between spills was a 40-second interspill period during which no beam
was delivered to the experiment.\footnote{These figures are for the 1999 run; 
the beam intensity during the 1997 run was typically no more than 
$7.5{\times}10^9$ protons/s in a spill 23\,s in length, with an interspill
period of 34\,s.} 

The primary-beam intensity was monitored by both an ion chamber (IC) 
and a secondary-emission monitor (SEM) located in the MC6 enclosure.
These devices integrated charge during the spill and were
calibrated to give an absolute measure of the total number of
protons traversing them per spill. 
Readings from each were recorded at the end of each spill. 
The IC began to saturate at instantaneous beam
intensities above about $1{\times}10^{10}$ protons/s,
and was fully saturated at about $1.3{\times}10^{10}$ protons/s.  
It was thus not accurate above the nominal operating intensity,
but worked well during low-intensity studies.  
The SEM was used as the primary beam-intensity monitor. 
It was calibrated (with an uncertainty of $\pm$5\%) by sending a 
known number of protons from the Switchyard to the Meson Center beamline, 
with all splits to other beamlines disabled.\footnote{Each SEM count
corresponded to $4.0{\times}10^7$ charged particles passing
through the device.}

The secondary beam was created in enclosure MC6 by steering the
primary protons onto a target centered on the entrance of a curved 
channel, or collimator.
The collimator was installed within a dipole magnet, MC6SW. 
(See Fig.~\ref{fig:pile} for an elevation view of the secondary-beam
forming area.)
To create an unpolarized hyperon beam (the usual mode of operation), 
protons were incident on the target at $0^\circ$ relative to the 
entrance axis of the collimator.
For about 10\% of the running time (``polarized'' running), 
nonzero horizontal targeting angles of ${\pm}$3.0\,mrad 
(and to a lesser extent ${\pm}$2.5, ${\pm}$2.0, and ${\pm}$1.0\,mrad)
were used.
The targeting angles were tuned by means of two ``angle-varying bend'' (AVB)
dipole-magnet pairs, horizontal (MC6EW1--2) and vertical (MC6UD1--2), 
located upstream of MC6SW. 
The largest possible horizontal (vertical) targeting angle was 
3.3\,mrad (1.5\,mrad).
The actual zero-degree horizontal targeting angles were $-0.52$\,mrad in
the 1997 run and $+0.09$\,mrad in the 1999 
run.\footnote{See Fig.~\ref{fig:precession_geo} for an explanation of 
the sign of the targeting angle.}
The actual zero-degree vertical targeting angles were very close to zero.
The rather large targeting angle in the 1997 run was
due to imperfect positioning of the target relative to the entrance
axis of the collimator.
The maximum angular deviation from the nominal targeting angles, limited
by beamline apertures, was 0.48\,mrad (0.26\,mrad) in the horizontal
(vertical) direction.
The average angular divergence of the beam was considerably less.

The position and shape of the primary beam were measured 
using eight segmented-wire ion chambers (SWICs) positioned 
at various points along the beamline.
Each SWIC had one plane of vertical and one of horizontal wires,
each with 48 wires spaced 1.0\,mm apart, 
except for the two SWICs closest to the target (MC6WC1 and MC6WC2), 
which had a 0.5\,mm wire pitch. 
The integrated charge accumulated on each wire was digitized and 
read out every 4\,s during the beam spill and displayed graphically
on video monitors at the end of each spill to facilitate beam
monitoring and tuning.
The charge profiles from five SWICs (MC00WC, MC2WC, MC5WC, MC6WC1, and MC6WC2)
were also recorded spill-by-spill by the SlowDA 
(described in Sec.~\ref{sec:daq_slow}). 
The center of each SWIC with respect to the nominal beam centerline was
determined by survey.  
The position of the target with respect to the SWICs was determined 
empirically by moving the target (between spills) in steps, 
either up-down or left-right, and finding the target position 
at which the interaction rate was maximized.
The charge profiles from MC6WC1 and MC6WC2 (also called the target SWICs) 
were fitted offline spill-by-spill to find the $x$ and $y$ coordinates
of the beam centroid at the target $z$ center as well as the targeting angles 
of the beam with respect to the $z$ axis. 
The $z$ distance between the target SWICs was 2.5\,m, 
with MC6WC2 located 0.22\,m upstream of the center of the target.

Due to sharing of charge across adjacent wires, 
the SWICs did not accurately measure the true size of the beam. 
The beam size and shape were determined from target scans (as just described),
in which  beamline settings were held fixed while the target was moved in
horizontal or vertical steps.  
The fraction of the beam hitting the target was reflected in
the counting rates per incident proton of various counters in the spectrometer.
One such ratio is plotted versus target location in Fig.~\ref{target_scan}.
Since the target size was well known, these distributions allowed
the beam size and shape to be determined.  
Using this method it was found that the beam at the target was 
approximately Gaussian in shape, typically with $\sigma_x = 0.45{\pm}0.08$\,mm
and $\sigma_y = 0.38{\pm}0.08$\,mm.  
For comparison, the shape of the beam as measured by the most downstream
SWIC (MC6WC2) was Gaussian with $\sigma_x = \sigma_y = 1.8$\,mm.

\section{Target Assembly}
\label{sec:target}
To produce positive and negative secondary beams of comparable 
intensities, two targets of different lengths were used. 
To minimize acceptance differences and to reduce attenuation 
of the produced particles, the targets were kept short. 
Copper targets of 20\,mm and 60\,mm lengths were used to produce respectively
the positively and negatively charged secondary beams.\footnote{In the 1997
run the positive polarity target length was 22\,mm.} 
The transverse dimensions of both targets were 
2\,mm${\times}$2\,mm.\footnote{In the 1997 run special runs were
taken for cross-section measurements with Be, Cu, and W targets
of 4\,mm${\times}$4\,mm cross section and 20\,mm length.} 
The positive- and negative-secondary-beam rates per proton on target
differed by less than 5\% (see Fig.~\ref{fig:rate_comp_beam}).\footnote{Note
that the Beam Hodoscope used to measure the secondary-beam intensity
was slightly undersized and hence the rates given in 
Fig.~\ref{fig:rate_comp_beam} are somewhat underestimated.}

The targets were supported by an assembly (Fig.~\ref{fig:target}) 
constructed from two sets of 1-mm-thick Kyocera alumina-ceramic plates,
each containing laser-drilled~\cite{ceramic} circular or semicircular holes,
within which the targets were clamped.  
The two targets, located 24\,mm apart horizontally, were each secured 
at two locations along their length between a central ceramic plate and 
a ceramic edge strip.  
Additional holes served as ``empty'' targets used for background studies 
and (at an early stage of the experiment) to hold additional targets 
used in run-condition optimization studies and cross-section
measurements.
The ceramic plates were bolted onto horizontal 8-mm-thick aluminum 
bars at the top and the bottom of the target holder. 

The target holder was mounted on a precision manipulator that was remotely
controlled to move in the vertical and horizontal directions in very fine
steps.\footnote{A 1\,mm horizontal (vertical) displacement
corresponded to 1862 (1621) stepping-motor counts.} 
To prevent the wrong target from being used for a given beam polarity, 
signals from position sensors in the target system were input to the 
primary beam safety-interlock system.  
Once the proper target was in place, it could be moved from its 
nominal position in the horizontal direction within a range of 
only $\pm$3.5\,mm without disabling the beam.
The targets could be repositioned reproducibly to within a few $\mu$m.
During normal operation, the temperature of the target was kept
below 900\,K using a blower delivering 7\,m$^3$/min of forced air.

\section{Collimator and Hyperon Magnet}
\label{sec:col}
A 6.096\,m long curved collimator installed within the
Hyperon Magnet channeled a charged secondary beam of modest 
momentum spread from the target to the spectrometer (Fig.~\ref{fig:channel}).
The upstream face of the collimator was 0.292\,m downstream of the
target centers; the targets nominally were centered horizontally and vertically
on the entrance aperture of the collimator.
The collimator, Hyperon Magnet, and surrounding shielding
also served as the dump for those primary-beam protons 
that did not interact in the target. 
Due to the width of the Detector Hall and height of the primary beam above the
floor, the charged secondary beam had to be deflected 
upwards.\footnote{Horizontal deflection by the Hyperon Magnet would have 
required some detectors to extend beyond the walls of the MC7 enclosure, 
while vertical deflection in both the Hyperon and Analyzing Magnets would 
have required some detectors to extend through the ceiling or into the floor. 
Thus the only feasible configuration was for the Hyperon Magnet to 
deflect upwards and the Analyzing Magnets horizontally.}  
The entire dipole-magnet/collimator assembly was encased in massive iron and
concrete shielding to reduce radiation in the surrounding area.

A particle with a trajectory along the center of the collimator
channel (central-orbit trajectory) exited at an angle of 19.51\,mrad 
above the horizontal.
The center of the downstream aperture of the collimator exit defines the origin 
of the Spectrometer Coordinate System, which is the coordinate
system used to describe the spectrometer elements.
The $z$ axis of the Spectrometer Coordinate System coincides with the
central-orbit direction at the collimator exit, and hence is at 
an angle of 19.51\,mrad to the horizontal (see Fig.~\ref{fig:precession_geo}).
To describe the collimator itself, the Beam Coordinate System is used. 
It is position dependent, with its $z$ axis tangent to the 
central-orbit trajectory within the collimator, and hence horizontal at
the collimator entrance and inclined at 19.51\,mrad at its exit.

Although most of the data were taken with the primary proton beam
incident at an angle of zero degrees with
respect to the entrance axis of the collimator, a significant amount
of data at nonzero targeting angles was also recorded
in order to produce polarized \gXi\ hyperons.
These data were taken with the incident proton beam deflected 
in the horizontal direction and hence lying in the $x$--$z$ plane in the
Beam Coordinate System, as shown in Fig.~\ref{fig:precession_geo}.
A positive (negative) targeting angle is defined to have 
$\hat{p}_{pb}{\times}\hat{p}_{sb}$
in the $+\hat{y}$ ($-\hat{y}$) direction, 
where $\hat{p}_{pb}$ is the primary-proton-beam direction 
and $\hat{p}_{sb}$ is the secondary-beam direction at the 
entrance of the collimator.
Figure~\ref{fig:precession_geo} shows the direction of the \Xim\
polarization for the two targeting angles as well as the direction of
precession.
The polarization precessed around the magnetic field direction,
and at the exit of the collimator in the Spectrometer Coordinate System,
the precessed polarization of 
\Xim's produced with a positive (negative) targeting angle 
was at an angle of about $10^{\circ}$ from the negative (positive) $y$ axis,
in the $-z$ ($+z$) direction.

As shown in Fig.~\ref{fig:channel}, the collimator
was made up of five segments joined together with steel dowel pins. 
Each segment consisted of a block of brass or
tungsten with an attached cover plate. For ease of fabrication, within each
block the arc of the channel was approximated by a series of 
straight-walled milled slots, each such slot 304.8\,mm in length. 
The dimensions of the slots of each segment are  given in 
Table~\ref{tab:channel}.
The outer transverse dimensions of the segments were 46.0\,mm (width) 
by 95.2\,mm (height), with the corners along the beam direction rounded 
to fit into the beam pipe inside the Hyperon Magnet.
Segment B (Defining Collimator) and Segment E (Exit Collimator) 
were made of tungsten.
The Defining Collimator served as the primary beam dump.
At the design magnetic field, the 800\,GeV/$c$ proton beam was dumped 
3.6\,mm (6.4\,mm) below the bottom edge of the rectangular 
aperture at the upstream end of the Defining Collimator when a 
positive (negative) secondary beam was selected.
The Exit Collimator provided a second iris to clean up the secondary beam.
Segment A and the tungsten segments (B and E) were water-cooled.
With respect to the center of the entrance aperture  of Segment A, 
the solid angle subtended by the limiting (downstream) aperture of 
the Defining Collimator was approximately 5.38\,$\mu$sr.

To center the target on the entrance aperture of the collimator,
the target was stepped through a series of $x$ and $y$ positions, 
with the proton beam centered on the target at each position.
The \gXi\ momentum and position distributions in the spectrometer
for each target position were compared with those generated by Monte Carlo 
simulations.
It was not possible to directly check the resulting position
of the target relative to the collimator entrance, as the
target area was inaccessible.  The target center
should have been $x = 0.0$\,mm and $y = 65.16$\,mm in the
Spectrometer Coordinate System.  The actual target positions,
determined by tracing reconstructed \gXi\ trajectories
back through the collimator, are given in Table~\ref{tab:tgt_position}.
Alignment errors in 1997 caused the target to be mispositioned
by $+4.6$\,mm in $x$ and $-2.3$\,mm in $y$.  In 1999 the
target positioning was much improved, with the target center
at the nominal position within a fraction of a millimeter in both
$x$ and $y$.

The Hyperon Magnet within which the collimator was placed
was an 11\,455\,kg B2 dipole magnet \cite{magnets} fabricated at 
Fermilab for the Main Ring of the accelerator. 
Its exterior dimensions were 0.362\,m (height) by 0.641\,m (width) by 
6.071\,m (endplate to endplate). 
It was installed on its side to produce a field in the horizontal 
(${\pm}\hat{x}$) direction.The charge of the secondary beam was changed by 
reversing the Hyperon Magnet field direction:
when producing a secondary beam of positive (negative) polarity,
the magnetic field was in the $+\hat{x}$ ($-\hat{x}$) direction.
A current of 4193\,A produced a field of 1.667\,T.
At this field the central-orbit momentum was 156.2\,GeV/$c$.
After the collimator was installed in the beam pipe of the Hyperon Magnet, 
the beam pipe was sealed at each end with a 76.2-$\mu$m-thick Kapton window.
To minimize scattering of the secondary-beam particles, 
the beam pipe was filled with flowing helium gas at atmospheric pressure.

The acceptance of the collimator, defined as the fraction of charged 
particles from the target that cleared the Defining Collimator and
emerged from the exit of the channel, is shown as a function of 
momentum in Fig.~\ref{fig:chanacc}.\footnote{This calculation 
assumed a transverse-momentum ($p_T$) probability distribution 
of the form  $\exp{(-p_T^2/2\sigma^2)}$, with $\sigma= 1\,$GeV/$c$.}
Momentum spectra of the reconstructed $\Xi^-$ and $\overline{\Xi}{}^{+}$ 
hyperons from the 1999 run are shown in Fig.~\ref{fig:xivsxib};
the mismatch is due to the different mechanisms for producing a 
$\Xi^-$ and a $\overline{\Xi}{}^{+}$ in proton-nucleon collisions.

The magnetic field of the Hyperon Magnet was monitored
by two miniature LPT-141-20-S Hall probes~\cite{group3}. 
These were located 0.89\,m upstream of the exit of 
the collimator and centered in $y$ within the milled slot in Segment E 
shown in Fig.~\ref{fig:channel}. 
The Hall probes were read out by the SlowDA 
(described in Sec.~\ref{sec:daq_slow}) via serial CAMAC using 
a DTM-141 teslameter~\cite{group3}.
Figure~\ref{fig:b2field} shows the spill-by-spill average 
Hall probe readings, as well as the magnet currents,
for all of the 1999 positive- and negative-polarity beam spills.
The magnitudes of the currents were almost identical between the two 
polarities:  for both the 1997 and 1999 runs, the fractional difference 
(averaged over all spills) between the 
positive- and negative-secondary-beam magnet currents was just 
$0.48{\times}10^{-4}$.
The currents were quite stable:  the rms deviation
was 2.6\,A (2.8\,A) for the 1999 (1997) run.
Unfortunately, the two Hall probes did not exhibit consistent readings:
Probe~2 had a 15\,G (91\,G) higher reading than Probe~1 in the 1999 run
for positive (negative) secondary-beam polarity, 
and a 20.2\,G (59.2\,G) higher reading in the 1997 run. 
This is because both probes had a zero-field offset, 
and Probe~1 was also most likely somewhat misaligned
relative to the field direction.\footnote{Because the collimator
is still radioactive, direct confirmation of the misalignment has
not been possible.}
Because the Hyperon Magnet was buried under considerable shielding,
after the initial installation it was impossible to access the 
Hall probes to recalibrate and realign them.

Since both probes measured the same field --- B2 magnets have
a very uniform field --- correlations in the two probe readings were
used to measure the intrinsic ability of the Hall probes to follow
changes in the Hyperon Magnet field.  The rms resolution of the probes for
tracking such changes was thus found to be 0.8\,G in 1997 and
2.0\,G in 1999.

\section{Vacuum Decay Region}
\label{sec:decayvol}
The parent- and daughter-particle decays of interest were required by the 
analysis software to occur in a fiducial volume within an 
evacuated decay pipe (Vacuum Decay Region).
The 13.005\,m long pipe was made up of three straight sections joined 
together with flanges and a bellows. 
The upstream, aluminum pipe was  2.195\,m long and had an inner diameter of 
0.254\,m and a wall thickness of 19.05\,mm. 
Its upstream window was titanium, 0.102\,m in diameter 
and 76.2\,$\mu$m thick, and located 0.318\,m downstream of the collimator exit 
(about 45\,mm downstream of the Hyperon Magnet beam-pipe exit window).
The middle segment was a 7.849\,m long aluminum pipe of
inner diameter 0.305\,m and wall thickness 19.05\,mm. 
The last segment, 2.657\,m in length, was coupled to the middle segment by a 
0.305\,m long, 0.305\,m diameter bellows.\footnote{By removal of the bellows,
space could be made available at the downstream end of the 
vacuum pipe for installation of an optical discriminator~\cite{optical_trig}, 
as was done for special trigger studies during the 1997 run.}
This steel pipe had an inner diameter of 0.584\,m and wall 
thickness of 25.4\,mm.
The downstream window of the Vacuum Decay Region was made of 0.508\,mm thick 
Kevlar~\cite{Kevlar} supporting a 0.127\,mm thick Mylar sheet
coated with 0.100\,mm of Al.

The decay pipe was inclined at an angle of 19.51\,mrad to the horizontal,
so as to lie along the central-orbit line.
It was evacuated to below 1\,mTorr pressure. 
Monte Carlo simulations indicated that the hyperon decays of 
interest were contained laterally within the inner diameter of the pipe with 
essentially 100\% probability.

\section{Analyzing Magnets}
\label{sec:analmags}
The momentum of each charged particle passing through 
the spectrometer was determined using sets of multiwire proportional 
chambers situated upstream and downstream of a pair of dipole magnets 
(Analyzing Magnets).
Each Analyzing Magnet, of the Argonne National Laboratory BM109 
type~\cite{BM109}, weighed about 46\,000\,kg.
The exterior dimensions of the steel magnet cores were 2.387\,m (width) 
by 1.321\,m (height) by 1.829\,m (length).
Figure~\ref{fig:bm109_front} shows a front view of the upstream magnet.
The apertures of the Analyzing Magnets were 
0.610\,m wide by 0.259\,m (0.305\,m) high in the upstream 
(downstream) magnet.
(For convenience, the upstream and downstream magnets are also
referred to as Magnet~1 and Magnet~2, respectively.)
The magnets were tilted upward at the 19.51\,mrad central-orbit angle.
The two magnets were separated in the beam direction by approximately 
0.071\,m between the face plates and 0.67\,m between the steel cores.
The upstream (downstream) Analyzing Magnet was run at a current of
about 2450\,A (2469\,A), producing a magnetic field of 
1.345\,T (1.136\,T).

Before the multiwire proportional chambers (described next) were installed, 
the Analyzing Magnet field profiles were determined on a three-dimensional grid 
using the Fermilab ``Ziptrack'' system~\cite{ziptrack}.
The three components of the magnetic field were measured in 
${\approx}$25\,mm steps in $x$ and $z$ and ${\approx}$100\,mm steps in $y$ 
using three orthogonal Hall probes mounted on a computer-controlled cart 
that ran through the magnet apertures on an aluminum I-beam.
The magnetic field was sampled at 20\,382 lattice points.
The major component of the magnetic fields at $x = 10$\,mm, $y = 14$\,mm 
is plotted versus $z$ in Fig.~\ref{fig:byfield}.

The magnetic field of each Analyzing Magnet was measured during the run 
using the same-model Hall probes (LPT-141-20-S) as for the Hyperon Magnet;
unlike those in the Hyperon Magnet, these probes worked well.
The Hall probes were placed in a region of relatively uniform field near 
the entrance apertures.
During each beam spill, they were read out five times, 
and (except for a few-minute period following each reversal of field polarity) 
the magnetic fields were found to be quite stable.
To maintain equal acceptances for hyperon and antihyperon decays,  
when the charge of the secondary beam was reversed by reversing the 
direction of the  Hyperon Magnet field, the direction of the 
Analyzing Magnet fields was also reversed. 
When running with positive (negative) secondary-beam polarity,
the magnetic fields were oriented in the $-\hat{y}$ ($+\hat{y}$) direction.

It was important that the magnitudes of the magnetic fields be
the same for positive- and negative-secondary-beam running,
in order to minimize potential biases in the \CP-violation analyses.
Differences between the magnitudes of the Hall-probe readings
for the two secondary-beam polarities were indeed small,
as shown in Fig.~\ref{fig:ana1_field} and Fig.~\ref{fig:ana2_field}. 
The difference between the positive- and negative-secondary-beam 
polarity Hall-probe values for all 1999 runs, 
averaged on a spill-by-spill basis,
was $-3.3$\,G ($-0.9$\,G) for Magnet~1 (Magnet~2); 
in 1997 the same difference was $-2.6$\,G ($+2.6$\,G).
Spill-to-spill variations of the Hall-probe
readings were also small: the rms deviation of all spills
was 4.3\,G (4.7\,G) for Magnet~1 (Magnet~2) in
1999, and about a factor of two greater in 1997.

\section{Multiwire Proportional Chambers}
\label{sec:MWPC}
Charged particles emerging from the Vacuum Decay Region were tracked
by nine high-rate, narrow-pitch multiwire proportional chambers (MWPCs).
Four MWPCs (C1--C4) were deployed upstream of the Analyzing Magnets, 
five (C5--C9) downstream.\footnote{In the 1997 run
chamber C9 was not used.}
Narrow-pitch MWPCs were used to keep the individual wire rates in the
secondary-beam region less than 1\,MHz; the pitch was increased by
roughly 25\% after
every second MWPC in order to keep the maximum wire rate approximately the
same in each chamber.

To accommodate the increasing spread of the hyperon and kaon decay products, 
the nine chambers were of four successively larger sizes,
with the additional constraint that the MWPCs fit within
the dimensions of the preexisting experimental enclosure.  
Since two (and in one case, three) chambers were built from each design, 
and  each chamber was installed at a different $z$ position, 
some chambers were oversized;
consequently, not all wires were instrumented.  
Of 24\,096 total wires, 19\,680 were instrumented.

\subsection{Physical Description}
\label{sec:mwpc_des}
The MWPC mechanical data are given in Table~\ref{tab:pwc_param}.
Figure~\ref{fig:mwpc_c1} shows a front view of C1, which was identical to C2, 
and similar to the identical pair, C3 and C4, except smaller.
Figure~\ref{fig:mwpc_c5} shows a front view of C5, which was identical to C6, 
and similar to the identical triplet, C7--C8, except smaller.
The upstream chambers were all mounted on stands fabricated from
Unistrut~\cite{unistrut} by means of aluminum extensions attached to the
chamber front and back plates.  The chamber heights and orientation
were fine-tuned by means of nuts on threaded rods that connected the 
stand to the aluminum extensions.
Stands for the rear chambers had winches that allowed those heavy 
chambers to be hoisted up (Fig.~\ref{fig:c7_front_stand}).  
The chambers were hung from threaded rods.

All the chambers were constructed similarly.  Each MWPC had four anode-wire
planes sandwiched by foil cathode planes (see Fig.~\ref{fig:pwc_planes}).
The four anode planes had wires oriented in three stereoscopic views. 
The outer two anode planes ({X} and {X$^\prime$}) had vertical 
wires, which measured coordinates in the $x$ direction. 
To enhance momentum resolution, the X wires were offset from those in 
X$^\prime$ by one-half of the wire spacing. The two inner anode planes 
({U} and {V}) had wires inclined at $\pm 26.57^\circ$ 
($= \tan^{-1}(1/2)$) with respect to vertical.  
All wires were gold-plated W-Rh.
Two outer, grounded, foil planes terminated the field region and 
provided for balancing of electrostatic forces.  
Except in the case of C9, the foils were 25\,$\mu$m thick Kapton
with a 120\,nm thick vapor-deposited Au layer on both sides.\footnote{In 
C9, 23\,$\mu$m thick conductive Kapton was used.}
The stack of planes was clamped between 12.7\,mm thick aluminum jig plates.
In the larger chambers, C5--C9, the back jig plate was reinforced by a 
box beam.
The chamber gas volume was sealed by windows composed of a laminate of 
25\,$\mu$m Mylar and 25\,$\mu$m Al with a 12\,$\mu$m layer of adhesive.  
The total material in each chamber was about 0.07\% of an interaction length 
and 0.2\% of a radiation length.

The anode and cathode frames were made of fiber-epoxy laminates, 3\,mm thick:
Stesalit 4411W~\cite{stesalit} for the front chambers and
FR4 for the rear chambers.  The 4411W was produced 
with a 25\,${\mu}$m (${\approx}$1\%) tolerance on the nominal thickness.  
The FR4 was ordered oversized and then ground down to the desired 3\,mm 
thickness with the same 25\,${\mu}$m tolerance.  Copper-clad FR4
was used for the anode frames in the rear chambers.
In the upstream chambers the total wire tension did not
produce any appreciable distortion of the anode frames.
That was not the case for the downstream chambers, which had many more
wires, each with a greater tension.  Hence, to eliminate relaxation of the
anode wire tension due to distortion of the anode frame, 
in the downstream chambers the anode (and cathode) frames were mounted 
on dowel pins that were rigidly fixed to the back jig plate.

The anode wires were glued onto the anode frames using Shell 
Epon~815 Resin and V-40 hardener~\cite{shell}.
In the upstream chambers, pockets were milled out of the anode frames
into which circuit boards were glued.  The anode wires were soldered
to traces on the circuit boards that carried the signals to preamplifiers
that were plugged into connectors soldered to the circuit boards.
The X and U planes had preamplifiers facing the same direction, 
whereas the X$^\prime$ and V planes had preamplifiers facing the opposite
direction.
The copper-clad anode frames of the downstream chambers had
traces that were etched out and onto which
the wires and preamplifier connectors were soldered.  
Unlike the upstream chambers, all of the preamplifiers faced forward,
which was facilitated by having the X$^\prime$ and U anode planes 
extend beyond the X and V anode planes.  
Card cages attached to the front and back chamber plates held
circuit boards that distributed power to the preamplifiers.

Upstream of the Analyzing Magnets, the hyperon and kaon decay products 
occupied the same region in the MWPCs as the secondary beam, 
while in the downstream MWPCs the decay products progressively 
separated from the beam. 
To reduce the sensitivity to out-of-bucket beam tracks, and to mitigate
aging effects~\cite{Solomons}, C1--C4 were therefore filled with a 
``fast'' gas mixture of CF$_4$/isobutane in a 50/50 ratio
(see Fig.~\ref{fig:pwc_speed}). 
Since a somewhat poorer time resolution could be tolerated in  C5--C8, 
in the 1997 run those chambers were filled with a 50/50 mixture of 
argon/ethane bubbled through isopropyl alcohol at $-0.7^\circ{\rm C}$.  
In the 1999 run, the downstream chambers also were operated with the fast-gas 
mixture, with the exception of C7 which remained filled with 
argon/ethane since it could not tolerate the higher voltage needed 
for operation with the fast-gas mixture.  
Gas flow rates corresponded to approximately three volume changes per day.
The chamber cathode-plane voltages were provided by Fermilab ES-7109
(``Droege") power supplies.
Typical operating voltages are given in Table~\ref{tab:pwc_param}.

Due to concern about possible aging effects, the signal pulse heights of 
selected wires in the beam region of a few MWPCs were monitored 
periodically throughout the 1997 and 1999 runs with a $^{55}$Fe X-ray
source.  No significant changes were observed.

\subsection{MWPC Electronics}
\label{sec:mwpc_readout}
Due to the short time available after approval of the experiment 
in which to assemble the apparatus, the availability of a large, 
high-speed latch system from a previous experiment, 
and the desire to separate the preamplifier from the discriminator
for noise reasons,
the MWPC signal-processing and readout chain was implemented as 
three distinct units, rather than a fully integrated on-chamber system.

The high rates demanded a fast, high-gain, low-noise preamplifier for the
wire chambers.
Since only ${\approx}10\%$ of the total avalanche charge
is collected in 10\,ns, the typical shaping time of an amplifier
appropriate for such high-rate conditions,
the expected signal was only ${\approx}5\times 10^4 e$, 
assuming an avalanche gain
of $10^5$ and an initial ionization of $5\,e$ (${\approx}2$ clusters).
The threshold of the amplifier/discriminator combination was specified to be
less than one-fifth of this average signal charge, or $10,000\, e$,
and the equivalent noise charge (ENC) of the amplifier
had to be sufficiently low that the random (thermal) noise rate 
at this threshold would not cause excessive hit multiplicity.
After a lengthy evaluation period a commercially available preamplifier 
ASIC (LeCroy MQS104A \cite{LeCroy} with four preamplifiers per chip) 
was chosen because of its ability
to meet specifications, low cost, and small footprint. 
Characteristics of the preamplifier are given in Table~\ref{tab:preampspecs}. 
Since the LeCroy MQS104A chip was designed to be used in an 
on-chamber system, a cable driver for the preamplifier
card was designed (Fig.~\ref{fig:preamp_schematic}).
The preamplifier card was a compact four-layer surface-mount board
of 16 channels. 
A total of 21\,600 preamplifier channels were mounted on 1350 circuit boards.
Before the boards were loaded each channel of the LeCroy MQS104A
chips were tested for gain, noise, and time response.  
Approximately 5\% of the chips failed and were replaced by LeCroy.
Each channel of the assembled boards was tested again, and boards with
failed channels were repaired.  The results of these tests are shown in
Fig.~\ref{fig:preamp_test}.

The preamplifiers were mounted on the chambers at the ends 
of the anode fanout-circuit-board traces.  
The amplified analog signals were conveyed through ${\approx}$10\,m of 
twisted-pair flat cable to discriminator/delay cards located in racks 
in the experiment enclosure. 
Characteristics of the discriminator cards are given in 
Table~\ref{tab:discrspecs}.
About 100\,ns of lumped-constant delay was incorporated on the cards to allow 
for trigger latency.  
The binary output signals from the discriminators were conveyed on 
${\approx}$60\,m of flat cable to a high-speed latch system
situated in the Electronics Hall and described in Sec.~\ref{sec:daq_fast}.

In such a large system of several thousand high-gain amplifiers,
attached to sense-wire ``antennae'' and packed together in close proximity, 
the possibility of coherent oscillation is always a concern.  
Initially the system was found to be marginally stable.  
Oscillations could be initiated by a large number of amplifier channels 
``firing'' simultaneously, that is, due to a momentary 
burst of high-intensity beam or external electromagnetic interference. 
Attempts to control oscillations by shielding amplifier cards or 
output cables were not successful.
In the end stability was achieved by establishing a very-low-impedance 
ground connection using copper foils between the amplifiers and the 
aluminum jig plates. 
With this grounding procedure the entire system of 
19\,680 instrumented wires could be run at a threshold corresponding to an 
anode signal of about $15\,000\,e$.
After about nine months of operation several preamplifier cards on one of the 
large chambers became unstable again.  Eventually this was cured by adding  
ferrite cable clamps on the output cables of a handful of suspect amplifier 
cards.

\subsection{MWPC Performance}
\label{sec:mwpc_perfor}
The MWPCs performed well over the course of the experiment.
During the 1997 and 1999 runs only a handful of wires broke,
resulting in little downtime.
The average MWPC efficiencies were quite high; even in the
intense secondary-beam region they were about 99\% (see
Fig.~\ref{fig:pwc_eff}).
Differences in efficiency between negative and positive 
secondary-beam-polarity running were minimal.
Although the narrow pitch and use of the fast-gas mixture in the
1999 run resulted in good time resolution at the chambers,
timing differences within the cables used to carry the signals
to the latch system, as well as cable-to-cable timing differences,
degraded that timing by 10--20\,ns.
Nevertheless, most of the out-of-time tracks were 
confined to the secondary-beam region, where they caused
minimal trouble for the trackfinding program.

\subsection{MWPC Alignment}
\label{sec:mwpc_align}
Much effort was taken to carefully measure the positions and 
orientations of the wire chambers.
The data used for the chamber
alignment came from special chamber-alignment runs that were
taken periodically.  In these runs the calorimeter was moved
out of the envelope of the non-deflected secondary beam,
and the currents of the Analyzing
Magnets were set to zero in such a manner that any residual
magnetic field due to hysteresis effects was less than 1\,G,
as measured by the Hall probes.\footnote{In the 1997 run the
currents were set to directly to zero, resulting in larger
residual fields, and hence opposite-polarity straight-through
runs were combined for the alignment analysis.}
A special scintillation counter, S45, which covered the entire extent of the
secondary beam, was placed just downstream of the Vacuum Decay Region,
and was used for the alignment trigger.\footnote{The S45 counter
was 305\,mm wide by 254\,mm high and read out by photomultiplier
tubes on each end.}

The X and X$^{\prime}$ plane offsets were first measured.
The two planes were supposed to be offset by one-half the wire spacing,
but in fact were occasionally off from the nominal values.
The stereo angles of the U and V planes, nominally ${\pm}26.57^{\circ}$
with respect to the vertical, were also determined.
Finally, searches were made for discontinuities in the wire spacings.
These occurred in the downstream chambers because their anode planes 
were too large to have all the wires laid down at one time,
hence the wires were mounted in groups.
Imperfect positioning of the groups led to small discontinuities. 

After the internal chamber alignment was completed,
the chambers were all aligned relative to each other.
This was done using the chamber $z$ positions as determined by survey.
A special version of the track reconstruction program
(see Sec~\ref{sec:track}) was used in which the chamber in question was
taken out of the fit.\footnote{Only chambers C1--C8 were aligned
in this manner.  The C9 alignment was done, after the other chambers
were aligned, by projecting tracks to it from the other eight chambers.}
After the relative alignment was completed the chambers were
positioned properly in the Spectrometer Coordinate System as follows.
First they were moved together by the same amount, preserving their relative
alignment, in such a manner that at the $z$ position of the collimator
exit (the origin of the Spectrometer Coordinate System) the mean
track position was centered on zero in both $x$ and $y$.
Then chambers were rotated en masse, again preserving their relative alignment,
about the origin of the Spectrometer Coordinate System, so that:
(1) straight-through tracks had an average $x$ slope of zero,
and (2) 156.2\,GeV/$c$ momentum tracks (the central-orbit momentum)
had an average upstream $y$ slope of zero.

The chambers were mounted vertically, and hence at a 19.51\,mrad
angle to the $z$ axis of the Spectrometer Coordinate System.
Rotations from the nominal chamber orientation were also determined
in the alignment procedure and were used by the track finding program.
After the chambers were aligned, the $x$ and $y$ positions, 
relative to the wire chambers, of all of the scintillation counters, 
the proportional tubes of the Muon Stations, and the Hadronic Calorimeter, 
were determined using charged particle tracks.

\section{Triggers}
\label{sec:triggers}
The HyperCP triggers were designed to have high efficiency, to be simple and 
fast, and to have single-bucket (18.9\,ns) time
resolution.\footnote{Single-bucket resolution was achieved for all 
except the muon triggers, for which, due to the low rate, the timing 
requirement was considerably more relaxed.}
Although 16 distinct triggers were used in the experiment
(Table~\ref{tab:triggers}), most were for monitoring purposes. 
We confine our discussion here to the main physics triggers.

The main physics processes considered in designing
the triggers are listed in Table~\ref{tab:trig_phys}.
These all produce at least one decay particle of opposite charge to
the secondary beam and two decay particles with the same charge as 
the secondary beam, with the decay particles all having substantially less
momentum than particles in the secondary beam.
Hence the main physics triggers all required a basic
left-right coincidence of charged particles in
hodoscopes situated far enough back in the spectrometer
that the decay particles had been swept away from 
the intense secondary beam by the magnetic fields
of the Analyzing Magnets.
The major backgrounds to this Left-Right trigger (LR)
were due to interactions of the secondary beam with the
Vacuum Decay Region windows, material
in the spectrometer, and particle production in the walls of the collimator.
These backgrounds were sufficiently high --- with about 1.5\% of an
interaction length of material in the spectrometer
the interaction rate was about 200\,kHz --- that further
reduction of the LR trigger rate was needed.
For nonmuonic decay modes this was provided 
by requiring a minimum amount
of energy deposited in the Hadronic Calorimeter
situated on the right side of the spectrometer.
For muonic decay modes this was provided by 
requiring the presence of penetrating particles in the Muon Stations.

The triggers used to select events for the three {\em CP}-violation-search 
modes,
\Xim\ra\gL\pim\ra\gp\pim\pim, \Km\ra\pim\pip\pim, and \Omm\ra\gL\Km,
were kept as simple as possible in order to minimize potential biases.
Fortunately, these simple triggers provided adequate rejection, 
so that higher- (second- and third-) level triggers were not needed.
Such higher-level triggers would have been difficult to implement
given the severe time constraints involved, 
and would have had the potential for uncorrectable biases.
A consequence of the decision to use a somewhat
loose trigger was the need for a high-rate data-acquisition system.

The detectors used to form the triggers were
the Same-Sign (SS) and Opposite-Sign (OS) Hodoscopes,
the Hadronic Calorimeter, and the Left and Right Muon Hodoscopes.
The Hadronic Calorimeter, described in detail
in Sec.~\ref{sec:cal}, was situated on the right (OS)
side of the spectrometer just outside of the halo of the secondary beam.  
Its purpose was to reduce the trigger rates for
nonmuonic decays by requiring a substantial energy deposit ($>45$\,GeV).
The Left and Right Muon Hodoscopes, described in detail in 
Sec.~\ref{sec:muon}, 
were at the back of the Left and Right Muon Stations. 
Each was composed of two banks of scintillation counters, 
one with 15 vertical counters, 
the other with 10 horizontal counters.

\subsection{SS and OS Hodoscopes}
\label{sec:hodos}
The two hodoscopes that formed the basis of all the 
HyperCP physics triggers were placed on either side of the secondary beam.
The Same-Sign Hodoscope was located 41.1\,m from the exit of the collimator 
and covered $x = 0.215$ to $2.195$\,m (Fig.~\ref{fig:ss_hodo}). 
The Opposite-Sign Hodoscope was located 48.4\,m from the exit of the collimator
--- the usually more energetic OS particle took farther to separate from
the secondary beam ---
and covered $x = -0.123$ to $-1.267$\,m (Fig.~\ref{fig:os_hodo}).
Both hodoscopes were centered roughly vertically on the central-orbit line.

The SS and OS Hodoscopes each had 24 scintillators, 680\,mm long by 90\,mm wide.
The scintillators differed only in thickness:  the SS counters were 20\,mm 
thick and the OS counters 10\,mm thick.
Each hodoscope had two planes of 12 counters each, the SS (OS) planes
separated by 41.3\,mm (69.9\,mm) in $z$, with the counters staggered
such that gaps in one plane were centered on counters in the other.
In each plane of the SS Hodoscope, adjacent counters were separated 
from their neighbors by 70\,mm, giving an overlap of 10\,mm at each edge.
In each plane of the OS Hodoscope, adjacent counters butted up
against their neighbors, giving an overlap of one-half the counter width.
Zero gap widths for the OS Hodoscope counters were used because
only one opposite-sign particle impacted the OS hodoscope for 
the \CP-violation decay modes, where high trigger efficiencies were essential.
Larger gap widths for the SS Hodoscope counters were tolerated because
there was a large probability that both same-sign particles would
impact the hodoscope for the \CP-violation decay modes.
Hence, there was a measure of redundancy in both 
hodoscopes.\footnote{In the 1997 run the OS Hodoscope was identical
to the SS Hodoscope, except that it had 16 rather than 24 counters.
It was modified for the 1999 run to increase the trigger 
efficiency for protons from the \CP-violation decay modes and
to allow the counter efficiencies to be determined from 
reconstructed \gXi\ events.}
The SS and OS counters were sufficiently long that no particle coming from
the target and passing through the Analyzing Magnets would
pass either above or below them.

Both hodoscopes were mounted on stands that were open to the secondary beam,
and that rigidly held the counters in place.
The stands were fabricated largely out of Unistrut \cite{unistrut}, 
with each counter fixed precisely in position on an aluminum plate.
The SS hodoscope was mounted on a rectangular frame that
hung from a stand, and that was oversized in order to allow
the secondary beam and its decay products to pass through
undisturbed.\footnote{The support pole in
front of the SS hodoscope shown in Fig.~\ref{fig:ss_hodo} was
pre-existing and could not be removed.}
Rollers enabled lateral ($x$) positioning;
vertical ($y$) adjustments were done by means of threaded rods attached to 
the rollers.
The OS hodoscope was mounted on a C-shaped frame, open to the
secondary beam, that was fastened to
a wheeled carriage that provided lateral motion.
Limited vertical adjustment was done by the means of screws on the
fixtures that mated the C-shaped frame to the carriage.

Bicron BC-408 scintillator~\cite{Bicron} was used for the SS counters,
and Bicron BC-404, which is slightly faster (2.2\,ns versus 2.5\,ns 
full-width-at-half-maximum),
for the OS counters.  The scintillators were diamond-cut.
They were wrapped in Tyvek~\cite{Tyvek} --- except for the far end 
which was coated with black
electrical tape to eliminate reflections in order to improve
the time response --- and wrapped again in 
Tedlar~\cite{Tedlar}.
Each scintillator was glued to a 100\,mm long tapered light guide.
A 25\,mm diameter acrylic disk with a slot the thickness of the 
light guide was glued to the light guide at its narrow end.
For the SS scintillators a thin silicone ``cookie'' was placed
between the photomultiplier tube (PMT) and the acrylic
disk, the entire assembly held in place by tape.
The OS photomultipliers were glued to the the acrylic
disks with Bicron BC-600 optical cement.
Phillips XP2230 PMTs were used in the SS Hodoscope and Hamamatsu R329 
photomultipliers~\cite{Hamamatsu} in the OS Hodoscope.

A transistor-based voltage divider provided high anode-current capability
for the PMTs.
Voltage to the divider was provided by a LeCroy 1440 power 
supply~\cite{LeCroy} which served all of the photomultipliers in
the apparatus.
To reduce the output pulse width to about 3.5\,ns full-width-at-half-maximum, 
a twisted-pair cable (clip line) 0.30\,m long terminated with a 
30\,$\Omega$ resistor was attached to the anode output at the base.  
The signals from each counter were taken to the trigger-logic area 
in the Electronics Hall via RG-8 cables of 24\,m length.

With a 13\,MHz secondary-beam rate, the highest-rate counters
were SS1 and OS1 at 2.4\,MHz and 0.44\,MHz, respectively.\footnote{The
rates in the SS counters  fell off rapidly with distance from the beam:  
for example, SS2 had one-quarter of the SS1 rate.}
As can be seen in Fig.~\ref{fig:eff_ss_os}, efficiencies were high, 
typically 99.8\% for the SS counters and 99.9\% for the OS counters.  
(The lower efficiency for SS1 was due to its high rate, whereas the
lower efficiencies for the counters beyond OS15 were due to the fact that
they were in the shadow of the frame and stand of chamber C9.
Note that protons from \gXi\ra\gL\gpi\ra\gp\gpi\gpi\ only 
impacted counters OS2--15.) 
The differences in the efficiencies for 
positive and negative running were typically much less than 0.1\%
(Fig.~\ref{fig:eff_ss_os}).
The small inefficiencies coupled with the large probability of two
counters being hit for the {\em CP}-violation-search processes
resulted in overall trigger inefficiencies of $\mathcal{O}(10^{-4})$ for
each hodoscope.

\subsection{Trigger Logic}
\label{sec:trig_logic}
The trigger was formed using NIM-standard logic signals
and mostly off-the-shelf NIM and {CAMAC} modules.
Care was taken to have single-RF-bucket time resolution throughout.
The lowest-level trigger elements were formed as follows.
Signals from the individual SS and OS counters were 
required to surpass a 30\,mV threshold, using Phillips 710 discriminators
\cite{Phillips}
producing a signal of 15\,ns length.
The outputs of these discriminators were logically
summed using LeCroy 429 fan-ins, then synchronized to
the 53\,MHz RF frequency of the Tevatron using a 
LeCroy 365ALP logic unit to produce the SS and OS trigger signals,
again of 15\,ns length.
The first SS counter (SS1) was not used in the trigger.
The in-time coincidence of the SS and OS triggers formed
the Left-Right (LR) trigger.

In the Muon Hodoscopes the signals from each bank of scintillation
counters were input to LeCroy 4413 discriminators, 
with a threshold set at 30\,mV, providing an output pulse of 20\,ns width.  
(See Sec.~\ref{sec:muon_hodo} for a description of the Muon Hodoscopes.)
The current-sum outputs of the discriminators were
discriminated at two different levels, one consistent
with at least two counters being hit (75\,mV), the other 
consistent with at least one counter firing (30\,mV).
Subtriggers from the vertical and horizontal banks
were formed requiring one (1MULH, 1MULV, 1MURH, 1MURV) and two
(2MULH, 2MULV, 2MURH, and 2MURV) counters in each hodoscope bank.
The one- and two-muon triggers were formed from
coincidences of the vertical and horizontal components of these subtriggers:
1MUL = 1MULH{\cd}1MULV, 1MUR = 1MURH{\cd}1MURV,
2MUL = 2MULH{\cd}2MULV, and 2MUR = 2MURH{\cd}2MURV.

The signals from the calorimeter were fed into a
custom-built module that made a simple linear
sum of the eight calorimeter cells.
(See Sec.~\ref{sec:cal} for a description of the Hadronic Calorimeter.)
Both outputs of the summing module were fed into 
Phillips 715 constant-fraction discriminators~\cite{Phillips}, 
one with a threshold consistent with 
${\approx}99$\% efficiency for a 45\,GeV hadron (CAL(K) trigger), 
the other with ${\approx}99$\% efficiency 
for a 60\,GeV hadron (CAL(CAS) trigger).

The logic used to form the four main physics triggers
is shown in Fig.~\ref{fig:trig_logic_99}.
These triggers were:
(1) the Cascade trigger (CAS), a coincidence of the
   CAL(CAS) and LR triggers,
(2) the Kaon trigger (K), a coincidence of the
   CAL(K) and LR triggers,
(3) the Unlike-Sign Muon trigger (MUUS), a coincidence
   of the 1MUL, 1MUR, and LR triggers, and
(4) the Two-Muon-Like-Sign-Left trigger (2MULSL), a
   coincidence of the 2MUL and SS triggers.
All of the triggers were input into programmable
prescalers (PD-22) built by Fermilab.  
The prescaled outputs were fanned into a custom-built Trigger-OR module.
A pulse from any of the 16 possible triggers would
cause the Trigger-OR module to inhibit any further triggers
until the inhibit was released by a signal from the data-acquisition system after the event was read out.  
The outputs of the Trigger-OR module started the gates
used for the latch systems and the analog-to-digital converters (ADCs)
and signaled the data-acquisition system to read out the event.

Great care was taken in setting up, timing in, and 
monitoring the performance of the triggers and their components.
Every counter and logic element in the trigger was latched
and scaled.  Scaling was done
both with and without the data-acquisition-system-inhibit veto 
using LeCroy 2251 scalers, most of which were modified to have a 
48-bit range to handle the high rates and the long Tevatron spill.
For every run the trigger efficiencies were calculated
from the latched subtrigger elements, and the trigger-bit latching
efficiencies were also monitored.  
The efficiencies were all very high, typically 99.9\%.

In the 1999 run the fraction of the total trigger rate of each of the 
four main physics triggers for the positive- (negative-) polarity 
running was typically  49\% (55\%) for CAS, 
37\% (37\%) for K, 1.3\% (0.92\%) for MUUS, and 1.6\% (1.3\%) for 2MULSL.
There was of course considerable overlap between events 
satisfying the CAS and K triggers.
The geometric acceptance of the main physics trigger, CAS,
was very high:  about 95\% of the \gXi\ra\gL\gpi\ra\gp\gpi\gpi\ decays
occurring within the Vacuum Decay Region whose secondaries cleared 
the apertures of the Analyzing Magnets were accepted by the trigger.
Event yields for the CAS trigger were relatively high.
For negative running, for example, about 18\% of
the CAS triggers produced reconstructed \gL\ra\gp\pim\ events,
and 6\% produced reconstructed \Xim\ra\gL\pim\ra\gp\pim\pim\ events.
The numbers were smaller for positive running due to the 
smaller \aXim\ production cross section:  
4\% for {\agL}  and 1\% for {\aXim}.
There are two reasons for this disparity in \gL\ and \gXi\ yields:
first, there was a large source of directly produced \gL's from
interactions of the secondary beam with the walls of the collimator;
second, the CAS trigger was in fact a \gL\ra\gp\gpi\ trigger
(and hence was perhaps misnamed).
Collimator production was indeed substantial, with the largest yield
from the CAS trigger sample being \Ks\ra\pip\pim\ decays.

\section{Hadronic Calorimeter}
\label{sec:cal}
Most Left-Right triggers were due to interactions of the 
secondary beam with material in the spectrometer.
The purpose of the Hadronic Calorimeter was to reduce the number
of these background triggers by requiring a minimum amount of 
energy consistent with the lowest-energy proton or antiproton from 
\gXi\ra\gL\gpi\ra\gp\gpi\gpi\ decays,
or the opposite-sign pion from \Kpm \ra \pipm\pip\pim\ decays.
The minimum energy of protons from \gXi\ decays,
(all of which impacted the calorimeter)
was approximately 75\,GeV and about 40\,GeV
for those opposite-sign pions from \Kpm \ra \pipm\pip\pim\ decays
that impacted the calorimeter.
Note that the calorimeter was used solely 
for the trigger and {\em not} in event reconstruction or data
selection.

The calorimeter was situated behind the magnetic spectrometer,
with its front face, at $z = 54.38$\,m,
sufficiently far downstream of the Analyzing Magnets that
protons from \gXi\ decays were well separated from the secondary beam
(see Fig.~\ref{fig:beam_at_cal}).
Its lateral size was dictated by the distribution of those protons.
Although protons from \gOm\ra\gL\gK\ra\gp\gK\gpi\
decays were slightly more spread out than those from
\gXi \ra \gL\gpi \ra \gp\gpi\gpi\  decays, they too were 
contained within the calorimeter active area.
This was not the case for the opposite-sign pions from 
\Kpm \ra \pipm\pip\pim\ decays, 
but funding constraints precluded building a larger
calorimeter.

The calorimeter had to be fast, have good energy resolution,
have no cracks, and have excellent efficiency over its entire fiducial area.
Speed considerations required that the active medium be scintillator.
Since the calorimeter was not used in event reconstruction, 
good shower-position 
resolution was not necessary, so to minimize the number of 
readout channels and simplify calibration, the calorimeter segmentation 
was made as coarse as possible. 
For the two HyperCP runs, the estimated radiation dose 
was less than 5\,Gy, thus radiation damage was not an issue.
At the typical secondary-beam rate of about 13\,MHz, the
rate of particles incident on the calorimeter was about 100\,kHz.

\subsection{Physical Description}
\label{sec:cal_des}
The calorimeter specifications are given in Table~\ref{tab:specs}.
Side and back views of the calorimeter and stand, with the light-tight
enclosure and photomultipliers omitted, are shown in 
Fig.~\ref{fig:cal_front_side}.
The calorimeter was mounted on a stand with jacks that allowed
limited vertical movement.  Rollers allowed the calorimeter to 
be moved horizontally ($x$), which was done during special
magnet-off chamber-alignment runs.
A schematic of the interior of the calorimeter, showing fibers,
light guides, and photomultipliers is found in Fig.~\ref{fig:cal_front}.
The calorimeter was composed of 64 layers of 24.1\,mm thick Fe and 
5\,mm thick scintillator,
giving a sampling fraction of 3.5\% and
a total thickness of 88.5 radiation lengths and 9.6 interaction lengths.
Its active area was 0.990\,m wide by 0.980\,m high.
For readout purposes it was subdivided into four
longitudinally and two laterally, for a total of eight cells.

Kuraray SCSN-38 scintillator~\cite{Kuraray} 
was used because of its superior speed, high light output, and low cost.   
Each of the 64 sheets of scintillator had 32 keyhole-shaped
channels milled into it, the channels separated by 30\,mm. 
The scintillator edges were painted with titanium dioxide paint  
(Bicron BC620~\cite{Bicron})
and then wrapped, first in DuPont Tyvek reflective paper~\cite{Tyvek}, 
then DuPont Tedlar black paper~\cite{Tedlar}, and finally with
a 0.81\,mm thick Al sheet used to prevent physical damage. 
No provision was made to make each scintillator sheet light-tight; 
rather the entire assembly was placed in a light-tight box. 
Bicron BCF-92 (G2) waveshifting fibers~\cite{Bicron} with a 2\,mm
diameter brought the light out of the scintillator sheets.
The large fiber diameter was chosen to give high efficiency in capturing 
light from the scintillator and to provide a long attenuation length.   

The size of the calorimeter and the choice of 2\,mm diameter fibers dictated 
that the readout be divided into eight groups of $16{\times}16 = 256$ fibers 
each.   
One end of each fiber had an Al reflective coating.  
The other ends of the 256 fibers in each cell were  
potted in a special low-viscosity,  low-exotherm, slow-setting, 
two-component epoxy made for us by Master Bond~\cite{MBond}.    
The epoxy was opaque and hence removed all of the cladding light. 
After curing, the fiber--epoxy fixture was sanded and polished in situ.
A tapered, square, acrylic light guide was mated, via a 5\,mm thick 
silicone cookie, to the fiber--epoxy fixture, as directly mating 
the fibers to the photomultiplier would have resulted in 
spatial inhomogeneities in response. 
The light guide was $50.8{\times}50.8$~mm$^2$ at the fiber 
end, $31.8{\times}31.8$~mm$^2$ at the photomultiplier end,
and  136.5\,mm long.
The acceptance for all core light emanating from the fibers was 
estimated by Monte Carlo simulation to be about 80\%.    
Another 5\,mm thick silicone cookie was inserted between 
the light guide and the photomultiplier. 
The photomultipliers were hung by threaded rods from the frames that 
clamped the fiber--epoxy fixtures into place. Springs provided sufficient 
compression to press the silicone cookies securely against the light guides.

A nitrogen laser calibration system monitored the response of the 
calorimeter, the laser pulsed at about 1\,Hz~\cite{laser}.
The laser light impinged on 
a group of quartz optical fibers, with one fiber
attached to each calorimeter light guide.  
The output intensity of the laser was monitored by a PIN diode.

\subsection{Calorimeter Electronics}
\label{sec:cal_elect}
The calorimeter electronics had to be stable at rates up to  100\,kHz 
and have a large dynamic range due to the need to measure the energy 
of muons for calibration purposes as well as the energy of the 
highest-energy protons from \gXi\ decays (about 180\,GeV).

The light trapped by the waveshifting fibers was converted into  
electrons and amplified by eight 50.8\,mm, green-extended, 
Hamamatsu R329-02 photomultipliers~\cite{Hamamatsu}, 
a tube tested to have exceptional stability as a function of rate. 
A transistorized base was used, similar to those used in other Fermilab 
experiments~\cite{Kerns}. 
It allowed the tubes to be run at anode currents up to 100\,$\mu$A 
with no measurable gain variation, and up to 300\,$\mu$A with a 
5\% gain increase~\cite{Durga}. 
The maximum average current experienced by the calorimeter 
photomultipliers was about 70\,$\mu$A in the upstream cell closest 
to the secondary beam.

Figure~\ref{fig:cal_readout} shows how the calorimeter signals were 
conveyed to the trigger and data-acquisition electronics during 
the 1999 run. 
Approximately 70\,ns of RG-8 cable was used to bring the signals  
from the calorimeter to a custom-built linear fan-in/differential line 
driver situated in the Electronics Hall.
The fan-in/driver served two purposes: it summed signals from the 
eight photomultipliers, providing two prompt summed outputs, 
and it sent copies of the individual inputs, as well as the 
summed signal, on 300\,ns of 75\,$\Omega$ ribbon coaxial cable  
(providing delay for trigger latency) to a receiver/shaper which in 
turn fed an analog-to-digital converter (ADC).\footnote{In the 1997 run a 
passive splitter was used to send part of the signal to the trigger system, 
but the resulting 60\,Hz ground-loop noise compromised our ability 
to discern clear muon peaks.  
The fan-in/driver modules were implemented in the 1999 run to 
solve this noise problem.} 
Differential signals were sent on adjacent channels of the  ribbon 
coaxial cable. 
The two prompt summed outputs were sent to constant-fraction 
discriminators set at thresholds well below the minimum energy 
expected from those opposite-sign pions from \Kpm\ra\pipm\pip\pim\ 
decays impacting the calorimeter (the K trigger)  and protons from 
\gXi\ra\gL\gpi\ra\gp\gpi\gpi\ decays (CAS trigger).   
The delayed signals received by the receiver/shaper were compensated 
for the dispersion in the long cable and then fed  to high-rate 
(333\,ns digitizing time) 14-bit ADCs \cite{datel} having a  
75\,fC least count.  The ADC gate width was 75\,ns.

\subsection{Calorimeter Performance}
\label{sec:cal_perfor}
Because the calorimeter trigger was formed from the linear sum of 
the individual photomultiplier outputs, much care was taken to  
minimize cell-to-cell variations in the calorimeter response. 
Special runs (``muon runs'') were periodically taken with the 
Hyperon and Analyzing Magnets turned off and the proton beam dumped 
into the collimator.  Two scintillation counters, CMu-1 and CMu-2, 
situated side-by-side and mounted on the back of the calorimeter, 
formed the trigger.
The muon flux from these runs was sufficiently intense that clear 
minimum-ionizing peaks were discernible, allowing all the photomultiplier 
gains to be adjusted to give the same response. 
By measuring the response of the calorimeter to muons,
the number of photoelectrons detected from  each of the 64 layers 
of scintillator was estimated to be about 2. 

Another type of special run (``one-third field run'') was also
periodically taken, usually just after a muon run, to determine
the proper constant-fraction discriminator settings.  
In these runs the currents of the Hyperon and Analyzing Magnets 
were set to one-third their nominal values in order to select 
a lower-momentum secondary beam.
This allowed the calorimeter trigger threshold to be determined 
from protons using a very clean sample of low momentum
\gXi\ra\gL\gpi\ra\gp\gpi\gpi\ events:
at normal secondary-beam momenta none of the \gXi\ decays
produced protons whose momenta were at or below the nominal trigger
thresholds.
Data were taken with a variety of discriminator settings and
were analyzed to determine the CAL(CAS) and CAL(K) discriminator settings
for 99\% efficiency, respectively for 60 and 45\,GeV/$c$ protons.

With an energy threshold of about 60\,GeV (at  $\approx$99\% efficiency), 
the calorimeter trigger provided 
a rejection factor  of about six over the Left-Right trigger rate. 
The efficiency of the calorimeter trigger (with the CAS trigger 
energy threshold) for a typical 1997 run is shown in Fig.~\ref{fig:cal_eff}. 
It was determined by finding the fraction of protons and pions from
\gL\ra\gp\pim\ and \Ks\ra\pip\pim\ decays that set the CAL(CAS) trigger bit, 
the events taken from the LR trigger sample.
Detailed studies of the trigger efficiency in 1999 showed that it was 
typically 99.3\% for protons from \Xim\ decays and 98.7\% for antiprotons 
from \aXim\ decays.
The major source of inefficiency was due to a problem with the
Phillips 715 constant-fraction discriminator---a problem that was found 
after the data-taking was completed, and which was reproduced on the 
test bench.  
A modest amount of energy deposited in the calorimeter in a previous 
RF bucket prevented the discriminator from firing, irrespective
of the in-time-bucket particle energy.
Fortunately this electronic inefficiency was not a source of bias
for the \CP\ analyses.
The inherent calorimeter inefficiency due to its energy response fluctuating 
below the trigger discriminator threshold
was 0.24\%, both for protons and antiprotons from \Xim\ and \aXim\ decays.
The calorimeter efficiency, 
as well as the trigger energy threshold, were monitored throughout the 
data-taking on a run-by-run basis. 
The energy resolution of the calorimeter was roughly $\sigma/E = 9$\% 
for 70\,GeV/$c$ protons (the lowest-energy protons from 
\Xim\ra\gL\pim\ra\gp\pim\pim\ decays).

\section{Muon Detection System}
\label{sec:muon}
A muon-detection system at the rear of the apparatus 
was used to identify muons having momenta greater than about 20\,GeV/$c$.
Figures~\ref{fig:mudet_side} and \ref{fig:mudet_front} depict  elevation
and front views of the muon system.
It consisted of two similar Muon Stations, one to the 
left and one to the right of the secondary beam.  
Each station contained three layers of ${\approx}0.75$\,m thick iron absorber,
with each layer followed by planes of vertical and horizontal proportional 
tubes.
An additional 0.91\,m thick iron absorber was installed in front of the Left
Muon Station for the 1999 run in order to reduce hadron 
punch-through.\footnote{This absorber was roughly the same width as
the others, but only 0.91\,m high, rather than the 1.40\,m height 
of the other absorbers.}
The Hadronic Calorimeter provided an additional 1.57\,m thick iron
absorber for part of the Right Muon Station.
Vertical and horizontal scintillation-hodoscope planes 
were employed for triggering and for identifying 
in-time proportional-tube hits. 
In the 1997 run, the hodoscopes were all mounted behind the last 
layer of iron.
In the 1999 run, in order to reduce sensitivity of the muon triggers 
to hadron punch-through and secondary-beam halo, the vertical and 
horizontal muon hodoscopes were separated, with the vertical
hodoscope planes moved to behind the second layer of iron, 
and in front of the proportional tubes, as shown in 
Fig.~\ref{fig:mudet_side}.

\subsection{Muon Proportional Tubes}
\label{sec:muon_tubes}
To detect muon tracks, the Muon Stations employed aluminum 
proportional-tube modules built for a previous experiment.
Each module contained eight 1549\,mm long square tubes
with approximately 2\,mm thick walls and 25.4\,mm spacing between centers.
Anode wires, of gold-plated tungsten, 37\,$\mu$m in diameter 
and 1.62\,m in length, were strung through the center of each tube.
There were 8 (5) modules in each vertical (horizontal) plane.
A gas mixture of P10\footnote{The P10 gas mixture consists of 90\% 
Ar--10\% CH$_{4}$.  In the 1997 run, 80\% Ar--20\% CO$_2$
was used.  It was replaced to increase the drift speed.} and 
CF$_{4}$ in an 86/14 ratio at room 
temperature and atmospheric pressure flowed serially 
through all the tubes in a plane.
The measured time resolution with this mixture 
was 80\,ns with +2.5\,kV on the anode wires. 
The efficiency of the modules, at about 93\%, was limited mainly
by the tube wall thickness.

The proportional tubes had 16-channel preamplifiers 
mounted directly on the modules~\cite{Yarema}. 
The amplified signals were sent to Nanometric Systems
N-277 discriminators~\cite{Nanometric} located near the Muon Stations.
The differential-ECL outputs from the discriminators were connected 
to latch cards in the data-acquisition system. 
The total number of proportional-tube channels was 624.

\subsection{Muon Hodoscopes}
\label{sec:muon_hodo}
The vertical (horizontal) hodoscope plane 
in each Muon Station consisted of 15 (10) scintillation counters. 
Each counter was 102\,mm wide and 25\,mm thick.  
The vertical (horizontal) counters were 1.00 (1.52) m long.
Signals from these counters were sent to LeCroy 4413 
discriminators~\cite{LeCroy}, and the ECL outputs from the discriminators were 
recorded by latch cards in the data-acquisition system.
The sum of analog signals from each counter in the vertical
and horizontal planes was used to form various muon-trigger signals 
as described in Sec.~\ref{sec:trig_logic}.

\section{Beam Hodoscope}
\label{sec:beam_hodo}
For the 1999 run, a pair of hodoscopes (see Fig.~\ref{fig:beam_hodo}), 
collectively called the Beam Hodoscope, 
were installed behind the Muon Stations and centered on the secondary beam.
It was used to monitor the secondary-beam intensity and position. 
The Beam Hodoscope consisted of one plane each of vertical and 
horizontal scintillation counters. 
The vertical (horizontal) plane had 10 (14) counters, each
250\,mm long and 5\,mm thick. 
The first two (three) and last two (three) counters in the vertical 
(horizontal) plane were 50.8\,mm wide, while the remaining counter 
widths were 25.4\,mm. 
The area covered by both planes of the Beam Hodoscope was slightly less 
than that of the secondary beam.

The signals from these counters were sent to LeCroy 623 
discriminators~\cite{LeCroy}.
The logical OR of the discriminator outputs of the vertical
and horizontal counters were formed separately, the coincidence of which 
formed the BEAM trigger.
The NIM signals from the discriminators were converted to differential-ECL
by LeCroy 4616 ECL/NIM/TTL translators, the outputs of which
were routed to latch cards in the data acquisition system.

\section{Data Acquisition System}
\label{sec:daq}
The accumulation of such a large dataset required a high-speed
data-acquisition (DAQ) system, one capable of recording up to 
100\,000\,events per spill-second.
The HyperCP DAQ was designed to meet that requirement,
with a data-to-tape rate of at least 23\,MB/s,
a front-end deadtime less than 2\,$\mu$s per trigger,
and a small (${\approx}$500 byte) event size.
The 1999-run DAQ configuration is summarized here.
Differences between it and the 1997 configuration are too
numerous to point out here; however, both are documented in detail 
in~\cite{HyperCPDAQ,DAQ97,IEEE}. 
There were two components to the DAQ system:
a FastDA system, which read out event-by-event data at a very
high rate to magnetic tape,
and a SlowDA system, which read out spill-by-spill data,
such as magnet currents, at a much slower rate to disk.
We first describe the FastDA system and then turn our attention to the
SlowDA.

Data were taken during a 40\,s spill period, and then again during
a 20\,s interspill period, which started 10\,s after the end of the 
spill period, and ended 10\,s before the start of the next spill.
A trigger given by the prescaled 53\,MHz RF signal from the Tevatron
was used to record events during the interspill period, when no
beam was being delivered to the experiment.  It allowed the 
quiescent response of the apparatus to be monitored.

\subsection{FastDA System}
\label{sec:daq_fast}
Readout of the MWPCs, calorimeter, muon system, and trigger
hodoscopes was done using an adaptation of the Nevis
Laboratories MWPC Coincidence Register (CR) 
system~\cite{Sippach}.\footnote{In the 1997 run the muon system
was read out by a custom-built system described in \cite{DAQ97}.}
Having built and operated nearly 10\,000 channels of this 
system for Fermilab Experiment E789, we chose for reasons 
of readout speed, cost-effectiveness, and time limitations to
augment that existing system to meet the HyperCP needs rather than build a
new system from scratch. 
Note that commercial FASTBus and CAMAC-based systems were too slow to meet 
the HyperCP requirements.  The layout of the FastDA system is
shown in Fig.~\ref{fig:path}.

The HyperCP latch system consisted of 37 crates. 
Each crate had CAMAC-standard card cages housing inexpensive 
two-layer printed-circuit boards communicating via a custom ECL backplane. 
The crates each accommodated up to twenty-three 32-channel latch cards 
plus one Buffered Encoder card used as a crate master for readout,
sparsification, and buffering the data. 
Two types of latch cards were used: one read out data 
that were ``sparsified" (encoded), such as wire-chamber hits;
the other read out information whose format was to
be preserved (unencoded), such as the 10 calorimeter ADC channels,
4 time-to-digital converter (TDC) channels,\footnote{A special
TDC module was built with four channels, each with a 0.5\,ns least count.
It was used to measure the time of arrival relative to the 
trigger of the BEAM, OS, and
SS triggers, as well as that of a counter, S7, that was placed
in the secondary beam just in front of the calorimeter.}
and the trigger latch bits.
Only 52 bytes of nonsparsified data were read out.
In all, 654 latch cards were employed, for a total 20\,928 channels.
Table~\ref{tab:fastda_data} summarizes the numbers and types of data read out
by the FastDA.

The crates were allocated among the five parallel datapaths 
in a manner that balanced the data loads as well as possible, based on the 
average numbers of hits per crate per event. 
Each datapath consisted of a variable number of latch crates,  
chained serially to a Compressor module.
The Compressor module aggregated events into ``super-events'',
each typically with ten events, to allow more efficient
use of the VME-backplane bandwidth during event building.\footnote{The 
Compressor module also had a data compression feature that was not used.}

The aggregated events were sent, via an interface module, 
through five optical fibers to five buffer memories, 
called the VDAS buffers~\cite{VDAS}.
Optical fibers~\cite{optical} were used due to the long distance 
(${\approx}100$ meters; see Fig.~\ref{fig:site_layout}) between 
the Electronics Hall and the Control Room,
where the data were written to magnetic tape. 
The VDAS buffers accepted 32-bit words at speeds up to 
10\,Mwords/s using the RS-485 signal protocol. They
could store approximately one spill's worth of data, 
allowing the tapewriting to proceed continuously, 
both during and between spills, thus reducing
the needed data rate to tape by about 50\%.
Care was taken to have the tape drives always writing,
as there was considerable delay in restarting a stopped drive.
A sixth optical fiber returned a signal used to disable triggers 
when the VDAS buffers were nearly full.

The event fragments from each VDAS buffer were assembled into 
complete events by nine Motorola MVME167 single-board 
computers \cite{Motorola} in three standard 6U VME crates.
Within a crate, on command from an MVME167, 
an Event Buffer Interface module read one super-event
fragment from each of the five VDAS buffer memories and
stored the fragments in an internal buffer within the MVME167.
The super-event fragments were then assembled by the MVME167
to form a complete event, stored in internal data buffers,
and then written to an Exabyte 8705 tape 
drive \cite{Exabyte} via an MVME712 interface module \cite{Motorola}.
To insure that all event fragments were from the same trigger,
a four-bit event synchronization number (ESN) was encoded into
the data by the Encoder.  The ESN was checked first by the 
hardware and subsequently by the event-building software to
verify that all event fragments were from the same triggered event.
Note that no reconstruction or selection criteria were applied 
by the MVME167 computers before the events were written
to tape.

A fraction of the events written to tape were copied into 
a special buffer, called the Sample Buffer, which sent the 
events to a shared buffer on the Silicon Graphics Indy workstation
\cite{sgi} that managed the FastDA and SlowDA systems (see below).
A program running on a second Silicon Graphics Indy workstation,
or on offline DEC Alpha computers, grabbed the events
and fed them to the monitoring program, which reconstructed the events,
filled histograms, calculated detector and trigger efficiencies,
and displayed selected events.  The monitoring program was essentially
the standard offline processing package.
An example of an event display from the monitoring program is shown in
Fig.~\ref{fig:event_display}.

Both the FastDA and SlowDA systems were managed by the DART Run Control 
(DRC) program, which ran on a Silicon Graphics Indy workstation
and employed a user-friendly Tcl/Tk interface.
Data acquisition operations were streamlined so that only five user 
commands were needed at the DRC level: 
Initialize, Start, Pause, Resume, and Stop.
Underlying these commands were a series of functions and script files that
relayed information to the FastDA, SlowDA, trigger system,
and latch system. 
To simplify DAQ operations further and to protect
the user from inadvertent errors, the DRC program controlled
which commands were available at any given moment. For example,
the Start command was not available until after the Initialize
command had been executed.

\subsection{SlowDA System}
\label{sec:daq_slow}
Data not desired on an event-by-event basis, such as magnetic-field values,
scaler readings, chamber voltages, etc., were written to disk 
by the SlowDA process, both during the spill and the interspill periods.
Table~\ref{tab:slowda_data} summarizes the information written to 
the SlowDA log files.
The data were read through a Jorway~\cite{jorway} serial
CAMAC to SCSI interface via the Simple Acquisition of Data~\cite{SAD}
client/server application connected to the accelerator VAX cluster. 
The DRC program was used to send commands to the SlowDA.
The SlowDA process was launched from the DRC startup script. 
It began by opening a connection to the DRC, opening a connection to the
serial CAMAC chain, and initializing all the CAMAC crate controllers. 
The process then waited for commands from the DRC.

\subsection{DAQ Performance}
\label{sec:daq_perfor}
The performance of the DAQ met or exceeded expectations. The
observed maximum throughput to tape was 27\,MB/s, limited by the
number and speed of the Exabyte tape drives. The measured
deadtime per event, 1.7\,$\mu$s (limited by the ADC readout time),
produced a typical livetime of 80\% at trigger rates near 100\,kHz.  
In the seven-month data-taking period of the 1999 run, 
the DAQ wrote 82\,TB of data to 20\,421 8\,mm
tapes (compared with 38\,TB  on 8980 tapes in the
1997 run). The limiting factor in collecting these data was
readout deadtime, as additional data-to-tape headroom remained at the
chosen operating conditions. Without deadtime, the number of
events written to tape would have increased by 25\%. 
Little increase in data rate was possible at the chosen operating conditions.

\section{Trackfinding Algorithm}
\label{sec:track}
The trackfinding algorithm  involved determining the spacepoints in 
each chamber, reconstructing the upstream and downstream track segments,
and matching track segments to form complete tracks. 
The software for performing pattern recognition and track reconstruction 
was written in C++.

\subsection{Chamber Orientation}
\label{sec:track_orient}
The orientation of the wires in each MWPC plane is defined by 
three Euler angles.
The four planes in a given MWPC all shared two of the angles, 
defined by the orientation of the MWPC,
with the third angle defining the direction of the wires.

The coordinate measured by each wire plane was expressed in the
Local Coordinate System $(x_{c}, y_{c}, z_{c})$, 
defined by: an $x_{c}$ axis perpendicular to the wire 
direction (corresponding to the direction of measurement), 
a $z_{c}$ axis normal to the wire plane and pointing downstream,
and a $y_{c}$ axis completing the right-handed coordinate system.
The Local Coordinate System is related to the Spectrometer Coordinate System 
by
\begin{equation}
\label{eq:appA-1}
\left( \begin{array}{c}	
	x_{c} \\
	y_{c} \\ 
	0 	
	\end{array} 
\right) =
M
\left( \begin{array}{c}	
	x \\ 
	y \\ 
	z-z_{0} 	
	\end{array} 
\right),
\end{equation}
where $M$ is a $3{\times}3$ transformation matrix 
and $z_{0}$ is the $z$ position of the wire plane at $(x,y)=(0,0)$.
The matrix $M$ is a product of three rotation matrices:
\begin{equation}
M = R_{z}(\theta_{z})R_{y}(\theta_{y})R_{x}(\theta_{x})\,,
\end{equation}
where $R_{x}$ corresponds to rotating the Spectrometer Coordinate System 
$(x, y, z)$ about the $x$ axis by an angle $\theta_{x}$ to get 
$(x_{1}, y_{1}, z_{1})$;
$R_{y}$ represents a rotation of $(x_{1}, y_{1}, z_{1})$ 
about the $y_{1}$ axis by an angle $\theta_{y}$ to get 
$(x_{2}, y_{2}, z_{2})$;
and $R_{z}$ rotates $(x_{2}, y_{2}, z_{2})$ 
about the $z_{2}$ axis by an angle $\theta_{z}$ to get
$(x_{c}, y_{c}, z_{c})$ in the Local Coordinate System.

Defining $s_{x}=\sin{\theta_{x}}$, $c_{x}=\cos{\theta_{x}}$,
$s_{y}=\sin{\theta_{y}}$, $c_{y}=\cos{\theta_{y}}$, 
$s_{z}=\sin{\theta_{z}}$ and $c_{z}=\cos{\theta_{z}}$, the
rotation matrix $M$ is given by
\begin{eqnarray}
\label{eq:appA-2}
M &=&
\left(
\begin{array}{ccc}
c_{y}c_{z}  	& s_{x}s_{y}c_{z} + c_{x}s_{z} 	& -c_{x}s_{y}c_{z} + s_{x}s_{z} \\
-c_{y}s_{z}     & -s_{x}s_{y}s_{z} + c_{x}c_{z} & c_{x}s_{y}s_{z} + s_{x}c_{z}  \\
s_{y}		& -s_{x}c_{y}			& c_{x}c_{y}
\end{array}
\right),
\end{eqnarray}
where $\theta_{x}$ and $\theta_{y}$ specify the orientation of the MWPC 
and are common to all four planes, and $\theta_{z}$ is the ``stereo 
angle" of the wire plane and defines the wire direction.
The wire coordinate is then
\begin{equation}
\label{eq:wire-coor}
x_{c} = M_{11}x + M_{12}y + M_{13}(z-z_{0}).
\end{equation}

\subsection{Reconstruction of Space-points}
\label{sec:track_recon}
Since downstream tracks were fairly well separated, from each other
and from the secondary beam, they were unlikely to produce adjacent hits.  
This was not the case for the upstream tracks.
Hence, downstream wires were clustered and upstream hits were not,
with three wires the maximum number allowed in a single cluster.   

By combining the hits from the four wire planes within an MWPC, 
pseudo spacepoints could be found.
The very small angles $\theta_{x}$ and $\theta_{y}$  
were ignored when finding spacepoints.  
Thus, the wire coordinate was approximately given by 
\begin{equation}
x_{c} = \cos(\theta_{z})~x + \sin(\theta_{z})~y.
\label{eq:wire-coor-approx}
\end{equation}

Since the U and V wires were oriented at equal, but opposite, angles
such that $|(\theta_{z})_{U}| = -|(\theta_{z})_{V}|$,
a checksum criterion was used to associate the U and
the V hits with the X and/or X$^\prime$ hits:
\begin{equation}
|u + v - 2x\,\cos[(\theta_{z})_{U}]| <\eta_{\rm checksum}\,,	
\end{equation}
where $u$, $v$, and $x$ are the U, V and X/X$^\prime$
coordinates of the hits.
Neglecting the 6-mm gap between the wire planes within an MWPC and
using Eq.~(\ref{eq:wire-coor-approx}), the ($x, y$) coordinates 
of the space-point candidates were determined by minimizing:
\begin{equation}
\chi^{2} = \sum_{i} \frac{[\cos(\theta_{z})_{i}~x + \sin(\theta_{z})_{i}~y - 
                       m_{i}]^{2}}
{\sigma_{i}^{2}}\,,
\end{equation}
where $m_{i}$ is the hit coordinate for wire plane $i$
that is associated with the space-point candidate and 
$\sigma_{i}$ is the error of the hit coordinate. 
A cut was made on the value of $\chi^{2}$ 
to filter out poor-quality spacepoints.
Neglecting the gaps between the anode planes was later found to 
cause a loss of the lowest momentum (large deflection angle) tracks. 
This was corrected after the initial reconstruction.
Using Eqs.~(\ref{eq:appA-1}), (\ref{eq:appA-2}) and the reconstructed 
($x, y$) coordinates of the space-point, the $z$ coordinate of the wire plane 
associated with the space-point is given by
\begin{equation}
\label{eq:zpos}
z = -\frac{M_{31}x + M_{32}y}{M_{33}} + z_{0}    .
\end{equation}

\subsection{Reconstruction of Upstream and Downstream Tracks}
\label{sec:track_track}
Upstream and downstream tracks were reconstructed separately.  
In each case, the MWPCs were labeled from 1 to 4 by ascending $z$ position. 
(C9 was not used at this stage.)
To allow for inefficiencies, pairs of seed spacepoints were formed 
from MWPCs (1,4), (1,3) and (2,4).
Each such track candidate was then projected to the other MWPCs
to search for additional spacepoints associated with it. 
A track candidate was required to contain at least
three spacepoints. 

A three-dimensional track is uniquely defined by the slopes and intercepts
of lines in two orthogonal projections, $x$--$z$ and $y$--$z$, given by
\begin{eqnarray}
  \label{eq:xeq}
  x &=& a_{x}z_{i} + b_{x}\,, \\
  \label{eq:yeq}
  y &=& a_{y}z_{i} + b_{y}\,,
  \end{eqnarray}
where $a_{x}$ and $a_{y}$ are the slopes, $b_{x}$ and $b_{y}$ the intercepts, 
and $z_{i}$ the $z$ positions of the wire planes measured from the 
bend plane of the Analyzing Magnets. 
Substituting from Eqs.~(\ref{eq:xeq}) and (\ref{eq:yeq}) into        
Eq.~(\ref{eq:wire-coor}), the calculated hit coordinate is given by
\begin{equation}
\label{eq:sseq}
(x_{c})_{i} = M_{11}(a_{x}z_{i} + b_{x}) + M_{12}(a_{y}z_{i} + b_{y}) + 	M_{13}[z_{i}-(z_{0})_{i}]\,,
\end{equation}
where the subscript $i$ refers to wire plane $i$ and $z_{i}$ is given by 
Eq.~(\ref{eq:zpos}).
The track parameters were determined by minimizing: 
\begin{equation}
\label{eq:track_chi2}
\chi^{2} = \sum_{i} \frac{[(x_{c})_i - m_{i}]^{2}}{\sigma_{i}^{2}}\,.
\end{equation}
The $z_{i}$ were assumed constant.  
A cut was made on the value of $\chi^{2}$ to suppress poorly 
reconstructed tracks and fake tracks.

In order to have a high efficiency for reconstructing closely spaced 
upstream tracks, cuts used in finding the upstream track candidates 
were more relaxed than the corresponding cuts used in finding the 
well-separated downstream track candidates. 
The drawback was that upstream track candidates consequently included 
more ghost tracks. 
Most ghost tracks were filtered out by matching the downstream track 
candidates to the upstream track candidates at the bend plane of the 
Analyzing Magnets.  
The $z$ position of the mean bend plane, $z_{\rm bendplane}$, 
was taken as the mean of the distribution of the $z$ positions at which the 
upstream and downstream tracks had the smallest separation.

\subsection{Matching Upstream and Downstream Tracks}
\label{sec:track_match}
Each downstream track candidate was matched to
an upstream track candidate to construct a complete track candidate.
Matched tracks were accepted subject to cuts on the slopes and intercepts and
the $z$ position of the point of closest approach. In particular, 
cuts were made on
\begin{equation}
\chi^{2}_{\rm slope}=\frac{(a_{yu} - a_{yd})^{2}}
                              {\sigma_{a_{yu}}^{2}+\sigma_{a_{yd}}^{2}},
\end{equation}	
\begin{equation}
\chi^{2}_{\rm intercept}=\frac{(b_{cxu} - b_{cxd})^{2}}
                          {\sigma_{b_{cxu}}^{2}+\sigma_{b_{cxd}}^{2}} +
\frac{(b_{cyu} - b_{cyd})^{2}}
            {\sigma_{b_{cyu}}^{2}+\sigma_{b_{cyd}}^{2}}\,,
\end{equation}
	
\begin{equation}
|z_{\rm intersect}-z_{\rm bendplane}| < \Delta z_{\rm tolerance}\,,
\end{equation}
where $a_{yu}$ and $a_{yd}$ are the $y$ slopes of the upstream and
downstream track candidates, $b_{cxu}$, $b_{cyu}$, $b_{cxd}$,
and $b_{cyd}$ are the $x$ and $y$ intercepts of the upstream and downstream
track candidates at the $z$ position of  closest approach, 
and $\sigma_{a_{yu}}$, $\sigma_{a_{yd}}$, $\sigma_{b_{cxu}}$, 
$\sigma_{b_{cyu}}$, $\sigma_{b_{cxd}}$, and $\sigma_{b_{cyd}}$ 
are their errors which were taken from the diagonal
elements of the covariance matrix from the track fitting. 
Accepted tracks then underwent a global constrained fit to obtain a 
better estimate of the track
parameters $a_{xu}$, $a_{yu}$, $b_{xu}$, $b_{yu}$ with the constraints
\begin{eqnarray*}
b_{xu} &=& b_{xd}\,, \\
b_{yu} &=& b_{yd}\,, \\
a_{yu} &=& Ga_{yd}\,,
\end{eqnarray*}
where 
\begin{equation}
G = \sqrt{\frac{1 + a_{xu}^{2}}{1 + a_{xd}^{2}}}\,.
\end{equation}
An additional $\chi^{2}$ cut on this fit further eliminated poorly 
reconstructed tracks, while the smallest value of this $\chi^{2}$ 
was used to determine the best segment match.

\subsection{Momentum Determination}
\label{sec:mtm_recon}
Tracks were reconstructed in the single-bend-plane approximation.
Neglecting the small inhomogeneities in the magnetic fields,
and using the thin-lens approximation, the $z$ of the bend plane 
was first determined to be 22.593\,m.
Studies using Monte Carlo events, generated 
with the full field map but reconstructed with a single bend plane and a 
constant field integral, showed that on average this procedure did not bias the 
momentum measurement.
Furthermore, there was only a minor degradation in the mass resolution,
as is evident in Fig.~\ref{fig:fullvssingle}.
All the raw data were therefore processed in a first analysis pass 
(discussed below) assuming a constant total 
transverse-momentum ``kick'' of 1.426\,GeV/$c$ (for the two Analyzing 
Magnets taken together) and a single, constant bend plane. 
Events satisfying the requirements of this first-pass analysis were recorded on data-summary tapes (DSTs) for subsequent analysis passes.
The subsequent analysis passes used 
field-integral values scaled on a spill-by-spill basis by the 
Hall-probe readings.

\section{Processing the Raw Data}
\label{sec:process}
The 120\,TB of experimental data were stored on 29\,401 8\,mm, 
5\,GB Exabyte~\cite{Exabyte} tapes (see Table~\ref{tab:data_raw}).
The data were processed (the first analysis pass) with a 
track-reconstruction and event-selection program that reduced the 
data volume by a factor of eight
by imposing very loose trackfinding and invariant-mass requirements.
The processing was done at Fermilab on a PC ``farm''
which consisted of a Silicon Graphics workstation \cite{sgi}
with tape drives and disk storage used
as an input/output node and 55 dual-CPU PC nodes configured as 
a batch system cluster, all running the Linux operating system. 

Output from the farm went to various 
``streams'' depending on the nature of the events.
All streams contained both raw-event information 
and reconstructed, or summarized, event information (called DST information),
and for brevity the processed data were all called DST data.
The largest stream, called the Main Stream, which was written to 334
AIT2~\cite{ait2} tapes, each with 50\,GB capacity, 
contained all selected events.
A second stream, called the Prescaled Stream, had a random 
selection of two \gXi\ events from each spill, 
roughly one hundredth the number of Main Stream \gXi\ events.\footnote{In
order to save space the Prescaled Stream, unlike the other streams, 
included only those raw hits associated with the found tracks.}
The Prescaled Stream, which was stored on disk, allowed a quick 
look at the entire dataset.
A third stream, called the Muon Stream, included events with at
least one track in the Muon Stations, and was stored on
5\,GB Exabyte tapes.
The final stream, called the Monitor Stream, included events from
various triggers that were used to monitor the quality of the data,
and was also stored on 5\,GB Exabyte tapes.

Histogram, efficiency, and log files needed to monitor
reconstruction, detector, and farm performance were generated
automatically in the farm analysis.
This information was accumulated on a run-by-run as well as
spill-by-spill basis, comprising 462 histograms per run and
244 per spill, for a total of 35\,947\,234 histograms!
Programs were written that extracted
averages and other important information
from these histograms, enabling the quality of the data to 
be intimately monitored.

The farm processing was almost fully automated and
information from the analysis of each tape was kept in a database. 
Monitoring of the progress of the farm analysis
was accomplished via dynamically created web pages. 
The data were processed in 11 months at a 364\,GB/day average rate.  
The reconstructed event totals from the farm processing are given
in Table~\ref{tab:data_recon}.
Note that the \Ks\ mesons listed in Table~\ref{tab:data_recon} were 
produced in material near the collimator exit and
hence were reconstructed with both beam polarities.
Table~\ref{tab:data_polar} gives the total number of \Xim\ events
from the ${\pm}2.5$ and ${\pm}3.0$\, mrad targeting-angle runs.
A more detailed discussion of the farm analysis is found in~\cite{chep2001}.

To obtain datasets of manageable size for physics analysis required further 
data reduction.
This was achieved by imposing a tighter set of selection criteria on the 
DST data than those of the original farming. 
The resulting output was ``split'' off into substreams called DST substreams.
The Main Stream data were split at the Parallel Distributed Shared Facility
(PDSF) of the National Energy Research Scientific Computing Center 
(NERSC)~\cite{nersc} using a computing  cluster of PC nodes running Linux;
the output data were stored there on the High Performance Storage System (HPSS).
Two splits were done at PDSF. One used an improved version of the 
track-reconstruction code that had better efficiency for 
very-low-momentum tracks, with information from the old as well
as the new track-reconstruction code recorded on the DST output.  
The output data were split into four DST substreams,
one each for the \gXi\ analyses, the \gK\ra\gpi\gpi\gpi\ analyses,
the \gOm\ analyses, and the rare-decay analyses.
In the other split, five output DST substreams were written,
one each for the \gXi\ analyses, the \gK\ra\gpi\gpi\gpi\ analyses,
the \gOm\ analyses, the \gK\ra\gpi\gmu\gmu\ analyses,
and one with LR triggers for efficiency measurements.
Both splits kept the raw event information.
The substream splittings at PDSF took two months and reduced
the size of the \gXi\ event sample to 2--3\,TB.
Copies of output data samples were made on AIT2 tapes for distribution 
to the collaborating institutions. 

In addition to the splitting of the Main Stream at NERSC,
two other splits were done on the Fermilab fixed-target farm,
with the data stored on the Enstore facility at Fermilab.
The Monitor Stream was split into two DST substreams,
one with dimuon and the other with single-muon candidate events.
The Prescaled Stream was split a single DST substream containing
LR trigger data.  No further track reconstruction was done in
these Fermilab splits and the output data were stored in the same format 
as the input data.

\section{The HyperCP Monte Carlos}
\label{sec:MC}
Simulating particle decays in the spectrometer at the level of
precision needed to measure \AXiL\ to $10^{-4}$
required generating tens of {\em billions} events.
Using standard packages, such as \textsc{geant} \cite{geant}, 
was not feasible due
to speed considerations, and hence a home-built Monte Carlo
simulation package was used.  This Monte Carlo had been developed
over many years by several  experiments exploring
hyperon physics at Fermilab, but was much improved for HyperCP.

Three types of Monte Carlo simulation were used in the HyperCP analyses:
the External Monte Carlo, the Collimator Hybrid Monte Carlo, and
the Hybrid Monte Carlo.
The External Monte Carlo (EMC) generated parent particles at the 
target, traced them through the collimator, decayed them after
the collimator exit and traced their decay products through the spectrometer,
digitizing their hits where appropriate.
The Collimator Hybrid Monte Carlo (CHMC) was used for the \gXi\
\CP-violation analysis.
On an event-by-event basis, it took the \gXi\ momentum, direction, 
and position at the collimator exit from real events and 
tacked on a Monte Carlo--generated \gXi\ decay.
This served two purposes:  it greatly reduced the time needed 
to generate an accepted \gXi\ decay and it provided 
\gXi\ decays that closely resembled data.
Typically, five to ten CHMC events were generated for every real event.
Finally, the Hybrid Monte Carlo (HMC) was used to estimate the
proton acceptance in the \gL\ rest frame for the measurements
of the \Xim, \aXim, \Omm, and \aOmm\ alpha parameters and the
\Xim\ polarization and beta-parameter measurements.
Similar in philosophy to the CHMC, it took every parameter from
the associated real events except for the proton and pion 
directions in the \gL\ rest frame, which were generated randomly 
with a uniform distribution.  Typically ten HMC events
were generated for each real event.
More on the HMC method can be found in~\cite{hmc}.

The External Monte Carlo generated parent particles at the target, 
usually with a Gaussian distribution given by the observed 
primary-beam parameters, although other distributions could also be used.  
There was considerable flexibility in the choices of the 
parent-particle momentum and direction distributions, from parameterizations
based on $x_f$ and $p_T$ to lookup tables based on data.
No attempt was made to simulate any of the other particles associated
with the proton interaction in the target, as the chance of a second
particle from the same interaction successfully traversing the collimator 
was vanishingly small.
The parent was then traced through the collimator and allowed
to decay after the collimator exit.
Note that the EMC also allowed the target and collimator to be bypassed
altogether, with parent particles generated at the collimator exit.
Without reconstruction of the tracks, about twenty-five
\Xim\ra\gL\pim\ra\gp\pim\pim\ decays per cpu second were 
accepted from EMC \gXi\ events generated at the target, 
using a dual-processor Linux-based Pentium PC running at 600\,MHz.

Both the External Monte Carlo and the Collimator Hybrid Monte Carlo 
allowed wide flexibility in the numbers and 
types of decay modes.  Two- and three-body decays of
polarized or unpolarized parent particles could be generated, 
with the decay daughters themselves allowed to decay.
Multiple scattering was an option for all of the Monte Carlo simulations.
Monte Carlo events could be superimposed on real events from
which the hits associated with the found tracks had been removed.
Much time was spent carefully tuning the Monte Carlos to correctly
simulate multiple-hit probabilities in the MWPCs as a function
of track angle.
Wire-by-wire MWPC efficiencies and hodoscope-counter efficiencies, 
both calculated on a spill-by-spill basis, were incorporated.
The calorimeter efficiency was calculated over a ten-by-ten
grid on a run-by-run basis.
All of the Monte Carlos used the \textsc{ranlux} random number generator with
the ``luxury level'' set to four~\cite{ranlux}.
The digitized Monte Carlo data were stored in the same format as the
real data and reconstructed in exactly the same manner.
To indicate the quality of the simulation,
Fig.~\ref{fig:mass_mc_data} shows the proton-pion invariant masses 
from both data and External Monte Carlo \Xim\ra\gL\pim\ra\gp\pim\pim\ events.
The mass resolutions are comparable, with the Monte Carlo rms mass resolution
1.00\,MeV/$c^2$ and the data rms mass resolution 0.97\,MeV/$c^2$.

\section{Spectrometer Performance}
\label{sec:perform}
The spectrometer worked exceptionally well, recording 
the world's largest sample of hyperon decays in two
runs of about twelve months' total duration.

A typical 1999 run took about three hours.
During normal data-taking a run cycle of three positive-polarity runs 
followed by one negative-polarity run was used.
In 1999 (1997) a total of 71\,984 (73\,013) spills in
765 (248) runs were taken, in about 1600 (1200) hours
of sustained running.

The spectrometer acceptance for hyperons and kaons exiting 
the collimator was large.  Figure~\ref{fig:geo_acc} shows the
acceptance (without track reconstruction) for 
\gXi\ra\gL\gpi\ra\gp\gpi\gpi, \gOm\ra\gL\gK\ra\gp\gpi\gK, 
and \Kpm\ra\pipm\pip\pim\ decays, where in each case the acceptance
is for those parents (\gXi, \gOm, \gK) that decay within the 
Vacuum Decay Region. 
The decay daughters have been required to pass the various apparatus 
apertures and trigger criteria, 
and the \gL, in the case of the \gXi\ and \gOm\ decays, is
required to decay before the exit of the Vacuum Decay Region.  
The loss of \Kpm\ra\pipm\pip\pim\ acceptance at low momentum is
almost entirely due to the opposite-sign pion missing the hadronic
calorimeter.
The smaller \gXi\ and \gOm\ acceptances result mainly from
decay mesons not clearing the apertures of the Analyzing Magnet, 
and, at high momentum,  the \gL\ decaying after the exit of the 
Vacuum Decay Region.
As more than 90\% of the \gXi\ and \gOm's do indeed decay within the 
Vacuum Decay Region, the \gXi\ and \gOm\ curves effectively give the
acceptances for \gXi\ and \gOm\ particles exiting the collimator.
That is not the case for the \Kpm\ra\pipm\pip\pim\ decays because of
the long \Kpm\ lifetime.

The momentum resolution of the spectrometer was excellent.
Monte Carlo studies of the momentum resolution of the reconstructed 
pions and protons from \gXi\ra\gL\gpi\ra\gp\gpi\gpi\ decays
give a resolution $\sigma/p\,[\%] = 0.0034p + 0.097$,
as shown in Fig.~\ref{fig:mtm_res}. 
The \gL\ and \gXi\ decay-vertex resolutions were less than
1\,mm in both $x$ and $y$ at the exit of the collimator,
and about a factor of two less for decays occurring just 
before chamber C1 (Fig.~\ref{fig:vtx_res}).
The $z$-vertex resolution for both \gL\ and \gXi\ decays
was of course much worse, ranging from about 0.2\,m to 0.3\,m.
Three-particle vertices, such as those from \Kpm\ra\pipm\pip\pim\ decays, 
were better resolved.

An important constraint for all analyses was that the parent-particle 
trajectory originate at the target: this eliminated
collimator production as well as misreconstructed tracks.
Figure~\ref{fig:tgt_res} shows the excellent spectrometer
target-pointback resolution
from a set of Monte Carlo \gXi\ra\gL\gpi\ra\gp\gpi\gpi\ decays.
Note that the resolution is on the order of the target size.
The $x$ resolution was superior to that in $y$ since
the wires were oriented largely in the vertical direction.

The good spectrometer resolution, absence of significant backgrounds,
and similarity of the positive- and negative-secondary-beam data
is illustrated in Fig.~\ref{fig:farm_masses} which shows
the reconstructed parent masses from the 
\gXi\ra\gL\gpi\ra\gp\gpi\gpi,
\gOm\ra\gL\gK\ra\gp\gpi\gK,
\Kpm\ra\pipm\pip\pim,
and \Ks\ra\pip\pim\ data samples.  
The standard deviation varies from about 1.5 to 3.0\,GeV/$c^2$,
with multiple scattering and the position resolutions of the PWCs 
each contributing roughly equally.
The differences in mass resolutions and background levels 
between particle and antiparticle are very small. 
This excellent agreement in data quality between 
particle and antiparticle is essential for searching for
\CP\ asymmetries in these decays.

\section*{Acknowledgments}
We thank Fermilab's A. Baumbaugh, T. Carter, S. Hansen, K. Knickerbocker, 
J. Meadows, D. Slimmer, and J. Streets for their help in assembling and 
commissioning the DAQ system and M. Crisler for the calorimeter 
laser-calibration system. 
C.R. Dukes and J. Forsman provided invaluable help in fabricating the
spectrometer.
This work was supported by the US Department of Energy, the National Science 
Council of Taiwan, ROC, the Swiss National Science Foundation,
and the University of Virginia Institute for Nuclear and Particle Physics.

\clearpage

\section*{List of Figures}

\begin{figure}
\centerline{\includegraphics[width=14.0cm]{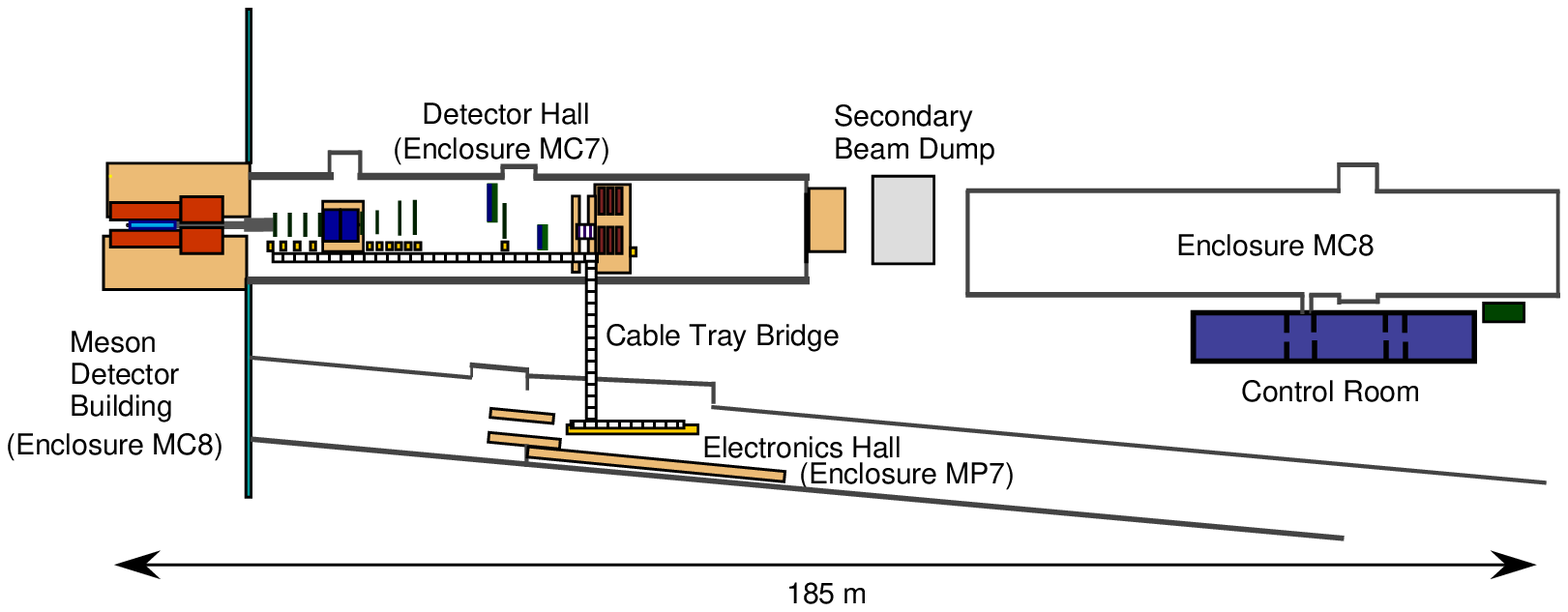}}
\mbox{}\\[0.5in]
\caption{}
\label{fig:site_layout}
\end{figure}

Fig.~\ref{fig:site_layout}.
Experimental layout: the target, Hyperon Magnet, and associated 
shielding were in enclosure MC6 in the Meson Detector Building;
the experimental apparatus in enclosure MC7;
the front end of the data-acquisition electronics
in enclosure MP7; and the back end of the data-acquisition
electronics in the Control Room alongside enclosure MC8, 
from which the physicists on shift controlled and monitored 
the experiment.

\begin{figure}
\centerline{\includegraphics[width=14.0cm]{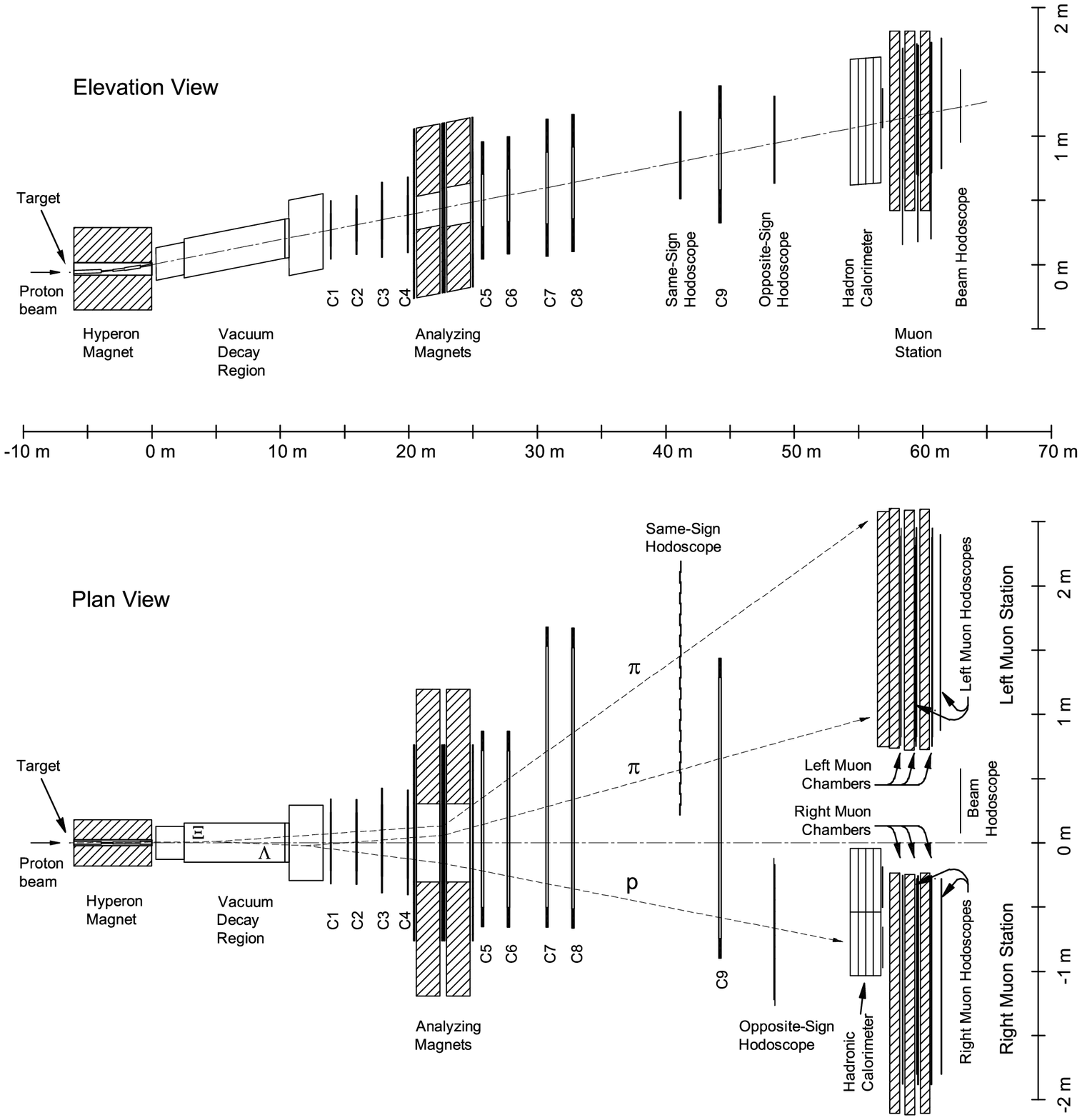}}
\mbox{}\\[0.5in]
\caption{}
\label{fig:HyperCP}
\end{figure}

Fig.~\ref{fig:HyperCP}.
Elevation and plan views of the HyperCP spectrometer.
The 1999 configuration is shown.  Superimposed on the plan view 
are the charged tracks of a \gXi\ra\gL\gpi\ra\gp\gpi\gpi\ decay.   
For clarity, enclosures, support structures, and helium bags 
are not shown.  Transverse dimensions have been exaggerated by a 
factor of ten.

\begin{figure}
\centerline{\includegraphics[width=14.0cm]{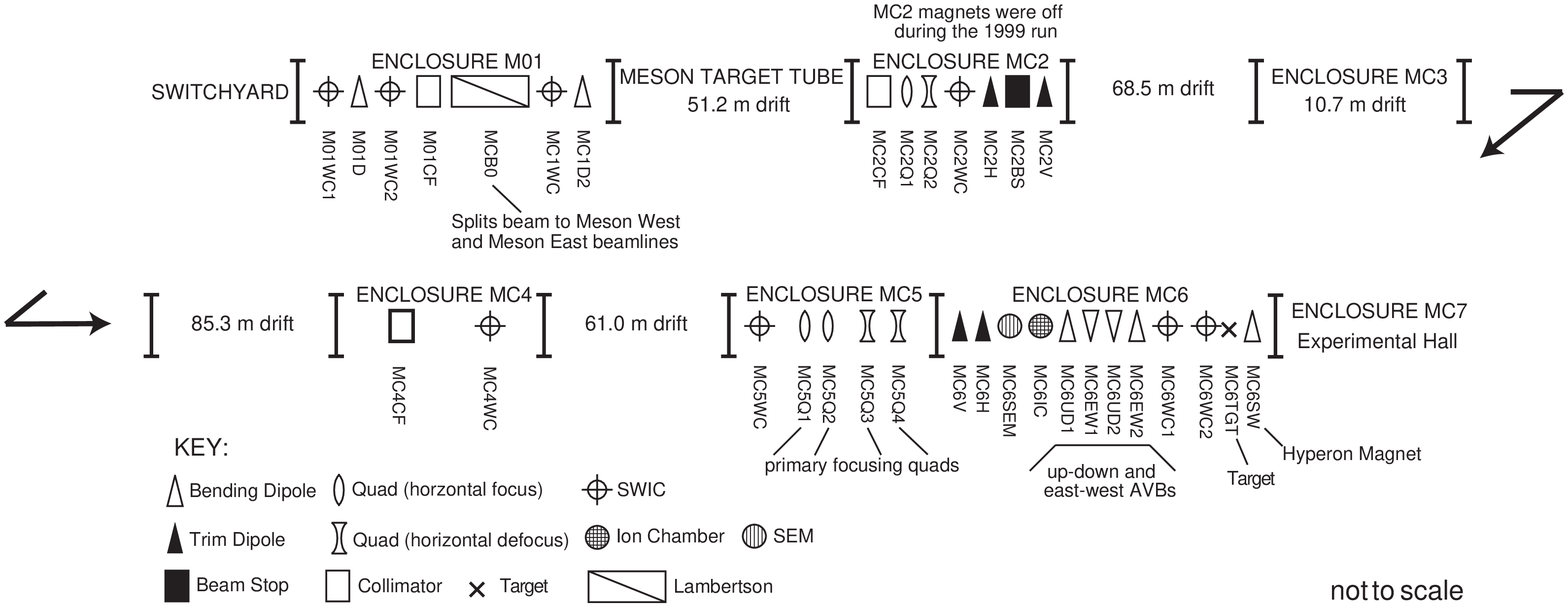}}
\mbox{}\\[0.5in]
\caption{}
\label{fig:beamline}
\end{figure}

Fig.~\ref{fig:beamline}.
Layout of the Meson Center beamline elements, starting from the M01 
enclosure (the Switchyard devices in M00 are not shown).
The last beamline element, MC6SW, is the Hyperon Magnet.

\begin{figure}
\centerline{\includegraphics[width=14.0cm]{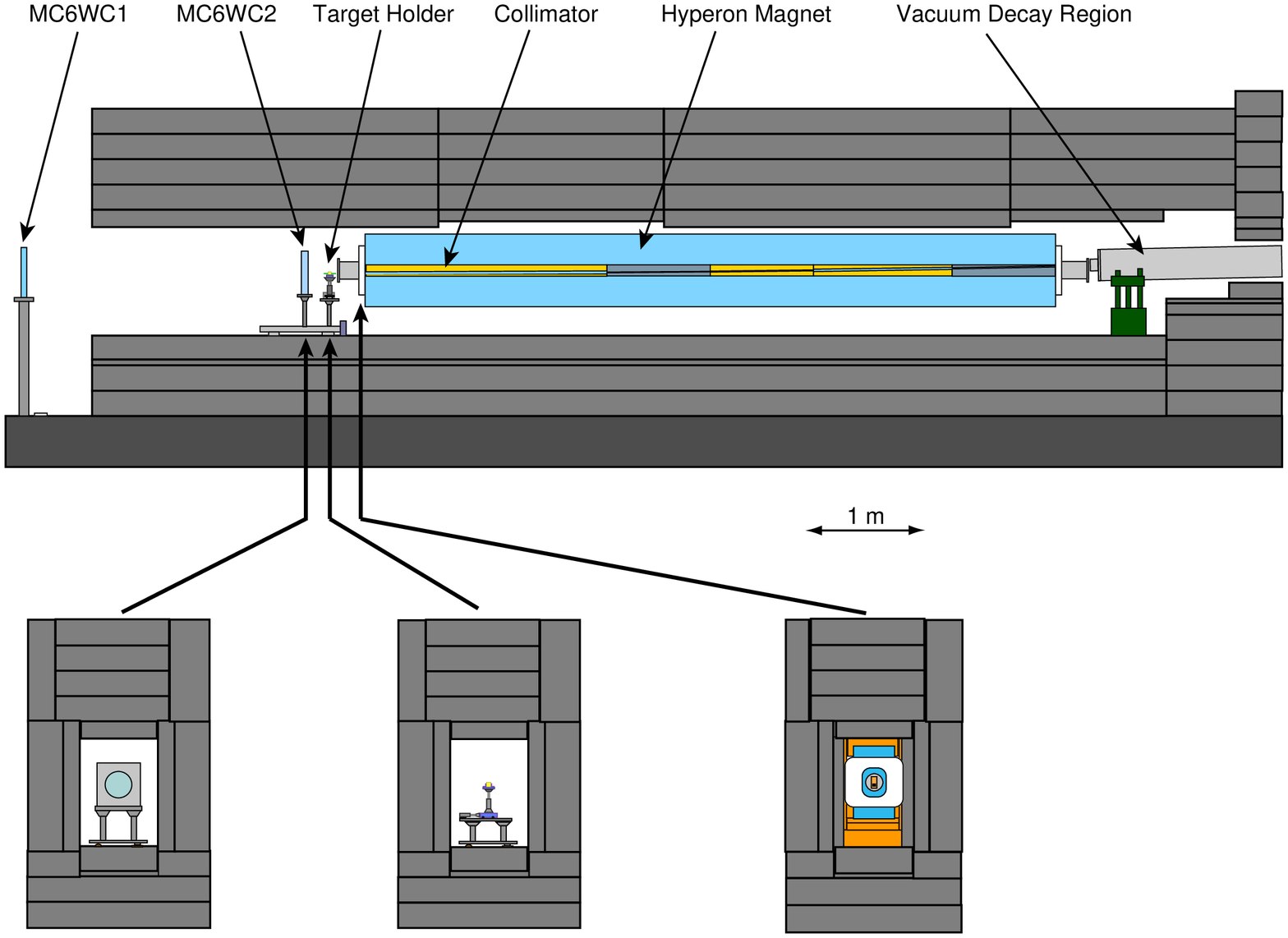}}
\mbox{}\\[0.5in]
\caption{}
\label{fig:pile}
\end{figure}

Fig.~\ref{fig:pile}.
Elevation view of the secondary-beam forming area, 
showing the two target SWICs, target holder, Hyperon Magnet, 
and part of the Vacuum Decay Region.  Cross sections at the 
downstream SWIC, target holder, and collimator entrance are shown.
The entire area was enclosed in iron and concrete shielding.

\begin{figure}
\centerline{\includegraphics[width=10.0cm]{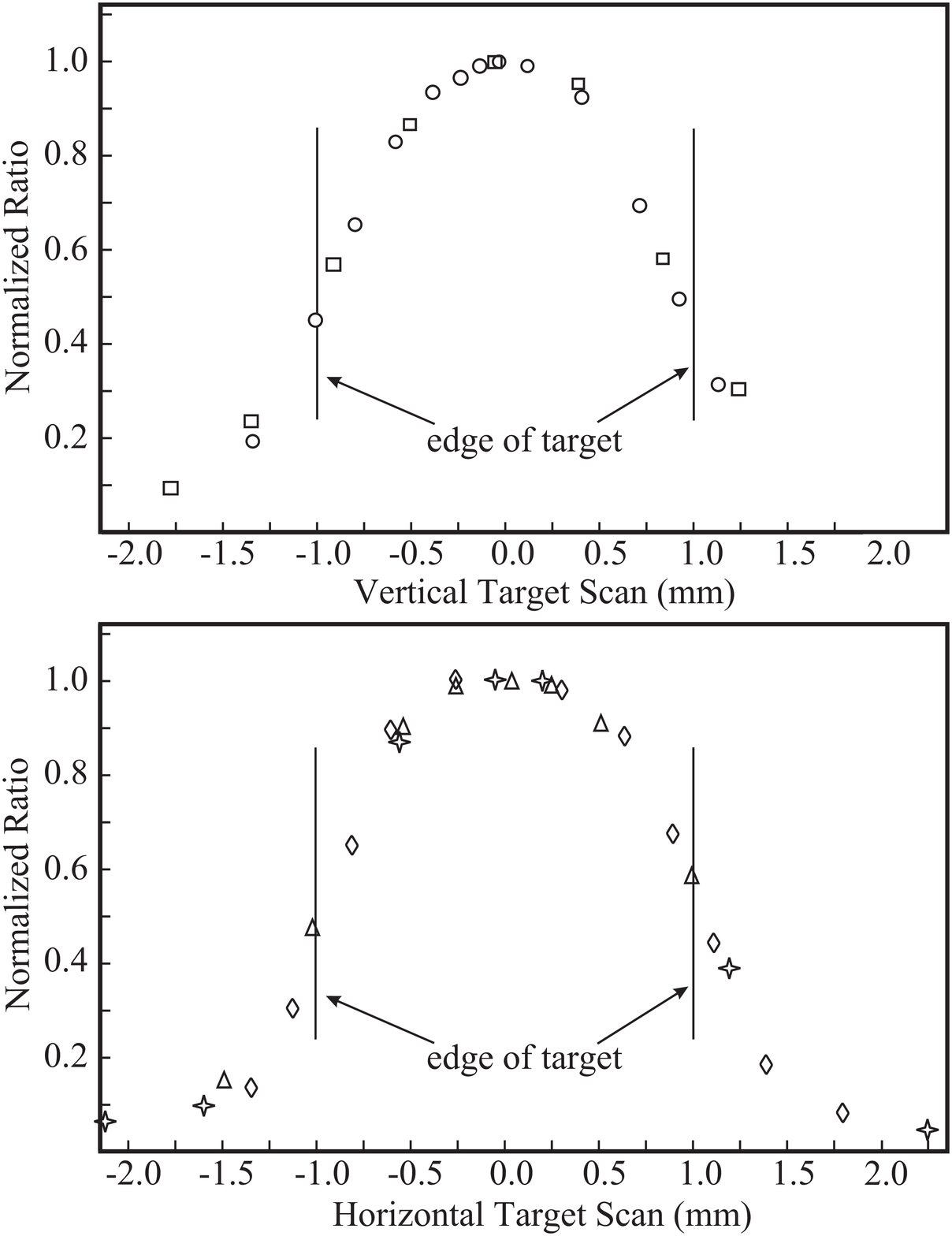}}
\mbox{}\\[0.5in]
\caption{}
\label{target_scan}
\end{figure}

Fig.~\ref{target_scan}.
Typical ratios of counting rates in the spectrometer per 
incident proton (normalized to a maximum value of unity) 
versus target position in the vertical (top) and horizontal 
(bottom) directions.  Symbols indicate various target scans.
From these distributions the shape of the primary beam was 
determined.  The target size is superimposed to set the scale.

\begin{figure}
\centerline{\includegraphics[width=10.0cm]{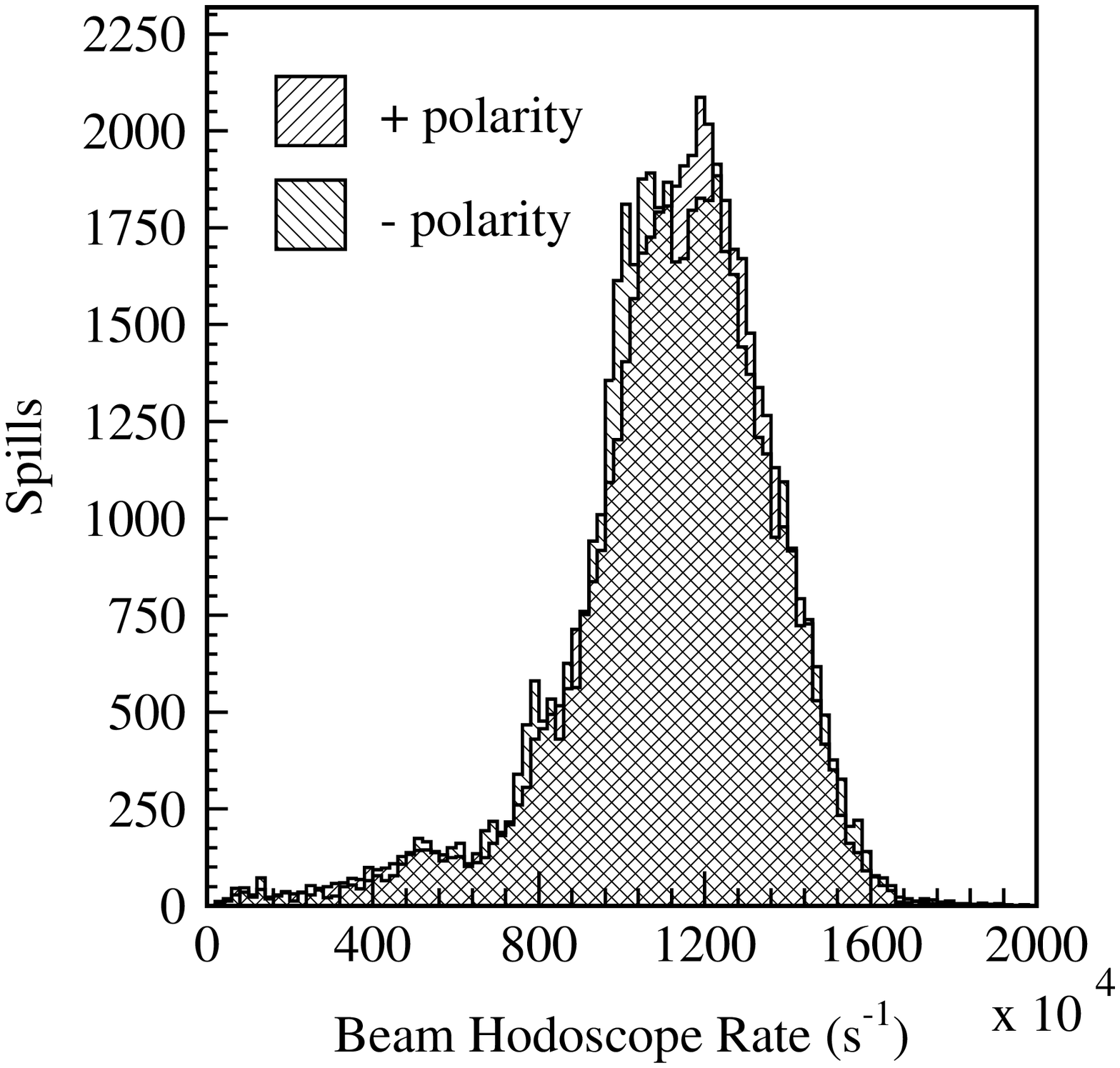}}
\mbox{}\\[0.5in]
\caption{}
\label{fig:rate_comp_beam}
\end{figure}

Fig.~\ref{fig:rate_comp_beam}.
Secondary beam intensity for all of the 1999 spills, as
measured by the Beam Hodoscope counters at the rear of
the spectrometer.  Spills without beam have been removed
and the number of negative-polarity spills has 
been scaled up to the number of positive-polarity spills.

\begin{figure}
\centerline{\includegraphics[width=14.0cm]{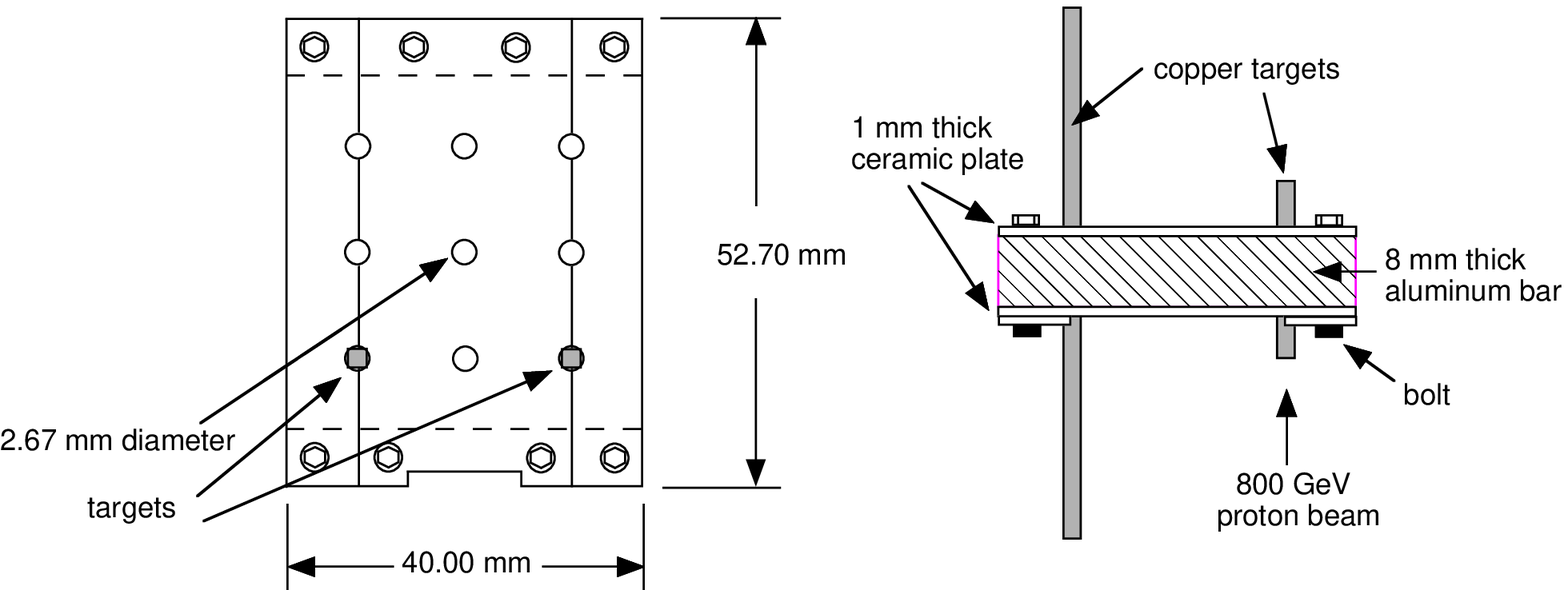}} 
\mbox{}\\[0.5in]
\caption{}
\label{fig:target} 
\end{figure}

Fig.~\ref{fig:target}.
Front (left) and top (right) views of the  target holder.

\begin{figure}
\centerline{\includegraphics[height=14.0cm,angle=-90]{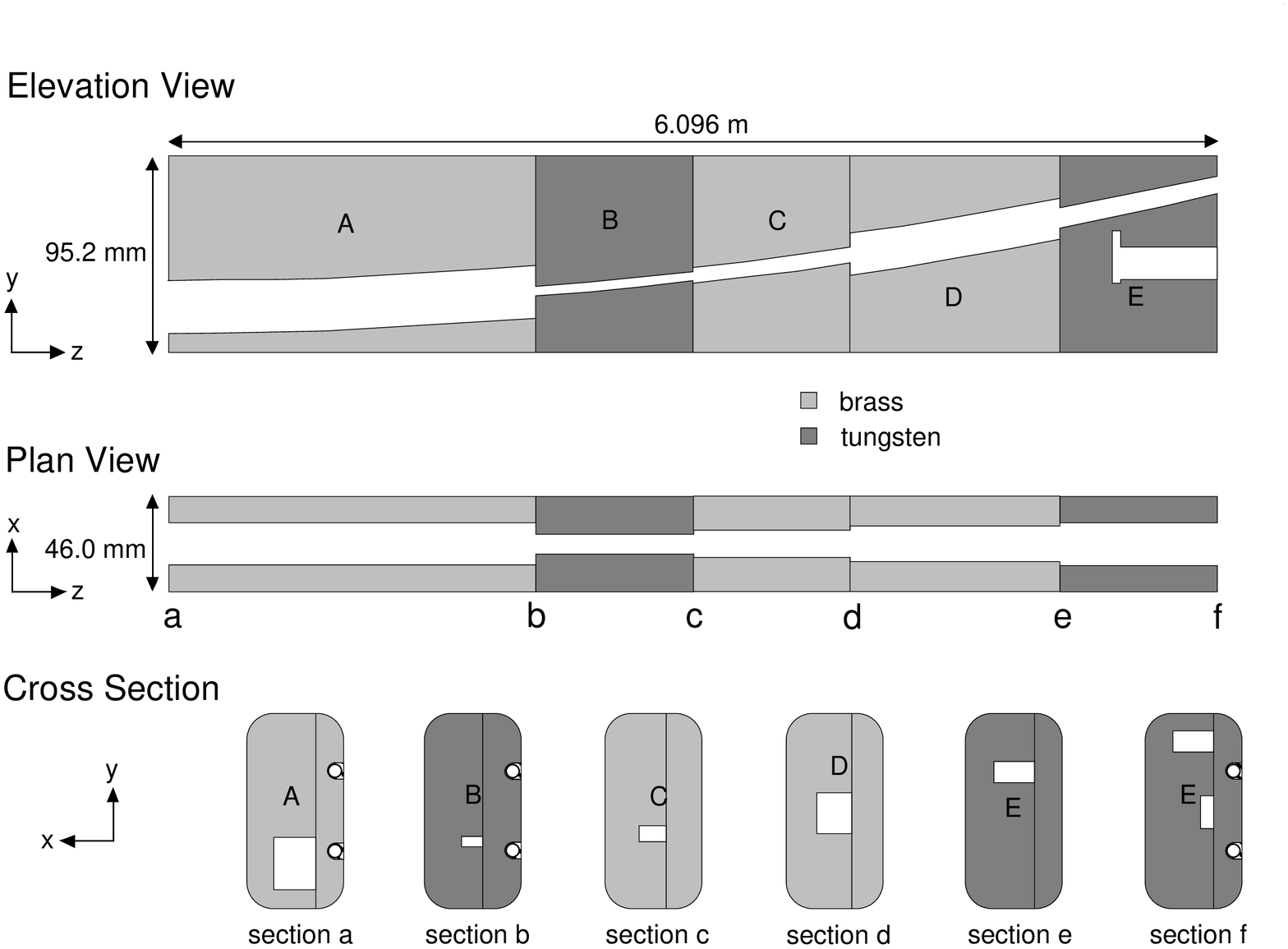}}
\mbox{}\\[0.5in]
\caption{}
\label{fig:channel}
\end{figure}

Fig.~\ref{fig:channel}.
Geometry of the collimator (transverse dimensions exaggerated).
The water-cooling pipes, indicated in the cross sections at locations a, b,
and f, cooled collimator segments A, B, and E directly; segments C and D
were cooled by conduction to the adjacent blocks. The milled slot in segment 
E accommodated two Hall probes.

\clearpage

\begin{figure}
\centerline{\includegraphics[width=14.0cm]{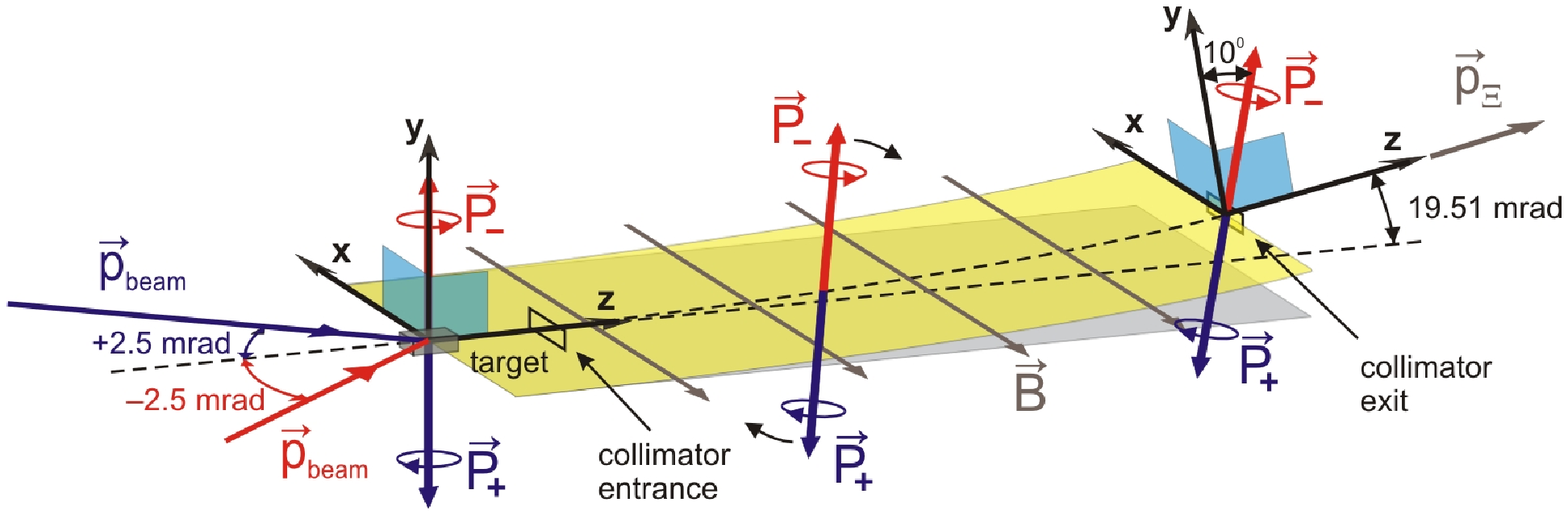}}
\mbox{}\\[0.5in]
\caption{}
\label{fig:precession_geo}
\end{figure}

Fig.~\ref{fig:precession_geo}.
Momentum and polarization directions at the target and
collimator for the nonzero-targeting-angle running.
Targeting angles of $+2.5$\,mrad and $-2.5$\,mrad
respectively produced the \Xim\ polarization vectors, 
$\vec{\rm P}_+$ and $\vec{\rm P}_-$, along the ${\pm}y$ axes.
The \Xim\ polarization precessed about the $x$ axis,
ending up about $10^{\circ}$ from the ${\pm}y$ axis
in the Spectrometer Coordinate System.

\begin{figure}
\centerline{\includegraphics[width=10.0cm]{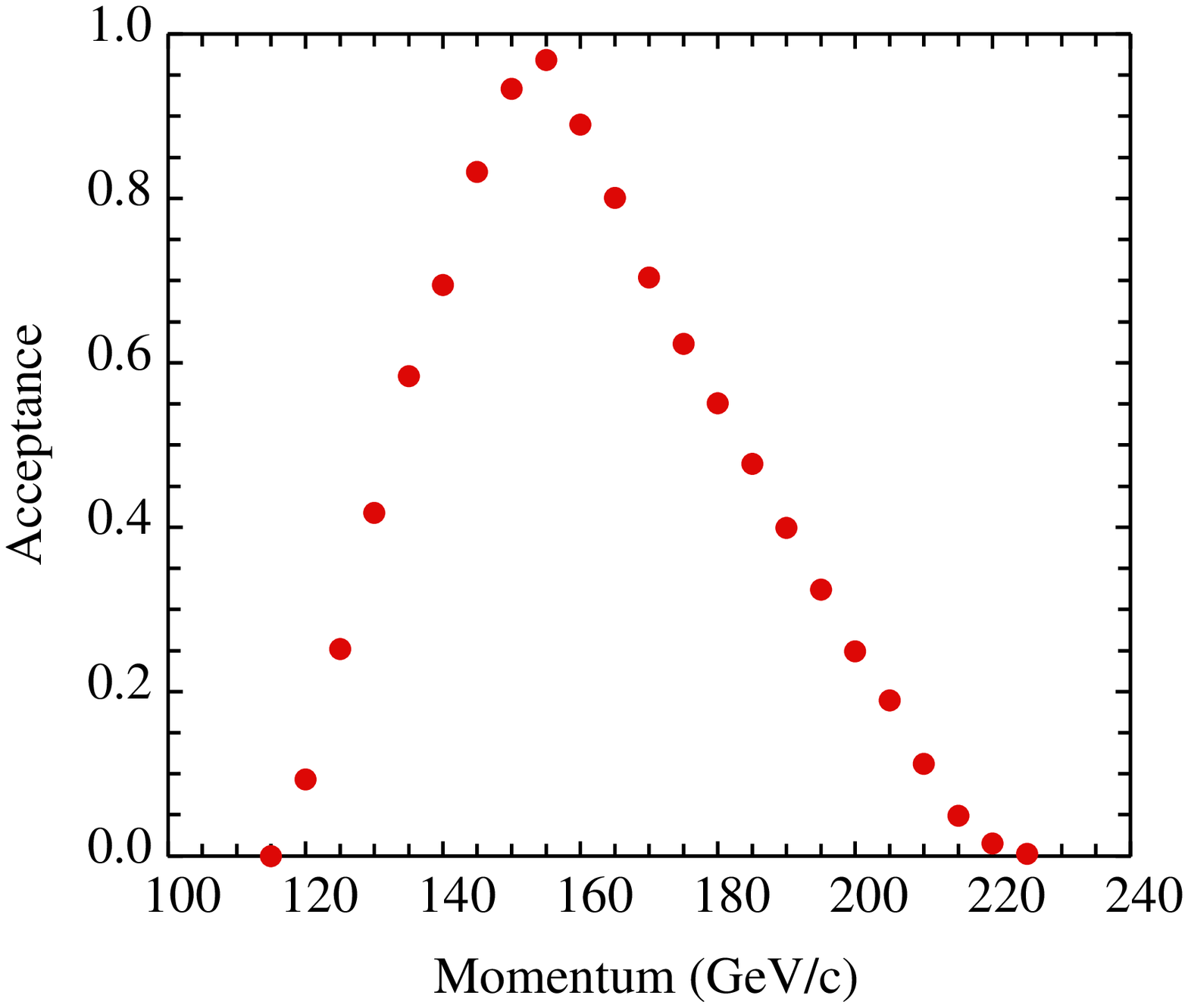}}
\mbox{}\\[0.5in]
\caption{}
\label{fig:chanacc}
\end{figure}

Fig.~\ref{fig:chanacc}.
Acceptance of the collimator at a magnetic field of 1.667\,T.

\begin{figure}
\centerline{\includegraphics[width=10.0cm]{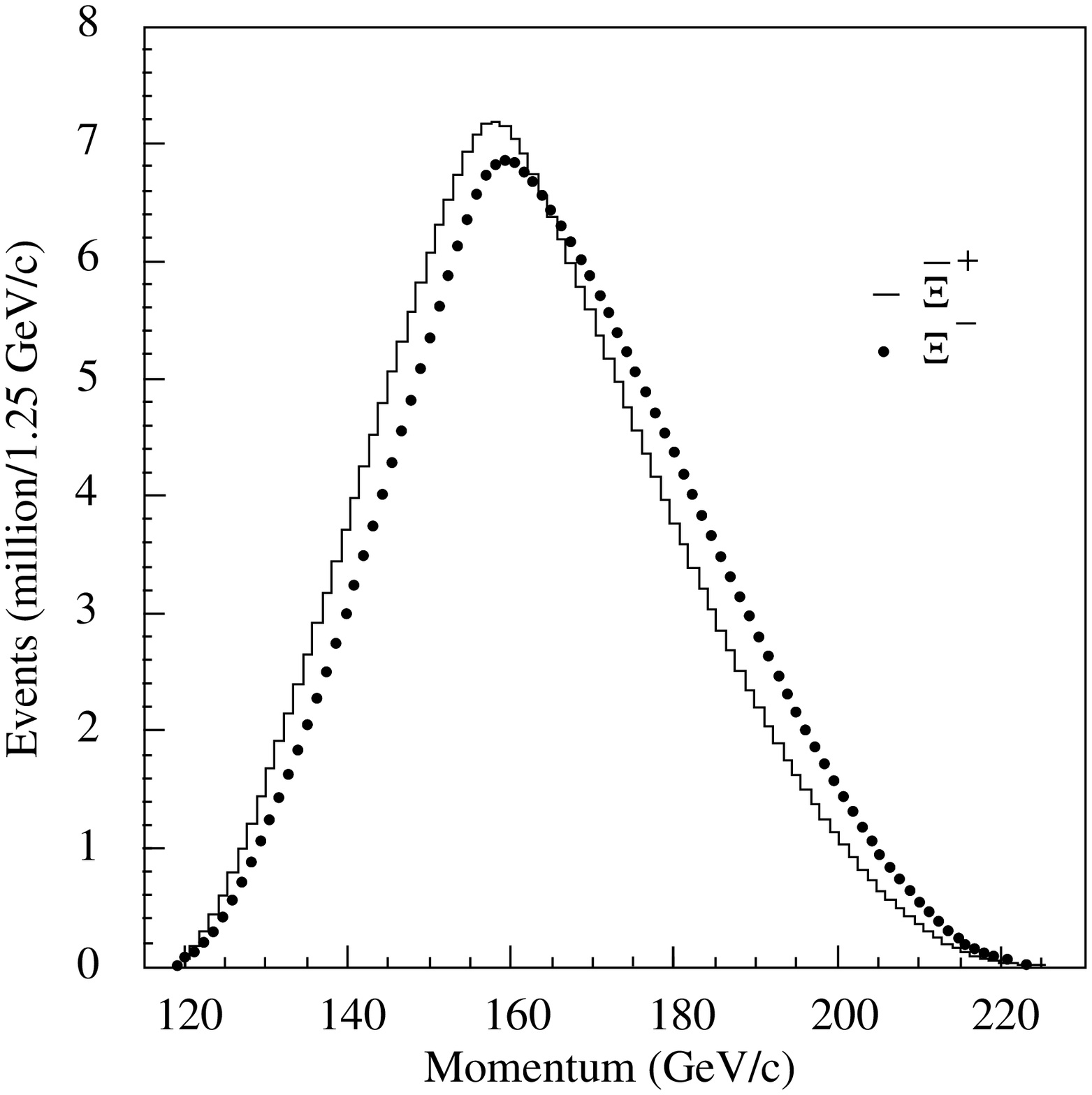}}
\mbox{}\\[0.5in]
\caption{}
\label{fig:xivsxib}
\end{figure}

Fig.~\ref{fig:xivsxib}.
Reconstructed momentum distributions of $\Xi^-$  
and $\overline{\Xi}{}^+$ events from the 1999 run.

\begin{figure}
\centerline{\includegraphics[width=10.0cm]{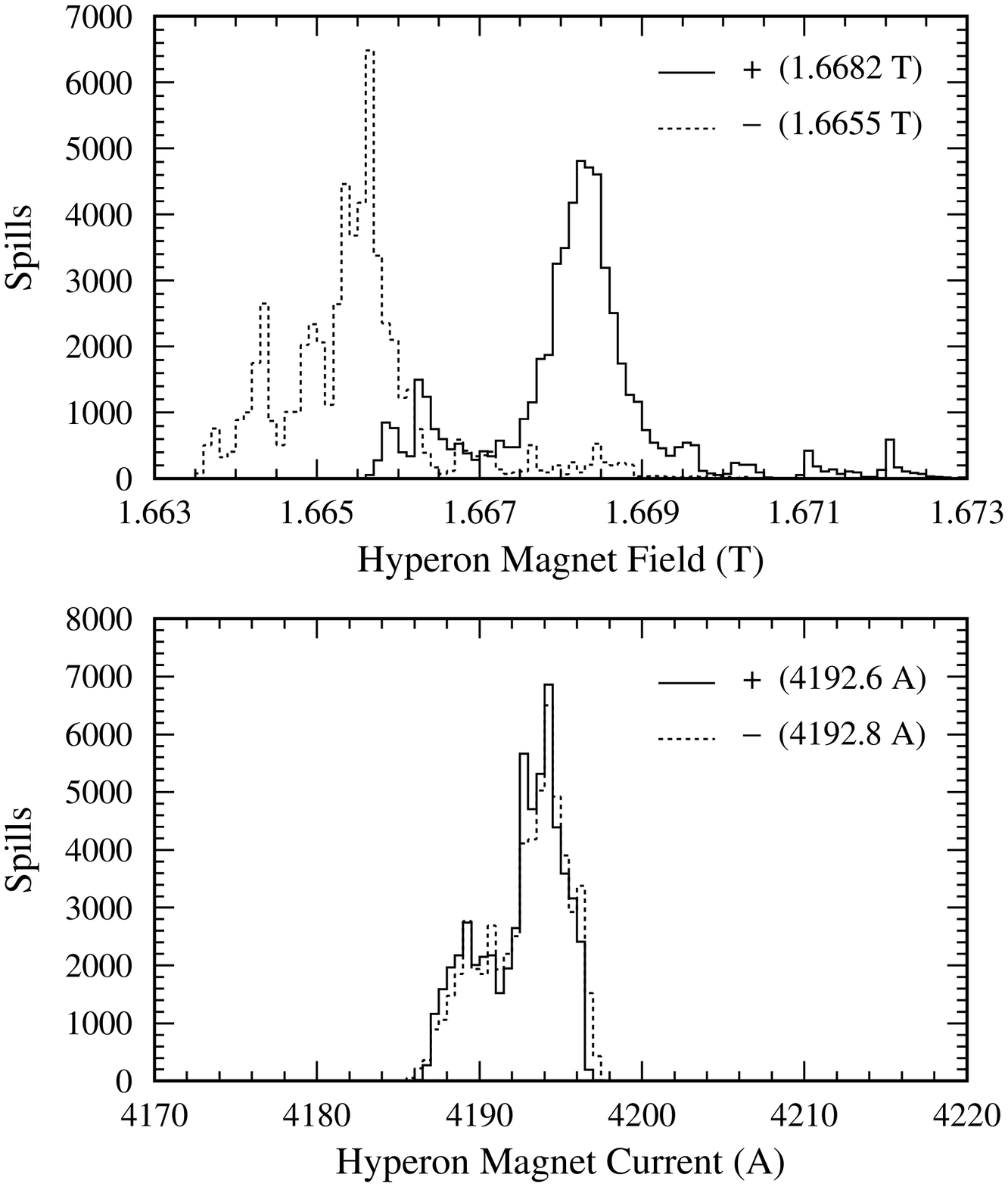}}
\mbox{}\\[0.5in]
\caption{}
\label{fig:b2field}
\end{figure}

Fig.~\ref{fig:b2field}.
Spill-by-spill values of the magnitude of the Hyperon Magnet 
magnetic fields (top) and currents (bottom) during the 1999 run.
Values for positive (negative) secondary-beam polarity are
indicated by solid (dashed) lines.  The average values are 
given in parentheses.  The top plot shows the average of the two 
Hall-probe readings.

\begin{figure}
\centerline{\includegraphics[height=14.0cm,angle=-90]{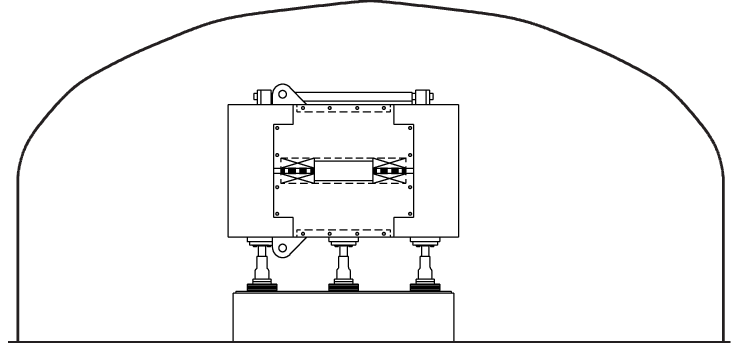}}
\mbox{}\\[0.5in]
\caption{}
\label{fig:bm109_front}
\end{figure}

Fig.~\ref{fig:bm109_front}.
Front view of the upstream Analyzing Magnet. The experimental
enclosure MC7 is shown.

\begin{figure}
\centerline{\includegraphics[width=10.0cm]{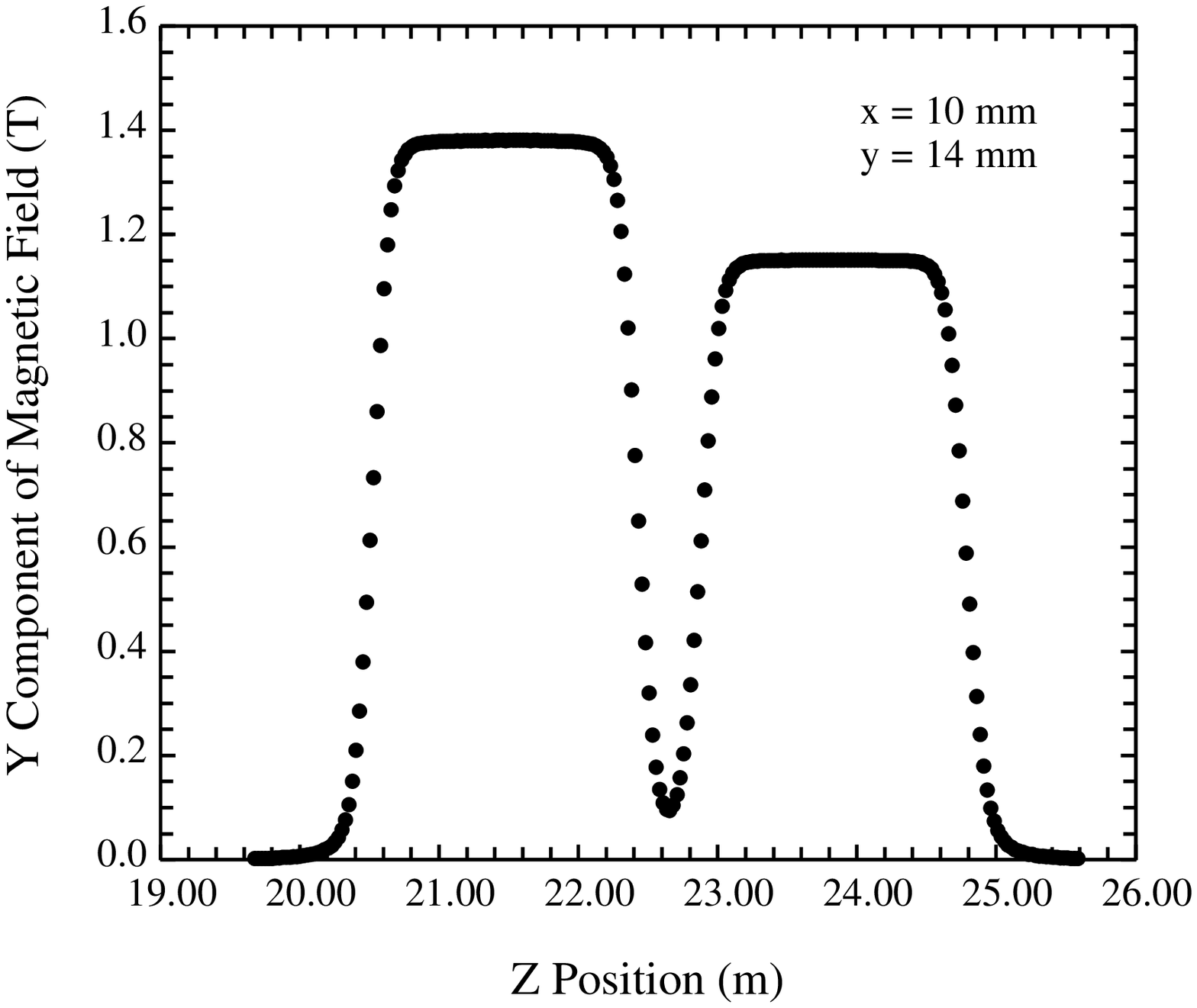}}
\mbox{}\\[0.5in]
\caption{}
\label{fig:byfield}
\end{figure}

Fig.~\ref{fig:byfield}.
The major ($y$) component of the magnetic fields in the two
Analyzing Magnets versus $z$ at $x = 10$\,mm and $y = 14$\,mm.

\begin{figure}
\centerline{\includegraphics[width=10.0cm]{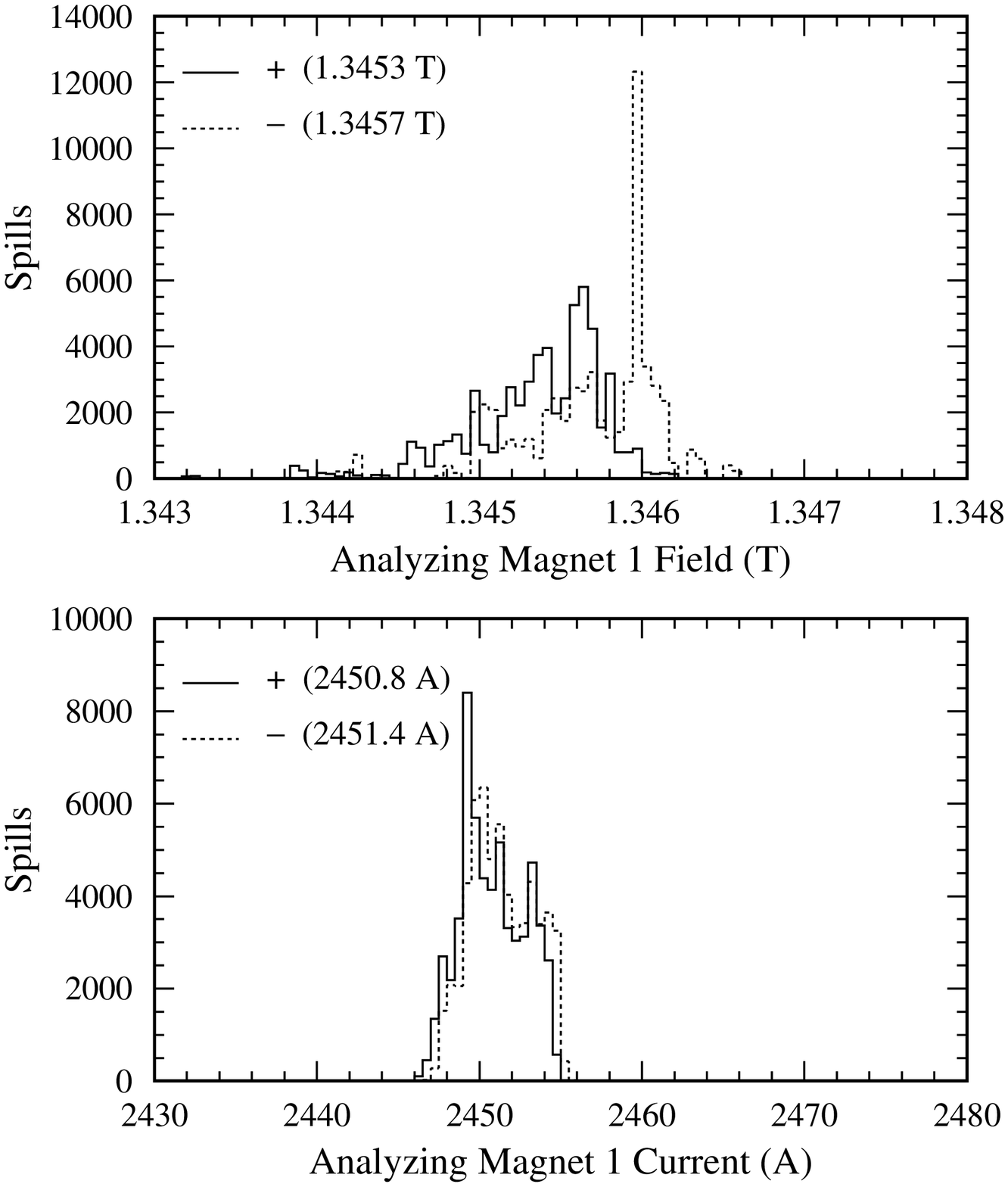}}
\mbox{}\\[0.5in]
\caption{}
\label{fig:ana1_field}
\end{figure}

Fig.~\ref{fig:ana1_field}.
Spill-by-spill values of the magnitude of the Analyzing Magnet 1 
magnetic fields (top) and currents (bottom) during the 1999 run.
Values for positive (negative) secondary-beam polarity are
indicated by solid (dashed) lines.  The average values are given 
in parentheses.

\begin{figure}
\centerline{\includegraphics[width=10.0cm]{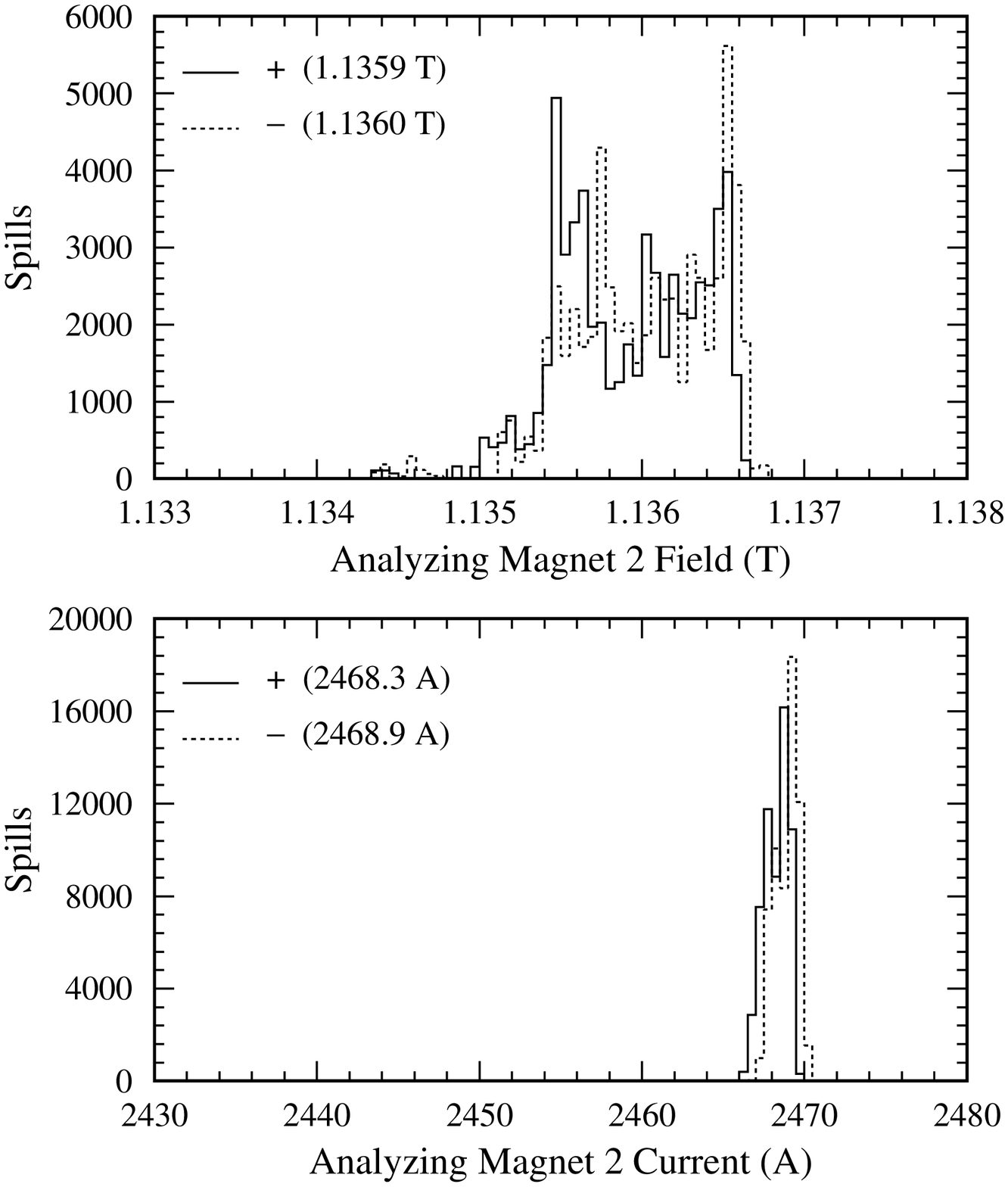}}
\mbox{}\\[0.5in]
\caption{}
\label{fig:ana2_field}
\end{figure}

Fig.~\ref{fig:ana2_field}.
Spill-by-spill values of the magnitudes of the Analyzing Magnet 2
magnetic fields (top) and currents (bottom) during the 1999 run.  
Values for positive (negative) secondary-beam polarity are 
indicated by solid (dashed) lines.  The average values are given 
in parentheses.

\begin{figure}
\centerline{\includegraphics[width=12.0cm]{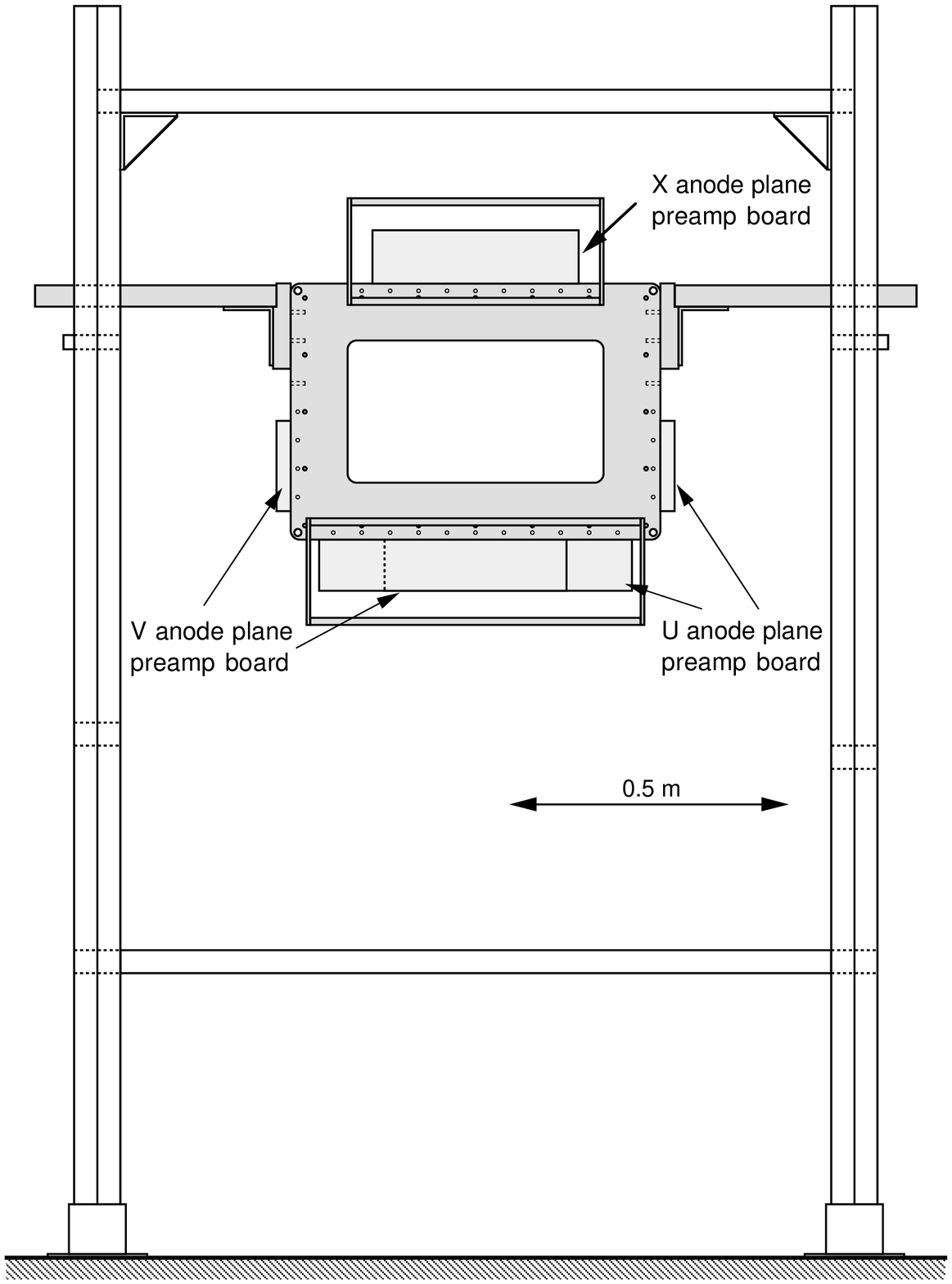}}
\mbox{}\\[0.5in]
\caption{}
\label{fig:mwpc_c1}
\end{figure}

Fig.~\ref{fig:mwpc_c1}.
Front view of chamber C1.  Chamber C2 was identical; C3 and C4 were 
similar but larger.

\begin{figure}
\centerline{\includegraphics[width=12.0cm]{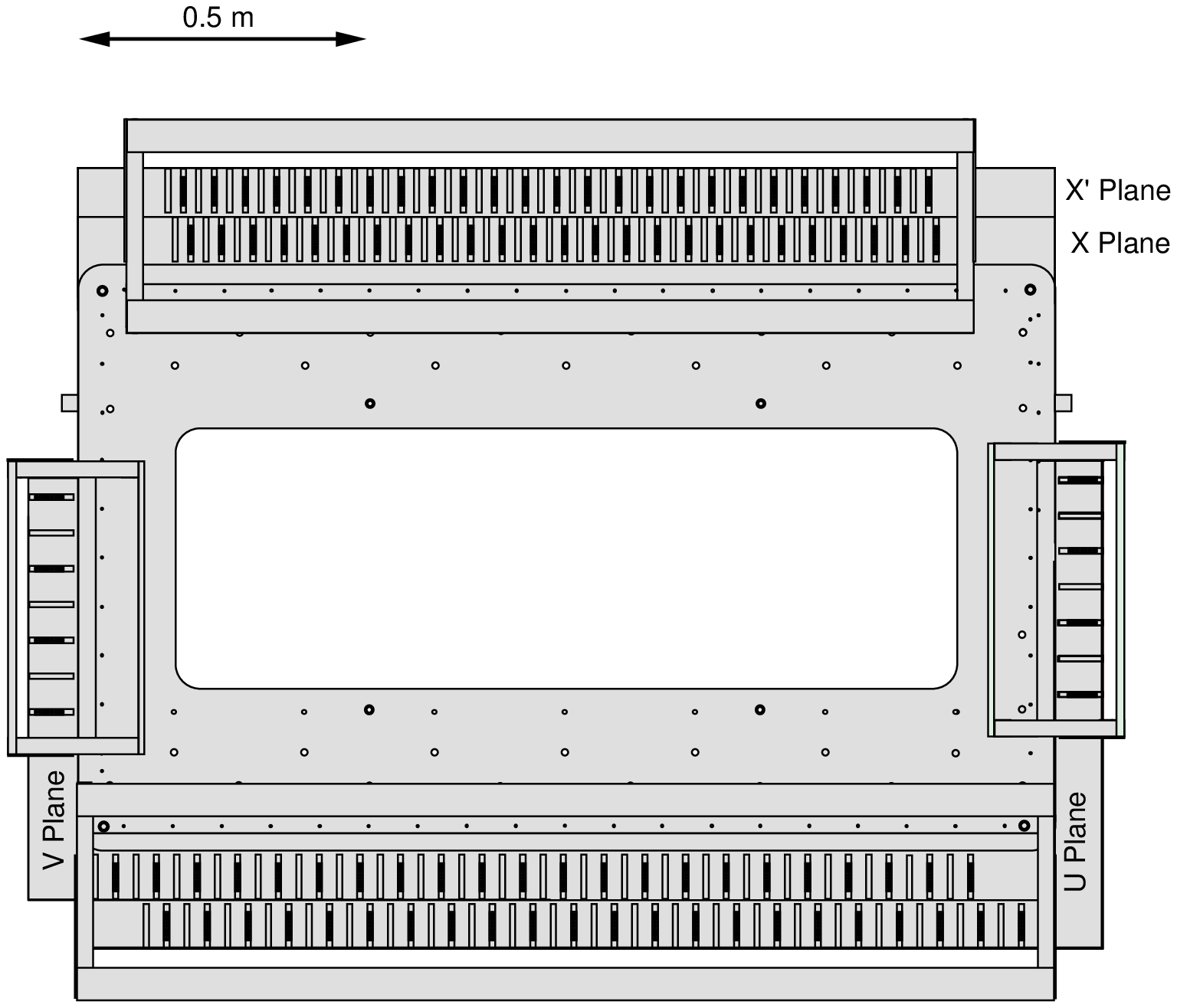}}
\mbox{}\\[0.5in]
\caption{}
\label{fig:mwpc_c5}
\end{figure}

Fig.~\ref{fig:mwpc_c5}.
Front view of chamber C5.  The chamber stand is not shown.  
Chamber C6 was identical; C7--C9 were similar but larger.

\begin{figure}
\centerline{\includegraphics[height=14.0cm,angle=-90]{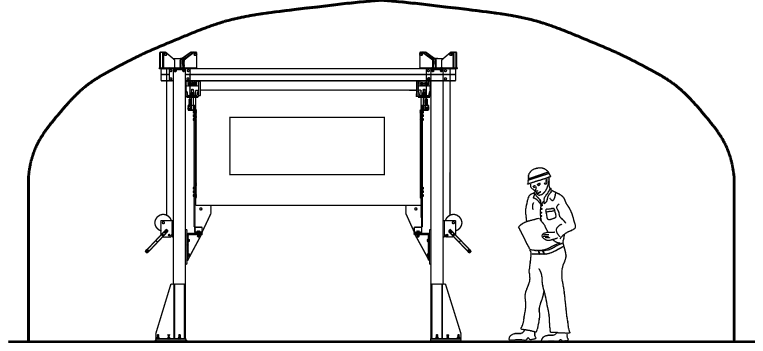}}
\mbox{}\\[0.5in]
\caption{}
\label{fig:c7_front_stand}
\end{figure}

Fig.~\ref{fig:c7_front_stand}.
Front view of chamber C7 and its stand.

\begin{figure}
\centerline{\includegraphics[width=8.0cm]{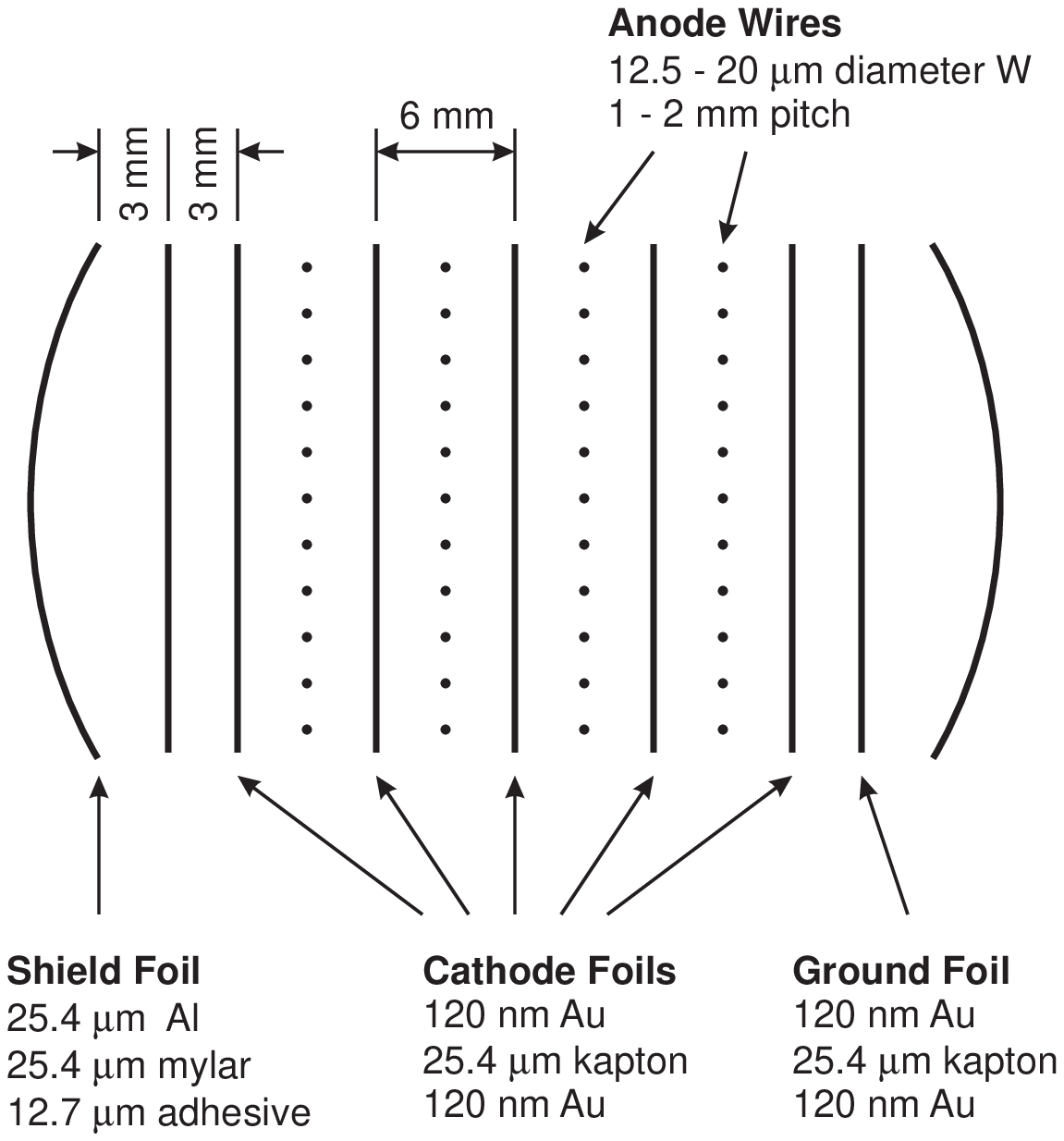}}
\mbox{}\\[0.5in]
\caption{}
\label{fig:pwc_planes}
\end{figure}

Fig.~\ref{fig:pwc_planes}.
Layout of planes in the multiwire proportional chambers.  

\begin{figure}
\centerline{\includegraphics[width=10.0cm]{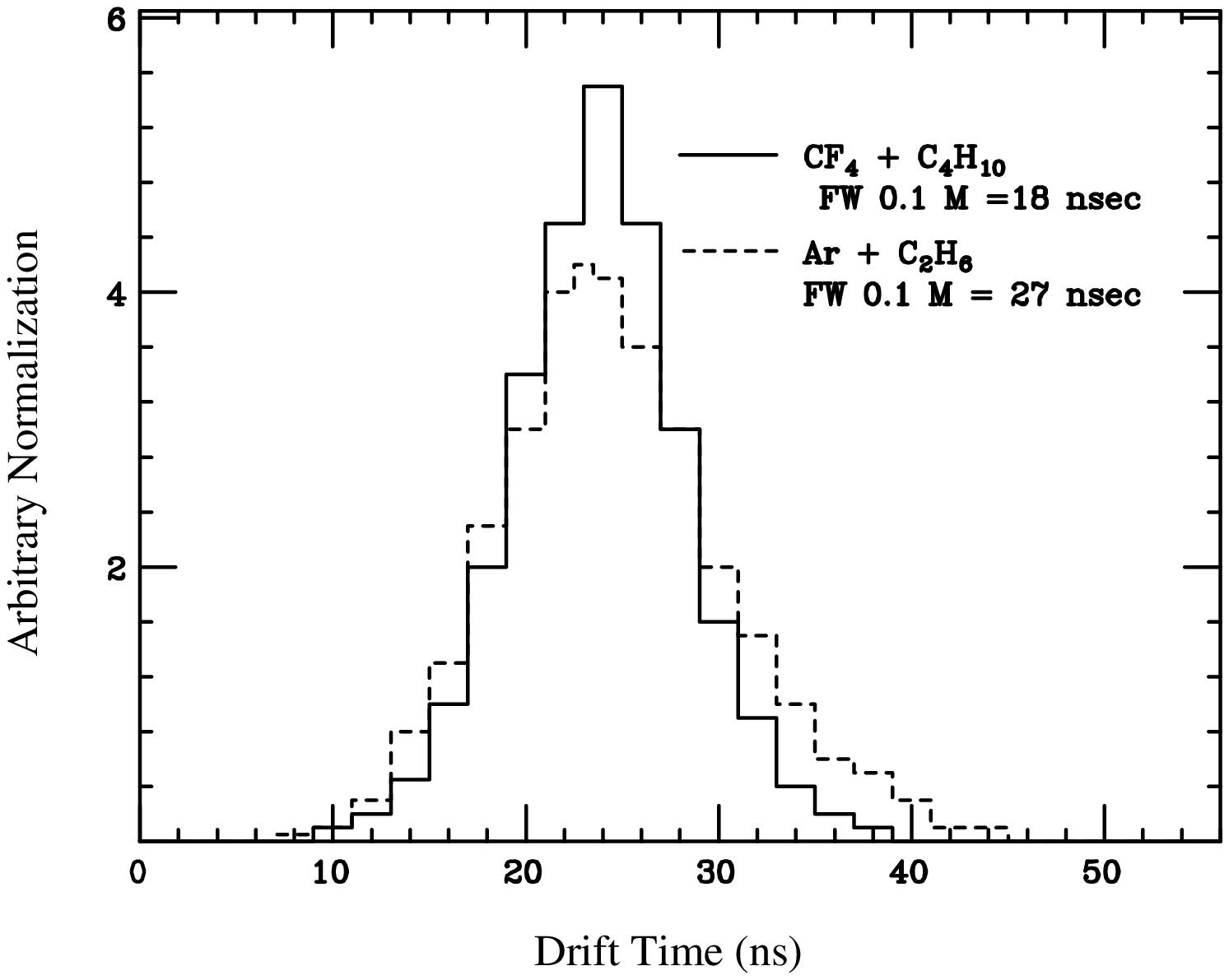}}
\mbox{}\\[0.5in]
\caption{}
\label{fig:pwc_speed}
\end{figure}

Fig.~\ref{fig:pwc_speed}.
Time response of C1 with argon-ethane and CF$_4$-isobutane gas
mixtures.  The full width at 10\%-maximum is 18\,ns for the 
CF$_4$-isobutane mixture and 27\,ns for argon-ethane.

\clearpage

\begin{figure}
\centerline{\includegraphics[width=14.0cm]{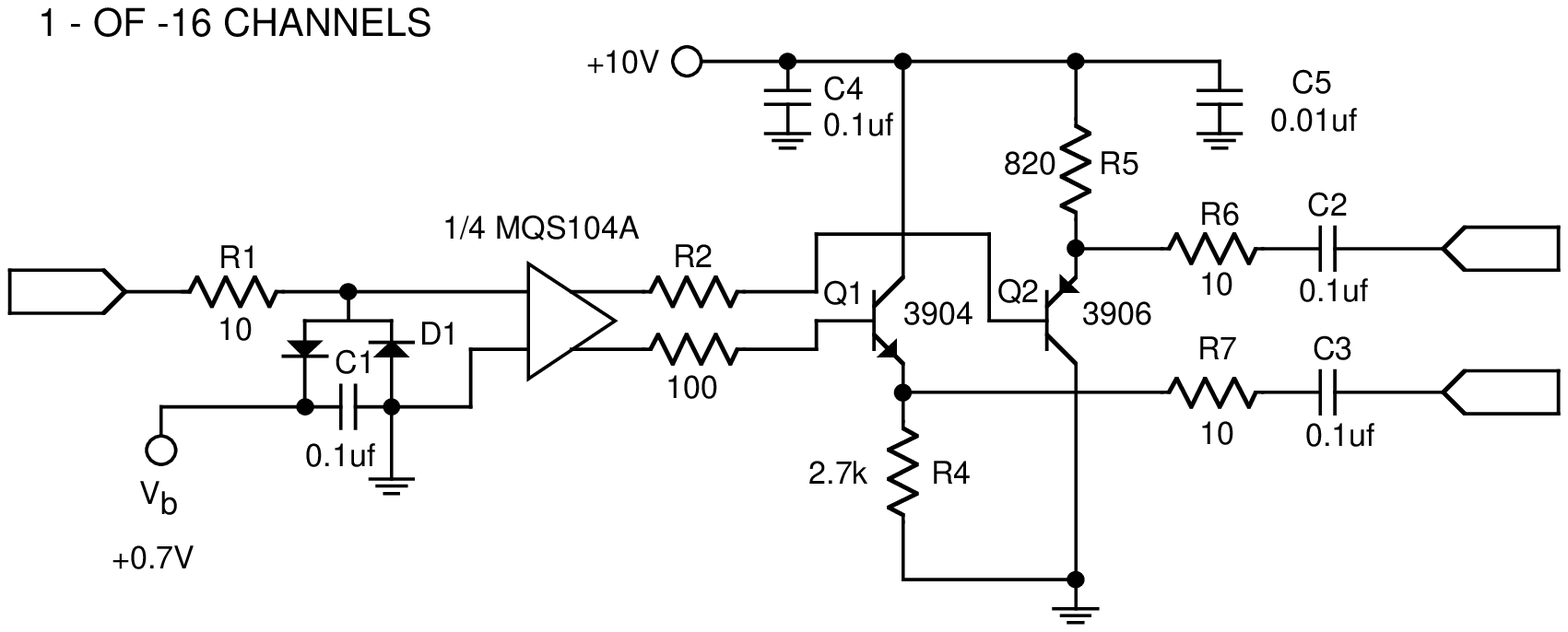}}
\mbox{}\\[0.5in]
\caption{}
\label{fig:preamp_schematic}
\end{figure}

Fig.~\ref{fig:preamp_schematic}.
Preamplifier schematic.

\begin{figure}
\centerline{\includegraphics[width=14.0cm]{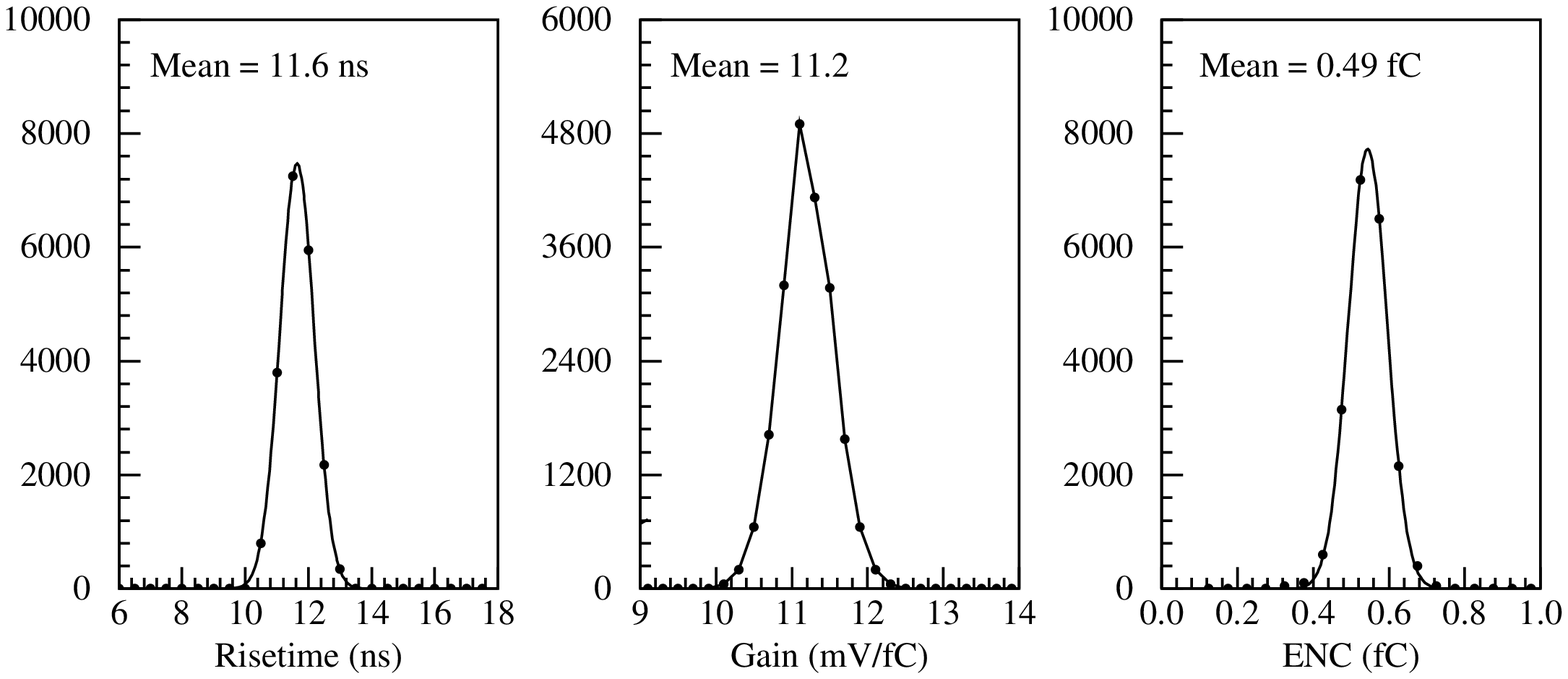}}
\mbox{}\\[0.5in]
\caption{}
\label{fig:preamp_test}
\end{figure}

Fig.~\ref{fig:preamp_test}.
Preamplifier test results.  Shown are the rise time,
gain, and equivalent noise charge.

\begin{figure}
\centerline{\includegraphics[width=14.0cm]{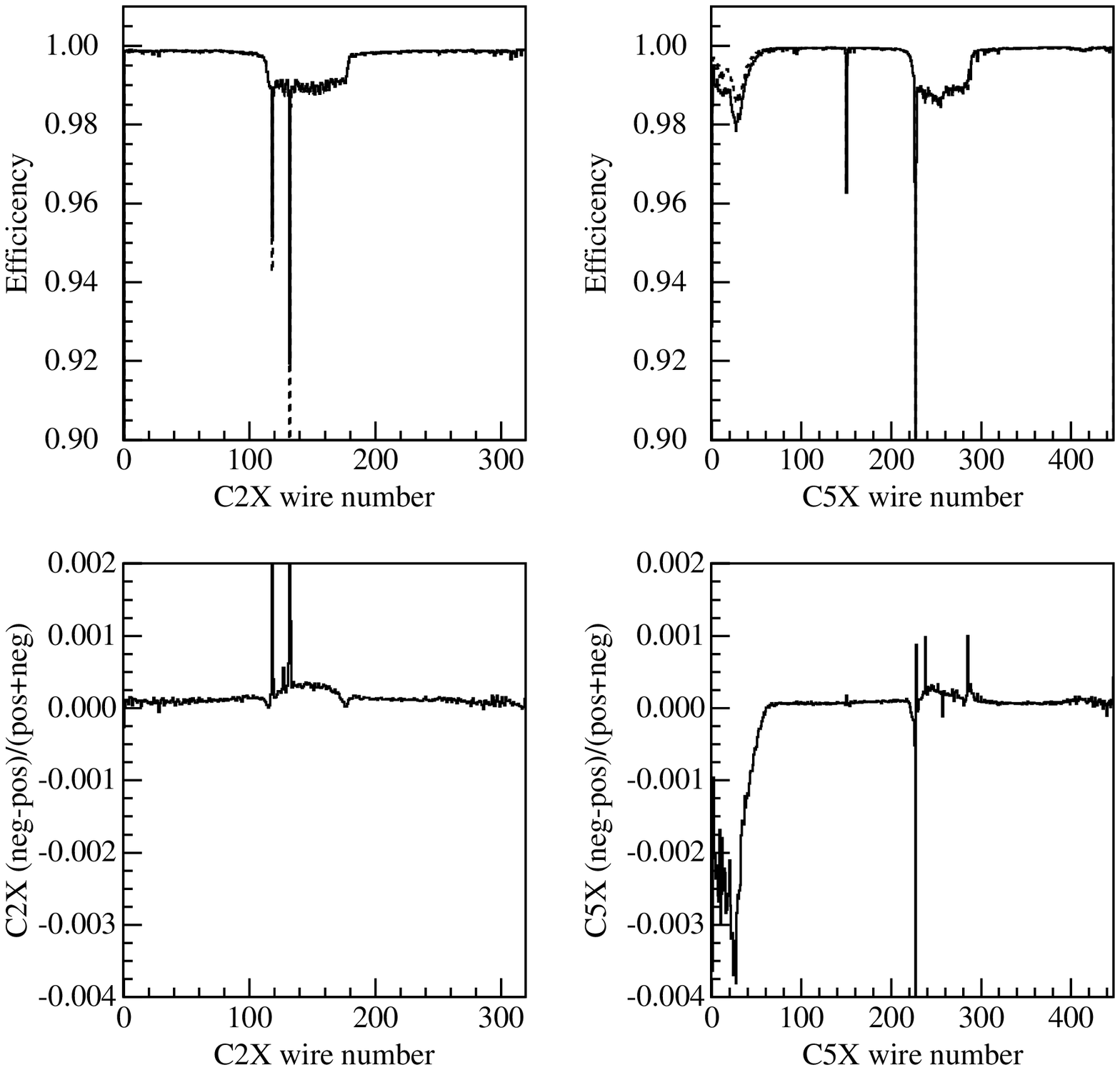}}
\mbox{}\\[0.5in]
\caption{}
\label{fig:pwc_eff}
\end{figure}

Fig.~\ref{fig:pwc_eff}.
Typical wire-by-wire MWPC efficiencies from a 1999 run.  The 
efficiencies of the X plane of one upstream (C2) and one 
downstream (C5) wire chamber are shown.  The top plots show 
the efficiencies: solid (dashed) lines are for negative (positive) 
secondary-beam polarity.  The bottom plots show efficiency 
differences divided by sums.  The secondary-beam region as 
well as several ``sick'' wires are clearly apparent.

\begin{figure}
\centerline{\includegraphics[width=14.0cm]{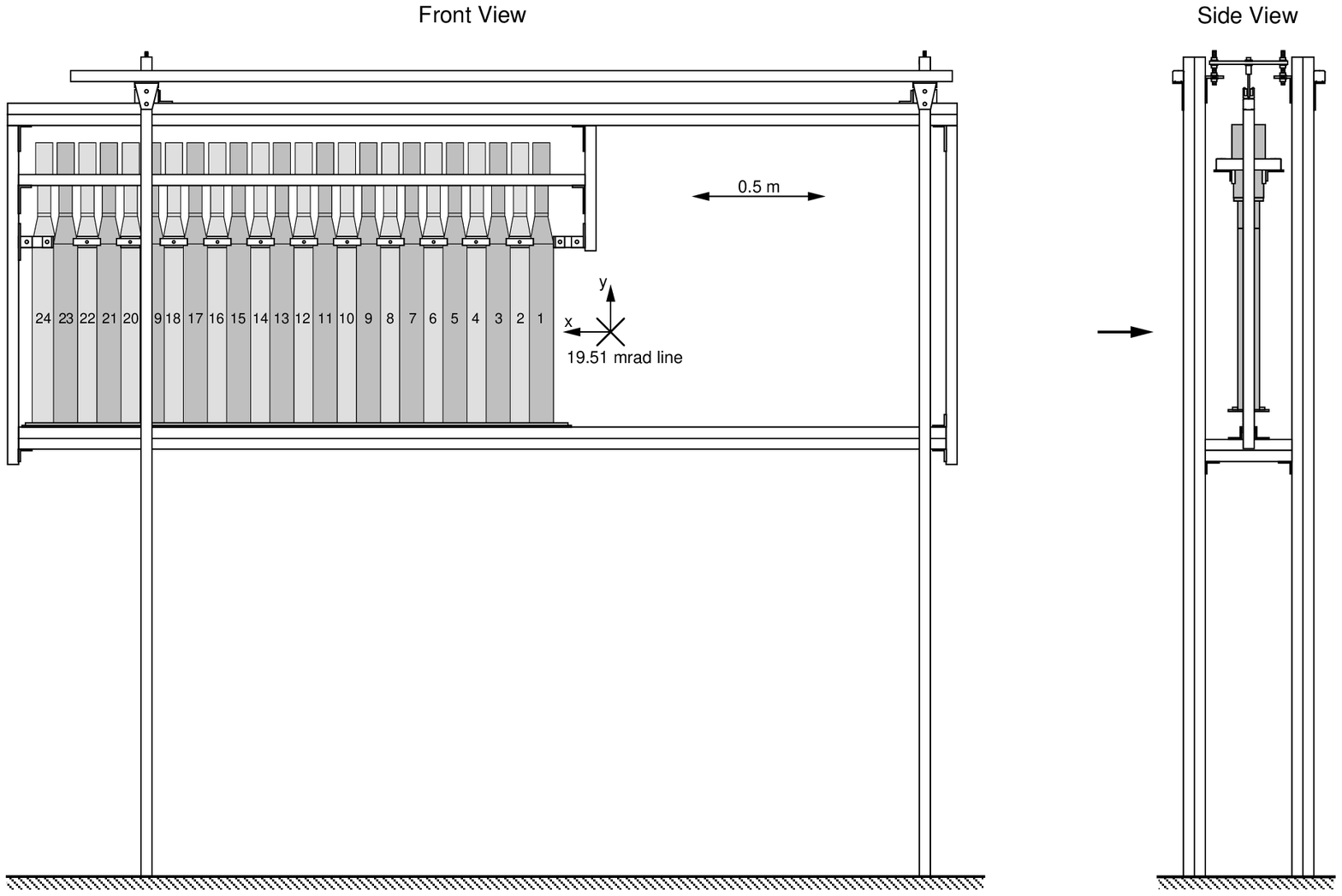}}
\mbox{}\\[0.5in]
\caption{}
\label{fig:ss_hodo}
\end{figure}

Fig.~\ref{fig:ss_hodo}.
Front and side views of the SS hodoscope.  Odd (even) numbered
counters formed the front (back) plane.

\begin{figure}
\centerline{\includegraphics[width=14.0cm]{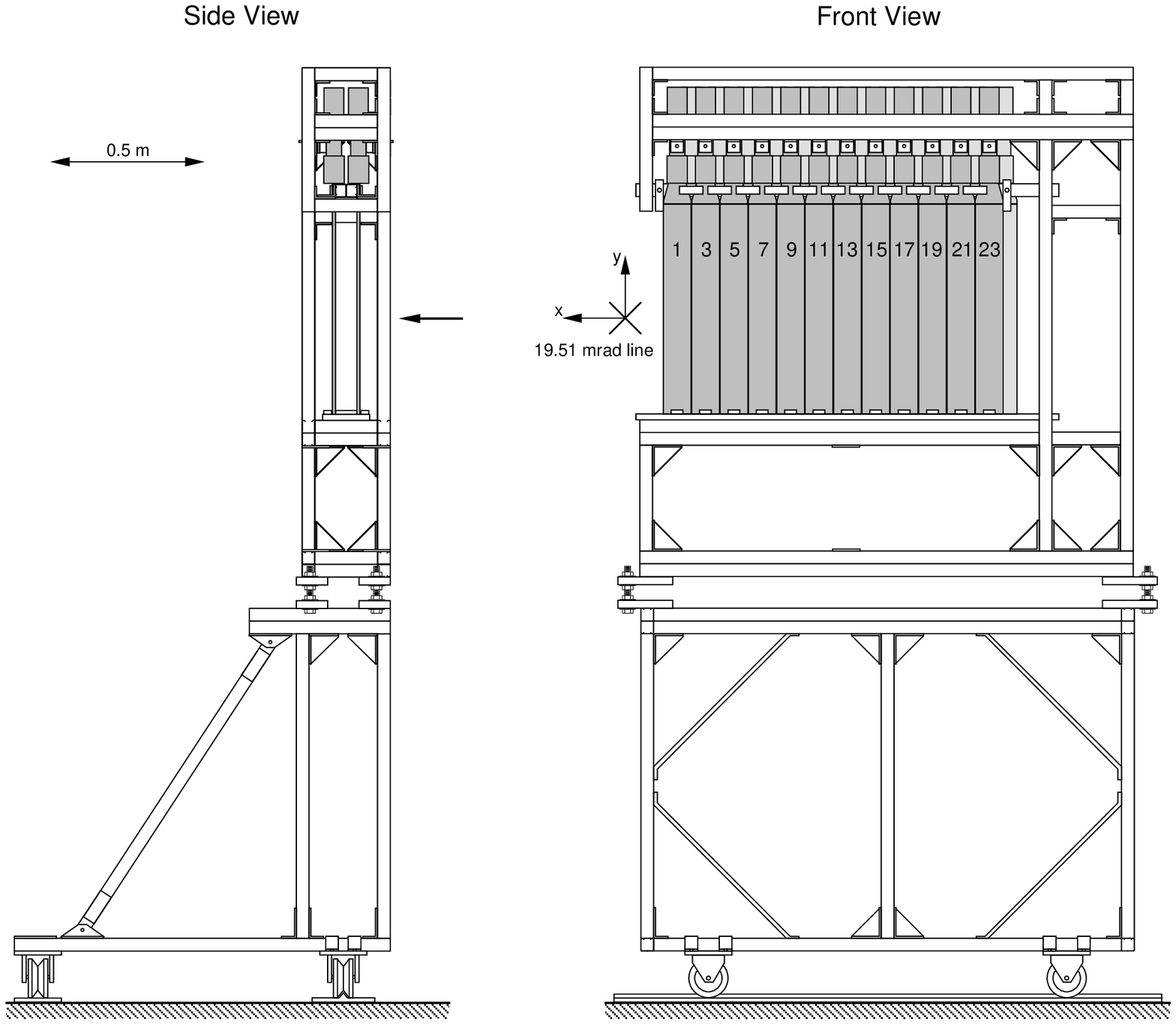}}
\mbox{}\\[0.5in]
\caption{}
\label{fig:os_hodo}
\end{figure}

Fig.~\ref{fig:os_hodo}.
Front and side views of the OS hodoscope.  Odd (even) numbered
counters formed the front (back) plane.

\begin{figure}
\centerline{\includegraphics[width=14.0cm]{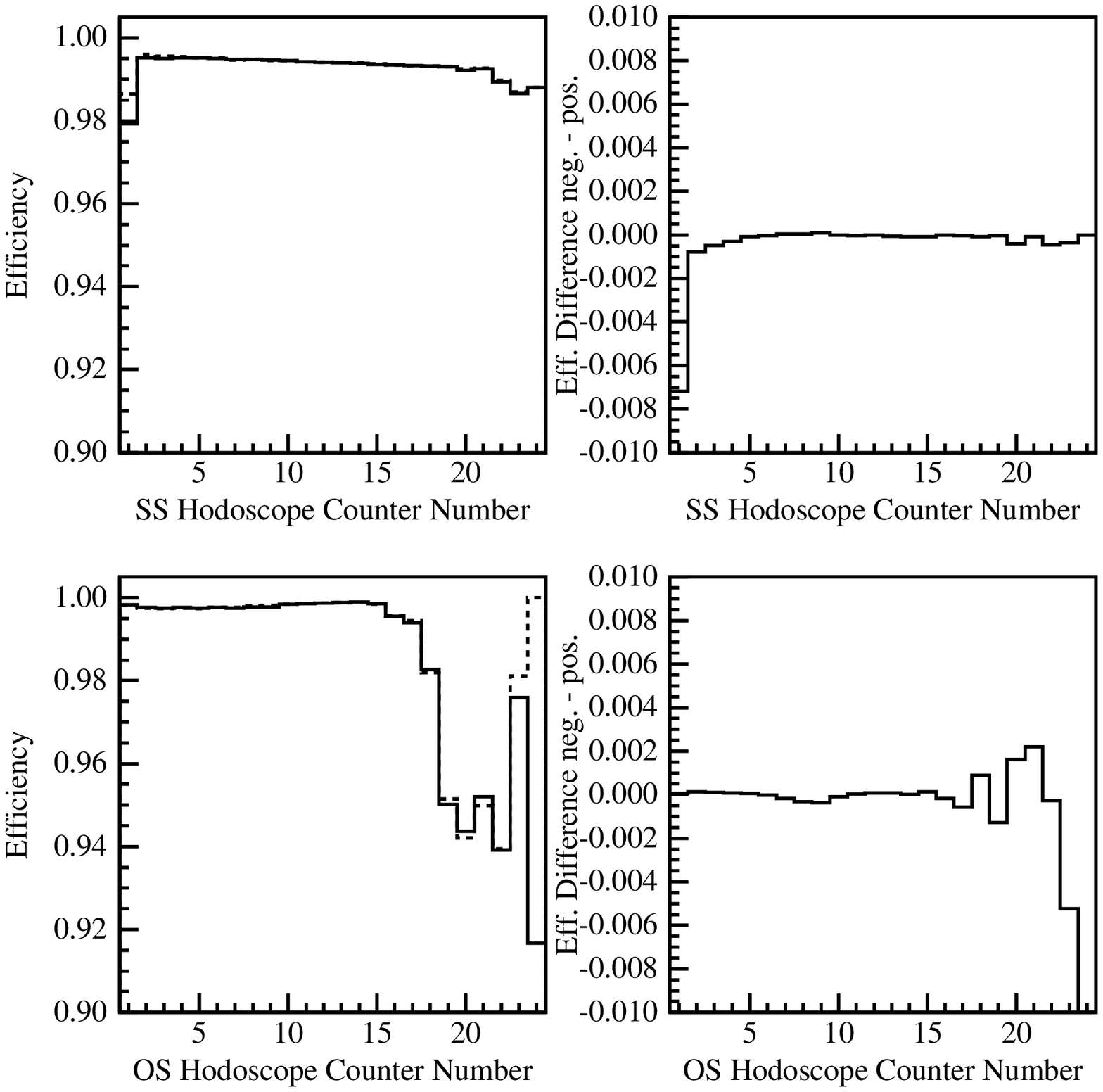}}
\mbox{}\\[0.5in]
\caption{}
\label{fig:eff_ss_os}
\end{figure}

Fig.~\ref{fig:eff_ss_os}.
Typical SS and OS Hodoscope efficiencies from a 1999 run.  The 
left-hand plots show absolute efficiencies, and the right-hand 
plots show differences between the positive- and negative-secondary-beam 
efficiencies.  The solid (dashed) lines are from negative (positive) 
data.  The low efficiencies for the high-numbered OS counters 
(OS16--24) were due to the fact that they lay behind the chamber 
C9 frame and stand.  The low efficiency for SS1 was due to its high rate.

\begin{figure}
\centerline{\includegraphics[width=14.0cm]{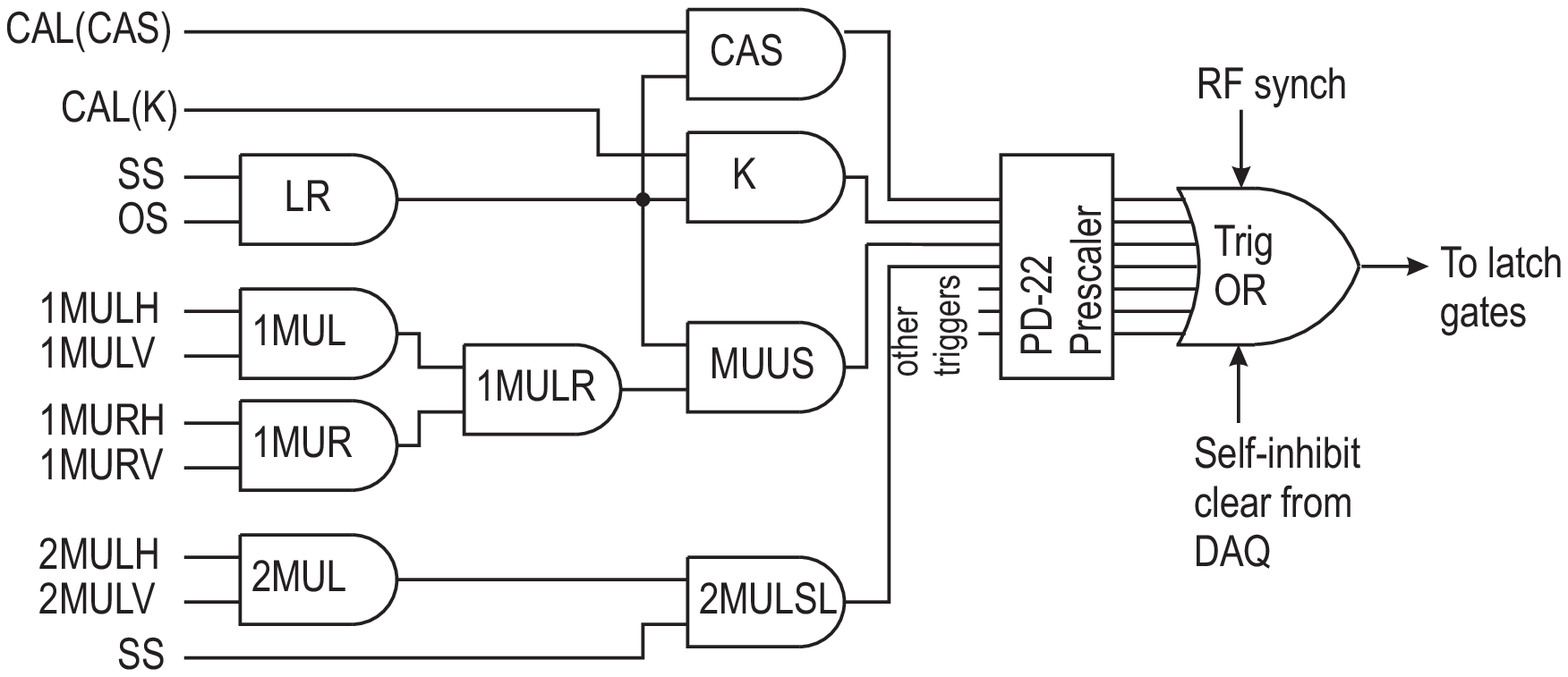}}
\mbox{}\\[0.5in]
\caption{}
\label{fig:trig_logic_99}
\end{figure}

Fig.~\ref{fig:trig_logic_99}.
Logic diagram of the four main physics triggers.

\begin{figure}
\centerline{\includegraphics[width=10.0cm]{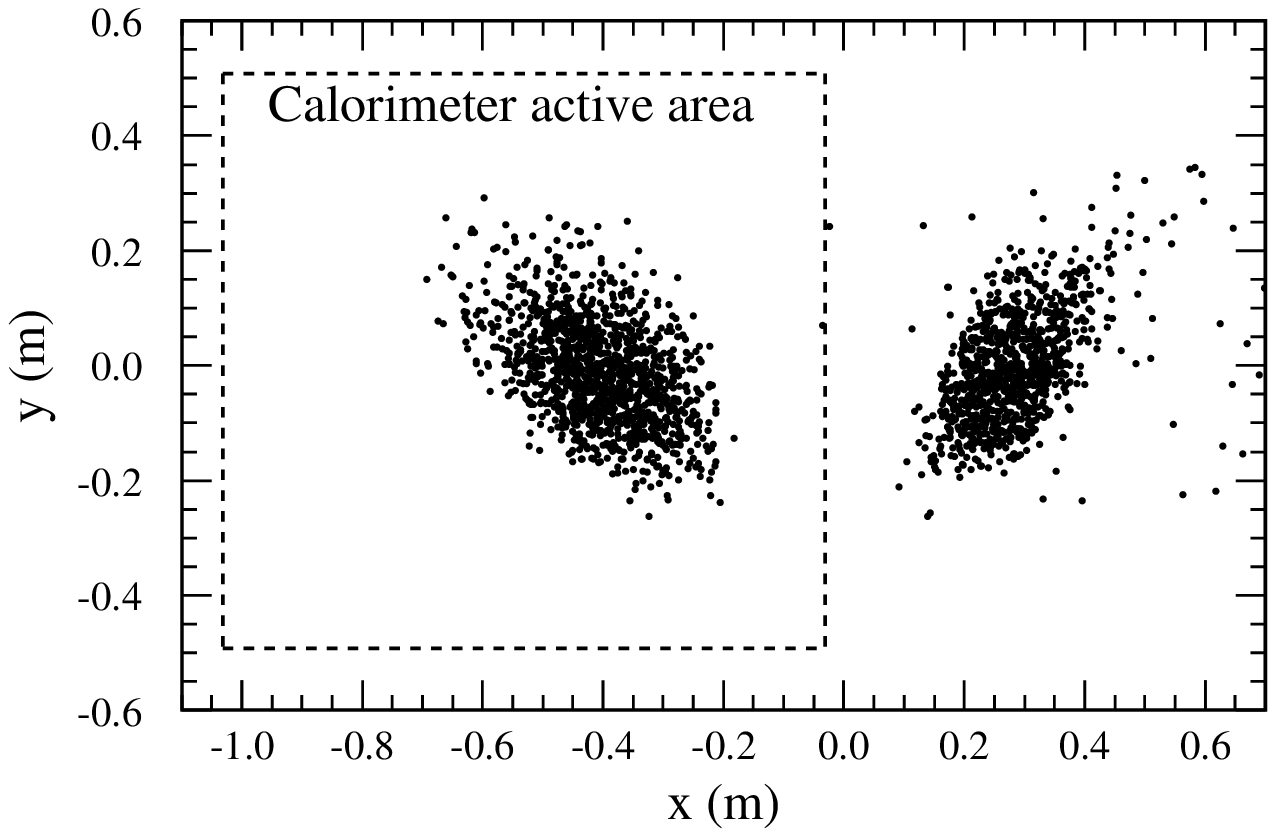}}
\mbox{}\\[0.5in]
\caption{}
\label{fig:beam_at_cal}
\end{figure}

Fig.~\ref{fig:beam_at_cal}.
The lateral distributions of protons from \gXi\ra\gL\gpi\ra\gp\gpi\gpi\ 
decays, and of charged particles from the secondary beam at the $z$ 
of the front face of the calorimeter.  The $x$-$y$ correlations arise 
from the orthogonal field directions of the Hyperon and Analyzing Magnets.

\begin{figure}
\centerline{\includegraphics[width=14.0cm]{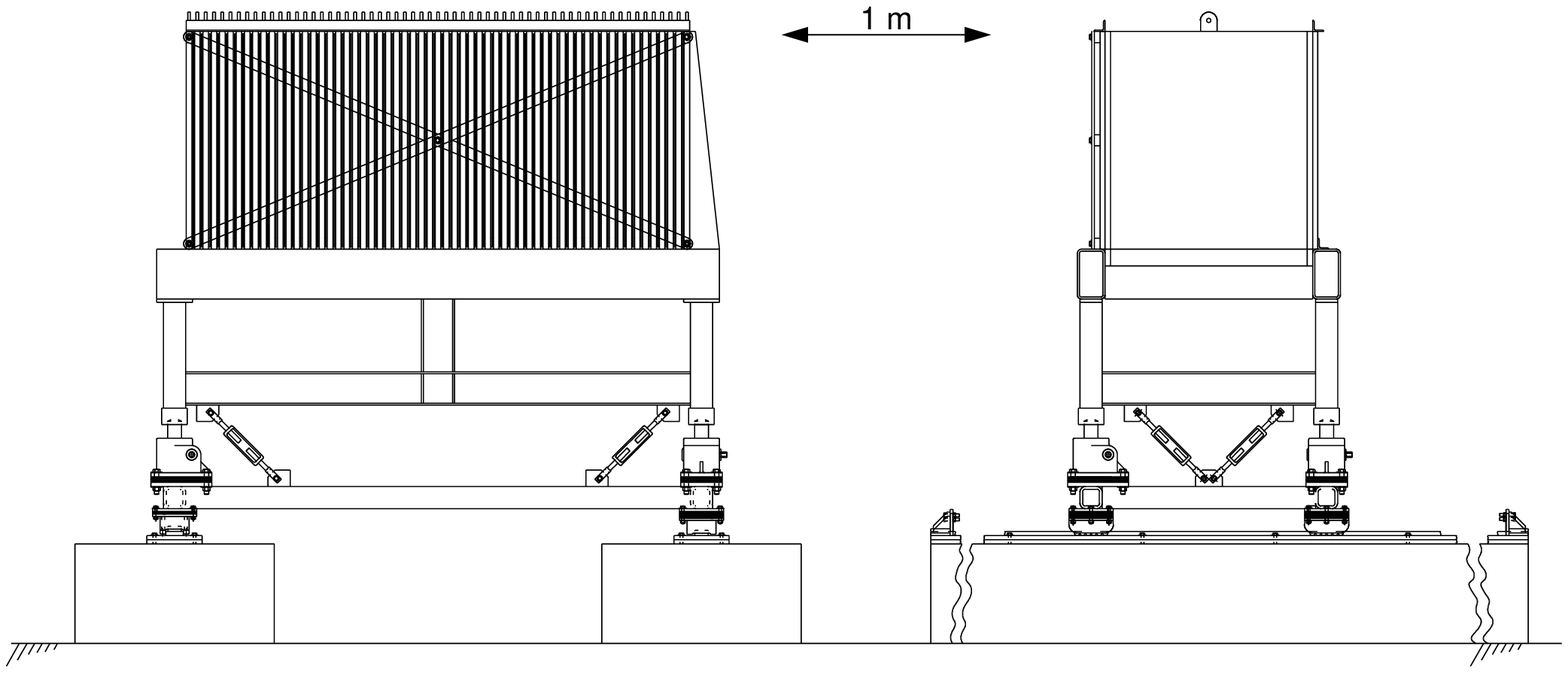}}    
\mbox{}\\[0.5in]
\caption{}
\label{fig:cal_front_side} 
\end{figure}

Fig.~\ref{fig:cal_front_side}.
Side (left) and back (right) views of the Hadronic Calorimeter and 
stand.  The light-tight enclosure, fibers, and photomultipliers are 
not shown.

\begin{figure}
\centerline{\includegraphics[width=8.0cm]{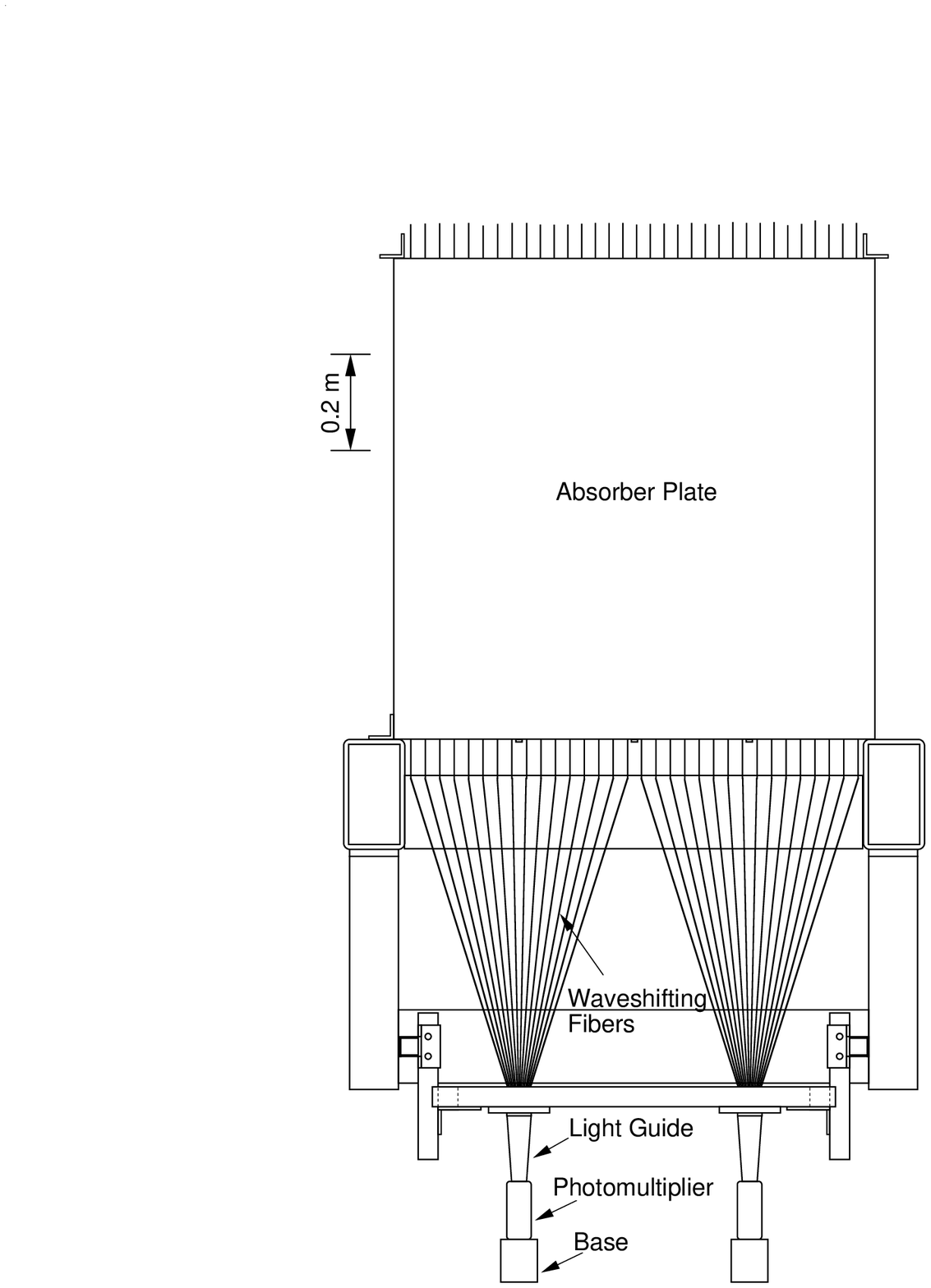}}    
\mbox{}\\[0.5in]
\caption{}
\label{fig:cal_front}
\end{figure}

Fig.~\ref{fig:cal_front}.
Front view of the inside of the Hadronic Calorimeter showing absorber
plate, fibers, light-guides, and photomultipliers.

\begin{figure}
\centerline{\includegraphics[width=14.0cm]{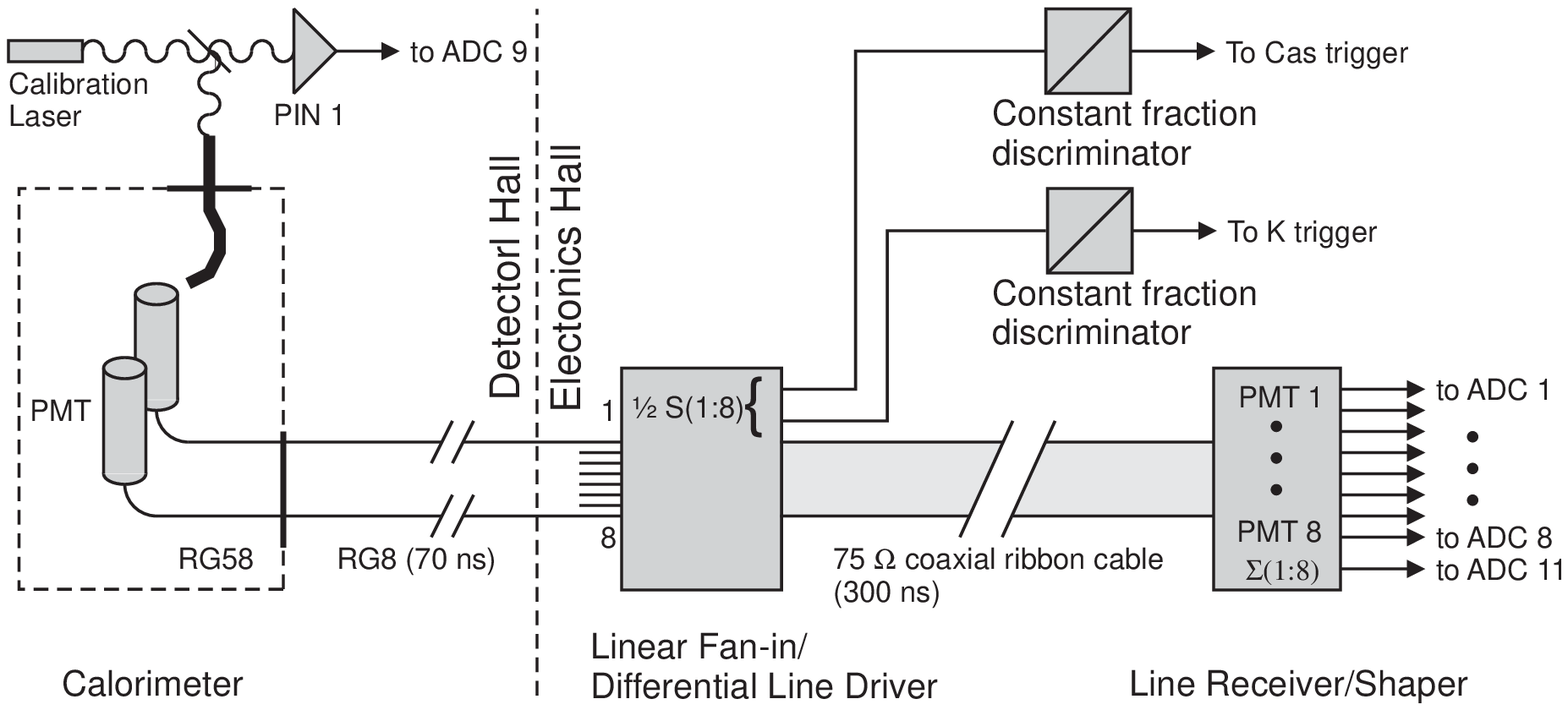}}    
\mbox{}\\[0.5in]
\caption{}
\label{fig:cal_readout} 
\end{figure}

Fig.~\ref{fig:cal_readout}.
Schematic of the calorimeter readout electronics.

\begin{figure}
\centerline{\includegraphics[width=10.0cm]{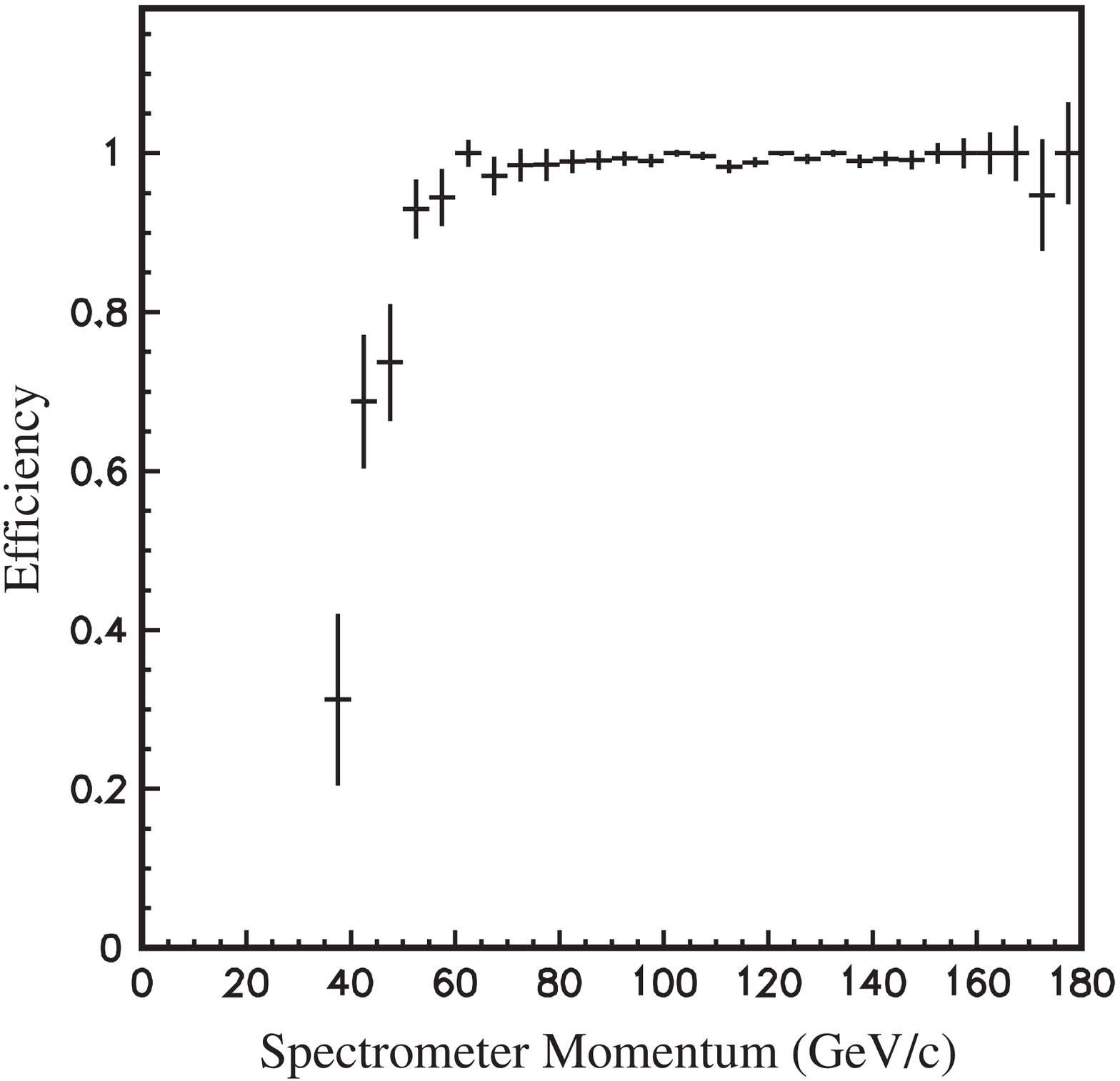}}    
\mbox{}\\[0.5in]
\caption{}
\label{fig:cal_eff}
\end{figure}

Fig.~\ref{fig:cal_eff}.
The calorimeter efficiency, with the CAS trigger threshold, versus
momentum, as measured using protons and pions from \gL\ra\gp\pim\
and \Ks\ra\pip\pim\ decays from a typical 1997 negative-polarity run.

\clearpage

\begin{figure} 
\centerline{\includegraphics[width=14.0cm]{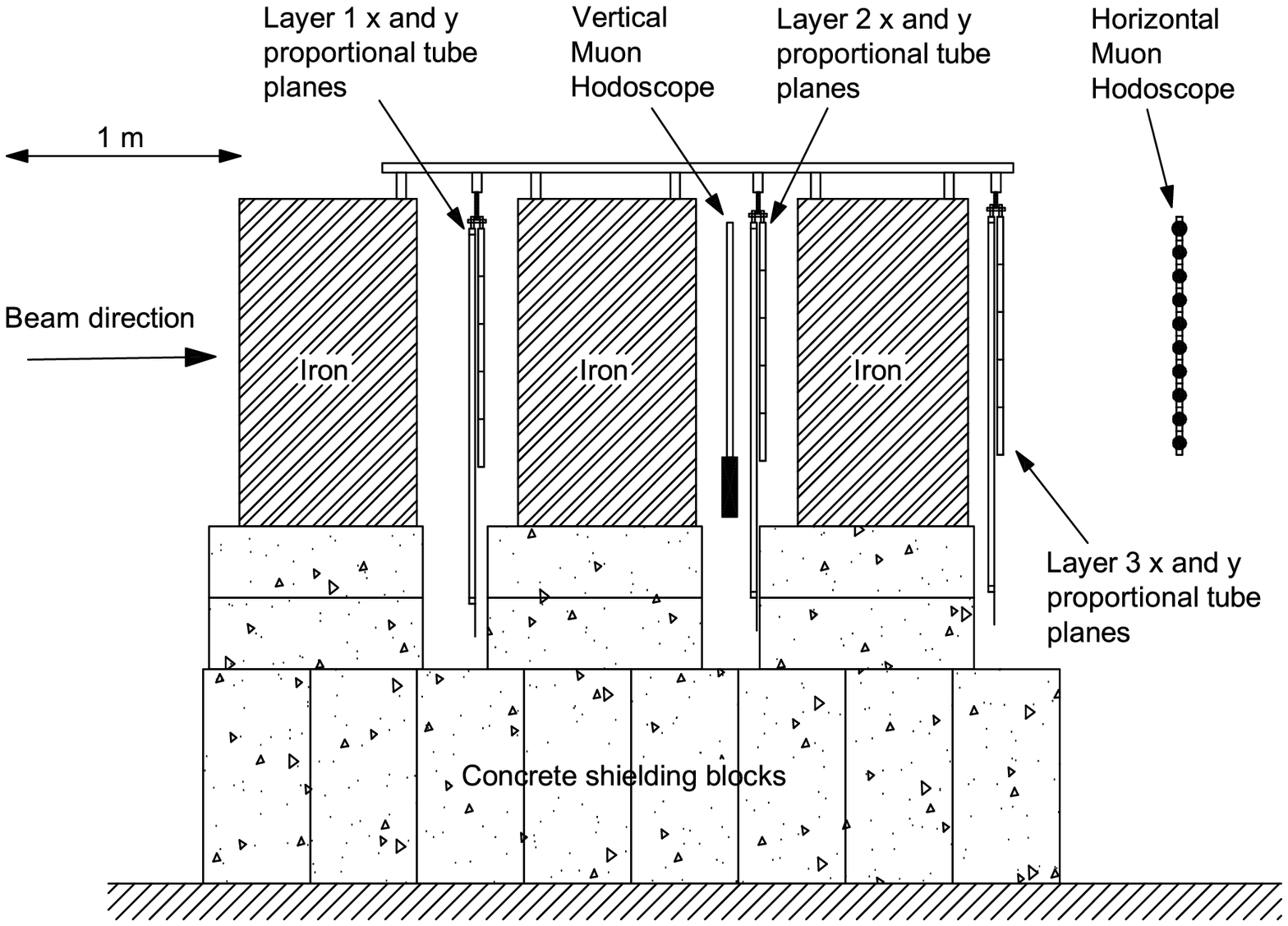}}
\mbox{}\\[0.5in]
\caption{}
\label{fig:mudet_side}
\end{figure}

Fig.~\ref{fig:mudet_side}.
Elevation view of the Right Muon Station.  The hodoscope stands
are not shown.  The Left Muon Station 
was similar except it had an extra front iron absorber.

\begin{figure}
\centerline{\includegraphics[width=14.0cm]{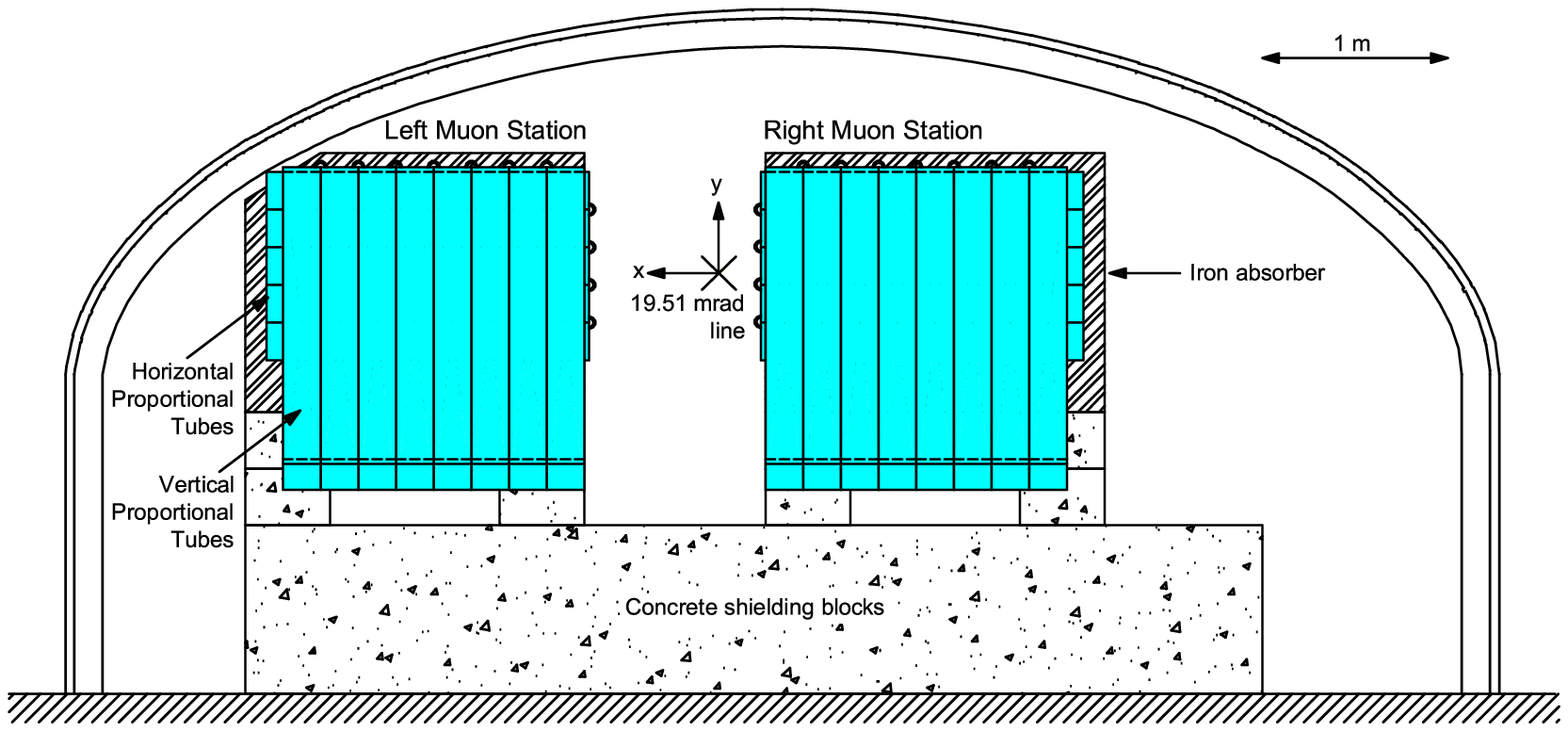}}
\mbox{}\\[0.5in]
\caption{}
\label{fig:mudet_front}
\end{figure}

Fig.~\ref{fig:mudet_front}.
Front view of the Muon Stations with the upstream iron
absorbers removed to show the vertical and horizontal 
proportional tube planes.  Enclosure MC7 is shown.

\begin{figure} 
\centerline{\includegraphics[width=10.0cm]{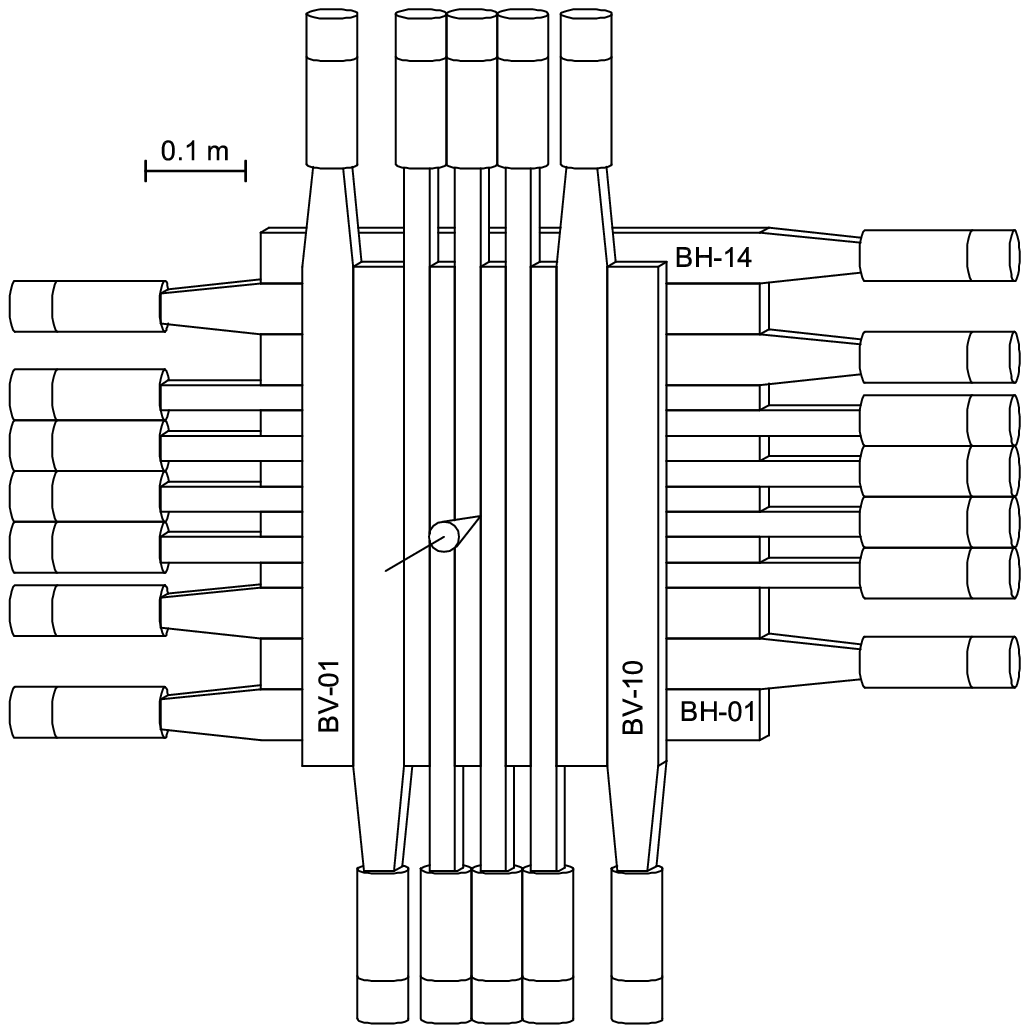}}
\mbox{}\\[0.5in]
\caption{}
\label{fig:beam_hodo}
\end{figure}

Fig.~\ref{fig:beam_hodo}.
Layout of the Beam Hodoscope counters.  The stand is not shown. 
The arrow indicates the beam direction.

\begin{figure}
\centerline{\includegraphics[width=13.5cm]{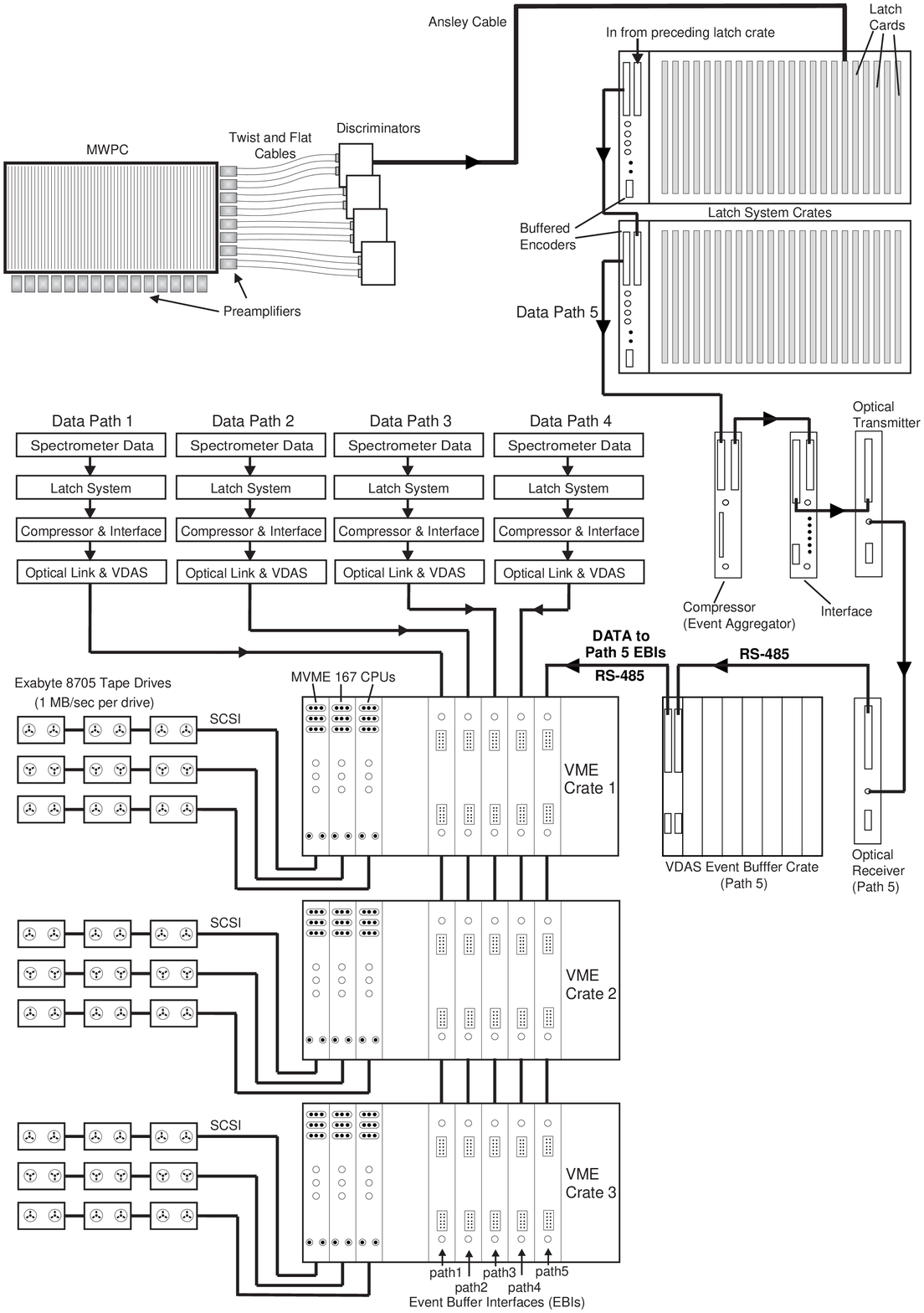}}
\mbox{}\\[0.5in]
\caption{}
\label{fig:path}
\end{figure}

Fig.~\ref{fig:path}.
Schematic of the HyperCP data-acquisition system.  Data were read out 
via five parallel datapaths, of which one (Path~5) is shown in detail.

\begin{figure}
\centerline{\includegraphics[width=14.0cm]{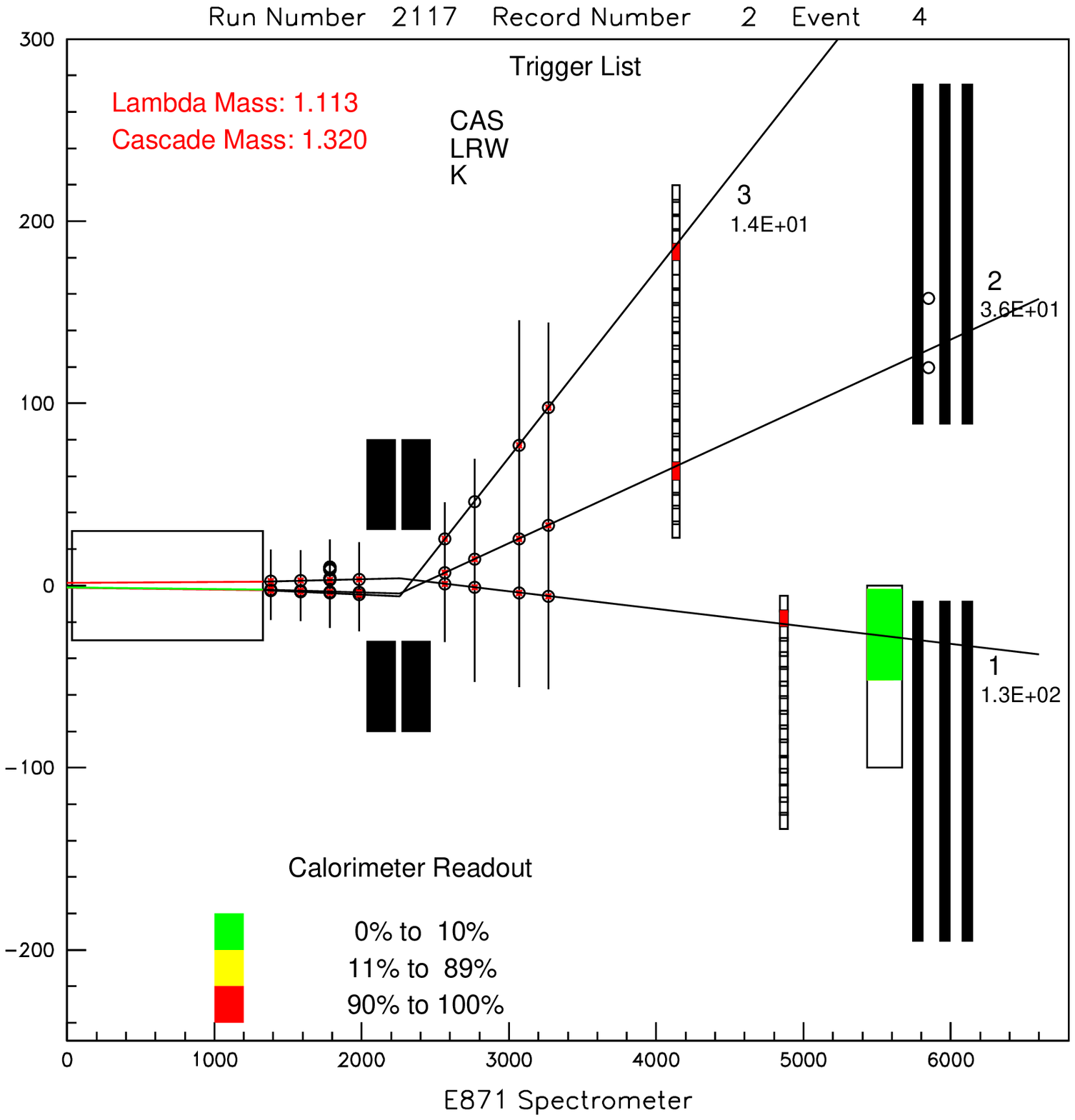}}
\mbox{}\\[0.5in]
\caption{}
\label{fig:event_display}
\end{figure}

Fig.~\ref{fig:event_display}.
Online event display, from a 1997 run, showing a plan view of the 
spectrometer with a \gXi\ decay that satisfied the CAS, LR, and K 
triggers, and that was reconstructed by the monitoring program. 
The \gXi\ and \gL\ masses are given, as well as the momentum of the 
decay products.

\begin{figure}
\centerline{\includegraphics[width=10.0cm]{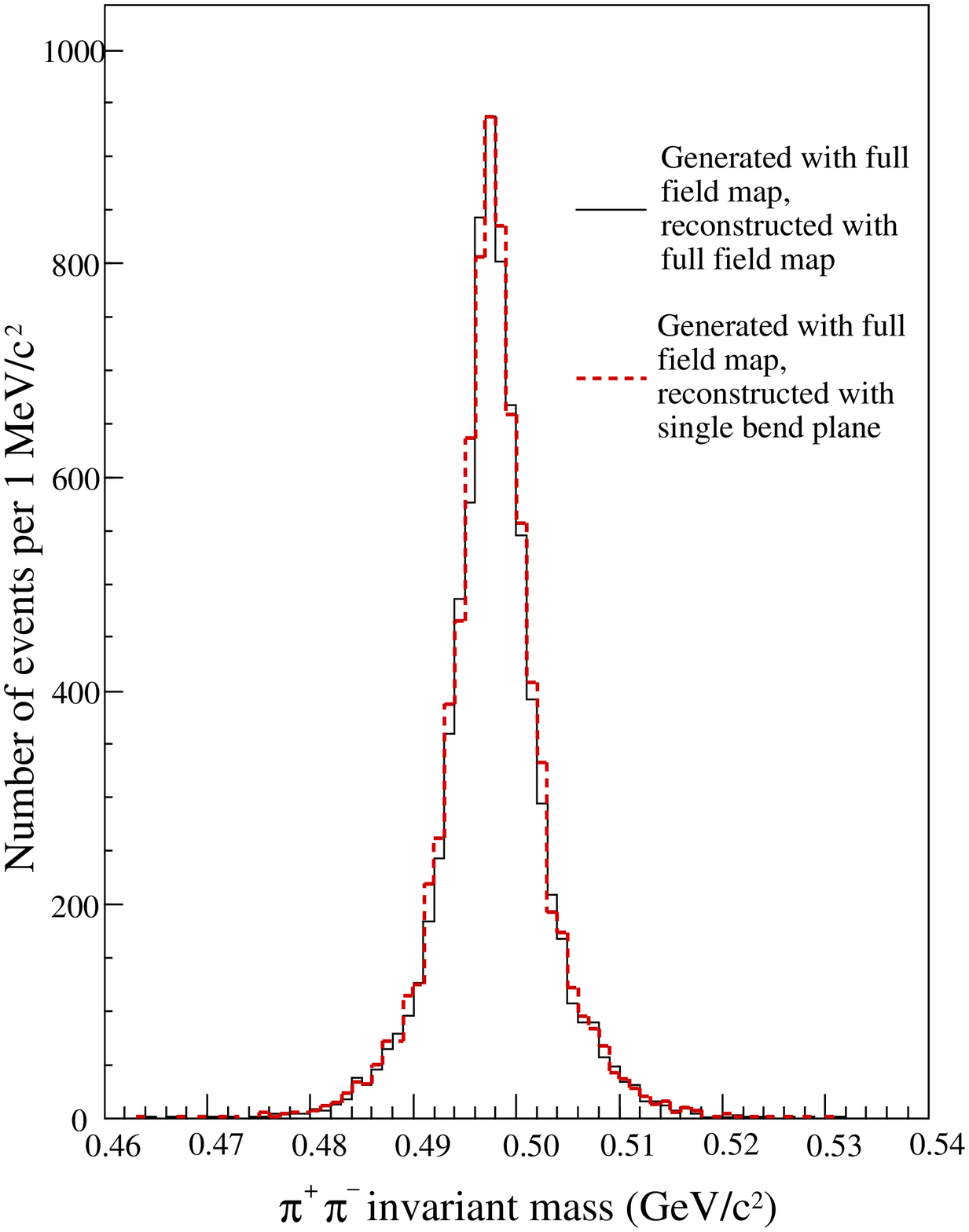}}
\mbox{}\\[0.5in]
\caption{}
\label{fig:fullvssingle}
\end{figure}

Fig.~\ref{fig:fullvssingle}.
Comparison of $\pi^{+}\pi^{-}$ invariant mass of Monte Carlo--generated 
$K_{S}$ events reconstructed using the measured field map (solid line) 
with that using a single bend plane and a constant field integral 
(dashed line).  In both samples, the generated tracks were traced 
through the Analyzing Magnets using the measured field maps.

Fig.~\ref{fig:mass_mc_data}.
The \gp\pim\ invariant mass from \Xim\ra\gL\pim\ra\gp\pim\pim\ decays.
Solid (dashed) lines are from real (MC) data.

\begin{figure}
\centerline{\includegraphics[width=10.0cm]{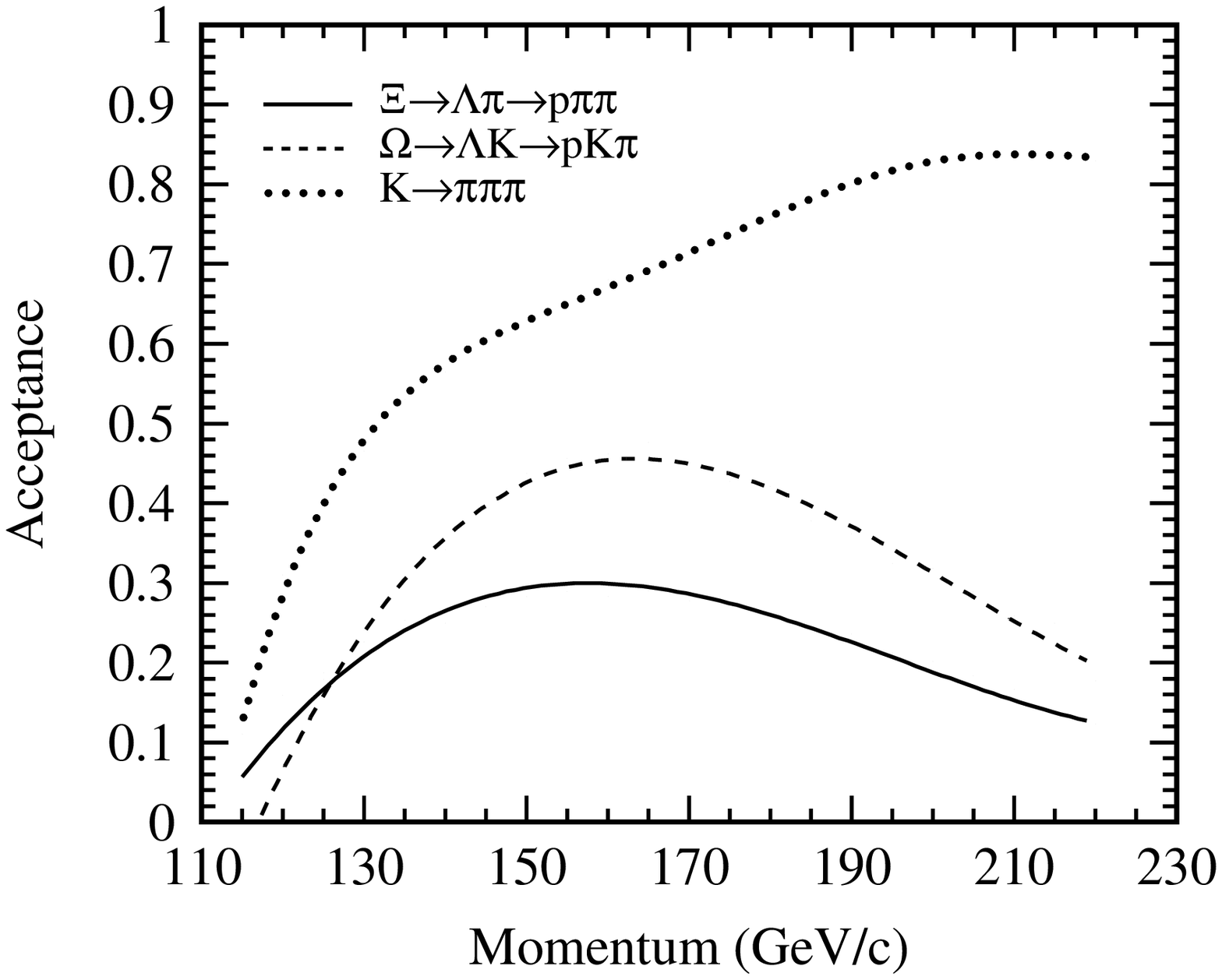}}
\mbox{}\\[0.5in]
\caption{}
\label{fig:geo_acc}
\end{figure}

Fig.~\ref{fig:geo_acc}.
The spectrometer acceptance for \Xim\ra\gL\gpi\ra\gp\gpi\gpi, 
\gOm\ra\gL\gK\ra\gp\gpi\gK, and \Kpm\ra\pipm\pip\pim\ decays,
where in each case the acceptance is for those parent particles
that have decayed within the Vacuum Decay Region.

\begin{figure}
\centerline{\includegraphics[width=10.0cm]{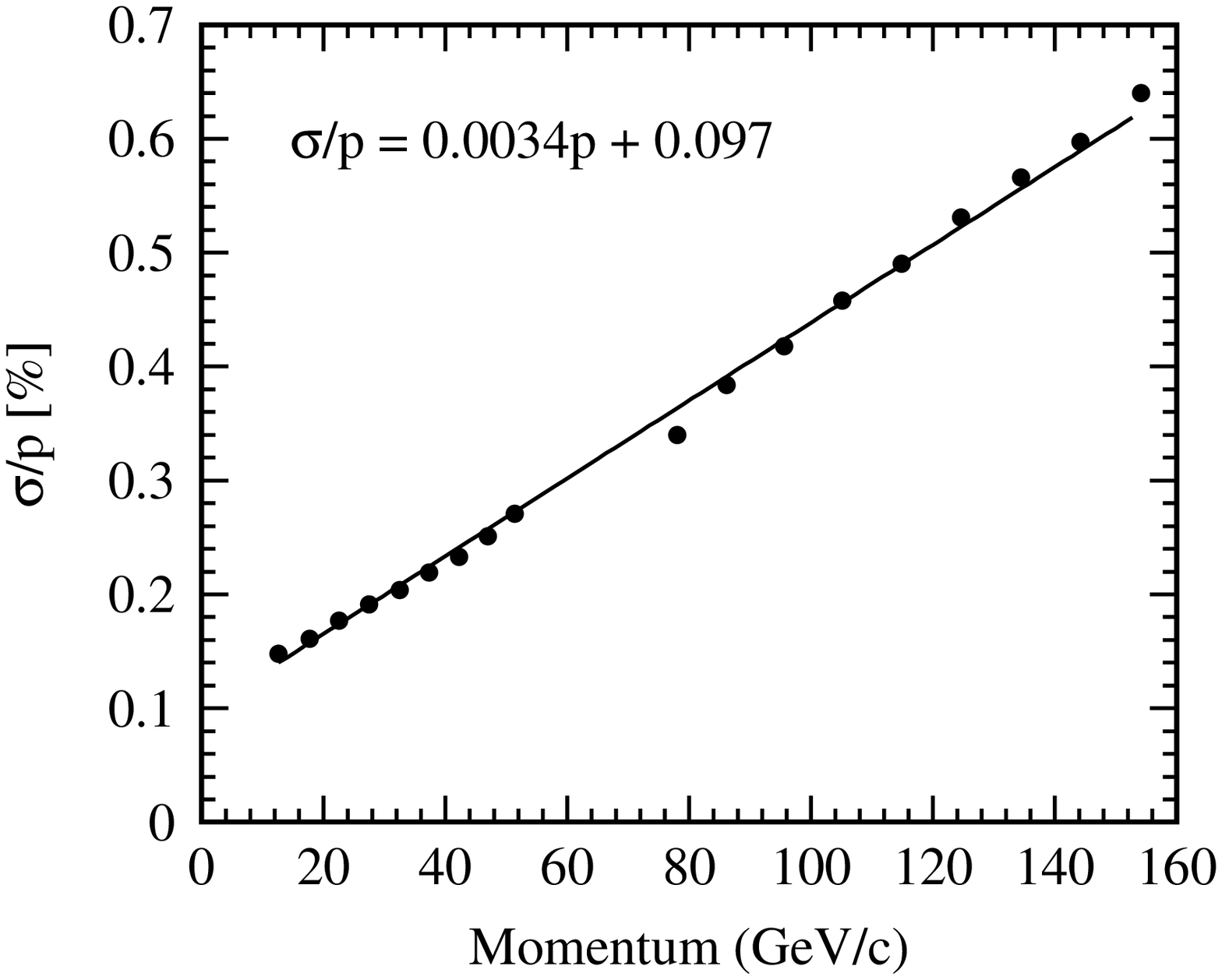}}
\mbox{}\\[0.5in]
\caption{}
\label{fig:mtm_res}
\end{figure}

Fig.~\ref{fig:mtm_res}.
The spectrometer momentum resolution, as determined by Monte Carlo, 
for pions and protons from \gXi\ra\gL\gpi\ra\gp\gpi\gpi\ decays.

\begin{figure}
\centerline{\includegraphics[width=10.0cm]{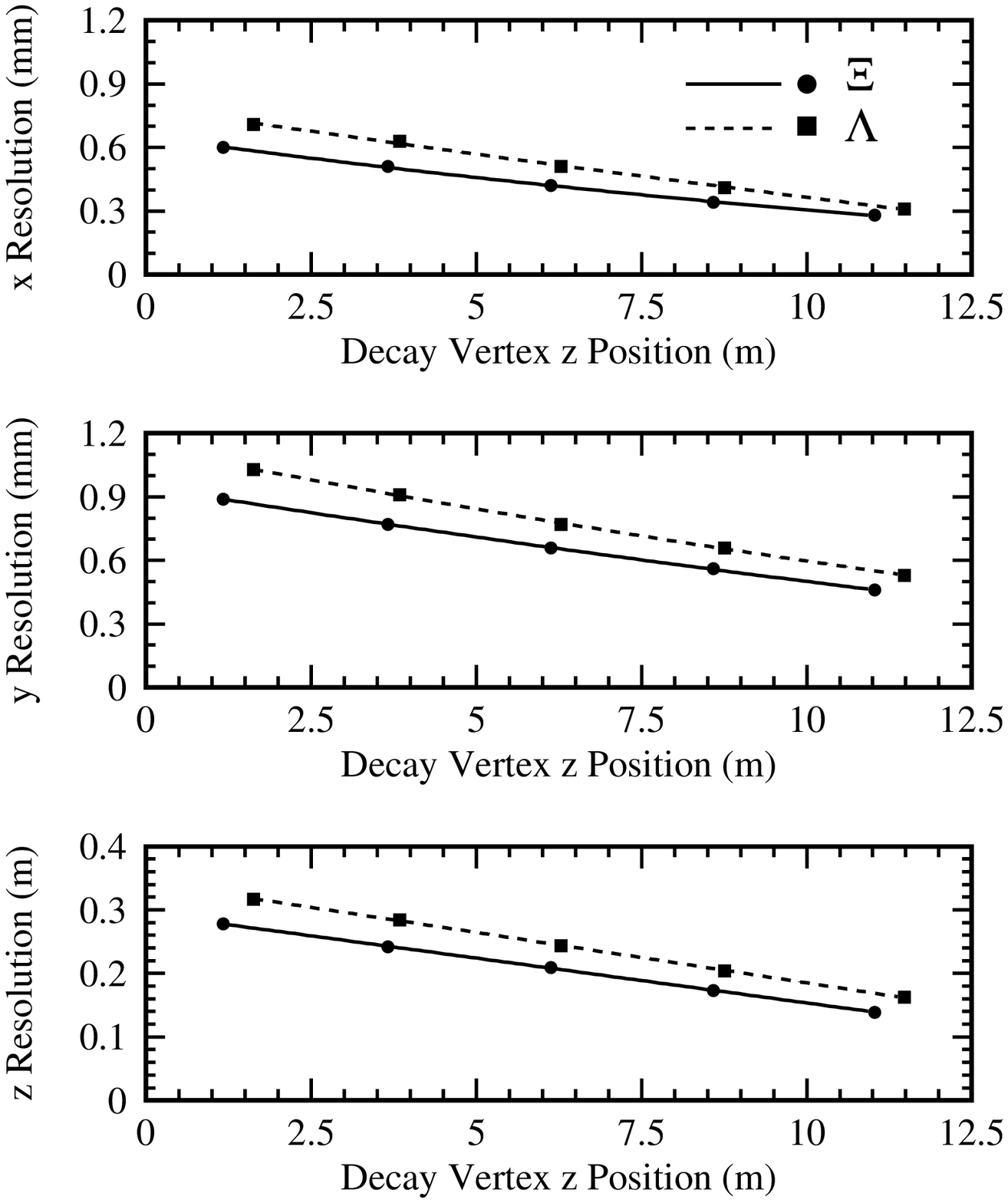}}
\mbox{}\\[0.5in]
\caption{}
\label{fig:vtx_res}
\end{figure}

Fig.~\ref{fig:vtx_res}.
Reconstructed rms \gXi\ and \gL\ decay-vertex resolutions, as 
determined by Monte Carlo, in $x$ (top), $y$ (middle), and $z$ (bottom).
The \gXi\ (\gL) resolution is represented by solid (dashed) lines.

\begin{figure}
\centerline{\includegraphics[width=8.0cm]{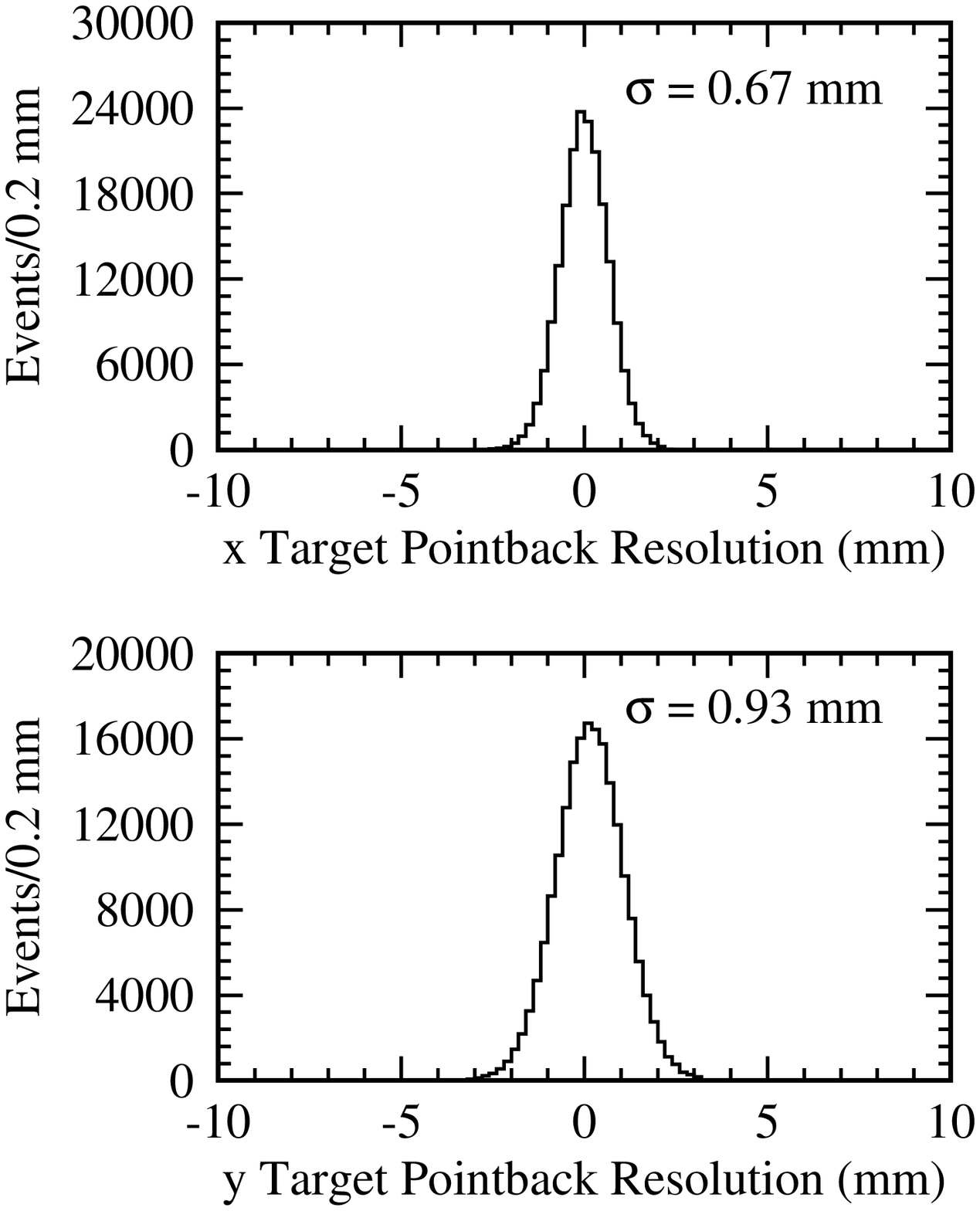}}
\mbox{}\\[0.5in]
\caption{}
\label{fig:tgt_res}
\end{figure}

Fig.~\ref{fig:tgt_res}.
The target-pointback resolution, as determined by Monte Carlo, 
in $x$ (top) and $y$ (bottom).

\begin{figure}
\centerline{\includegraphics[width=7.0cm]{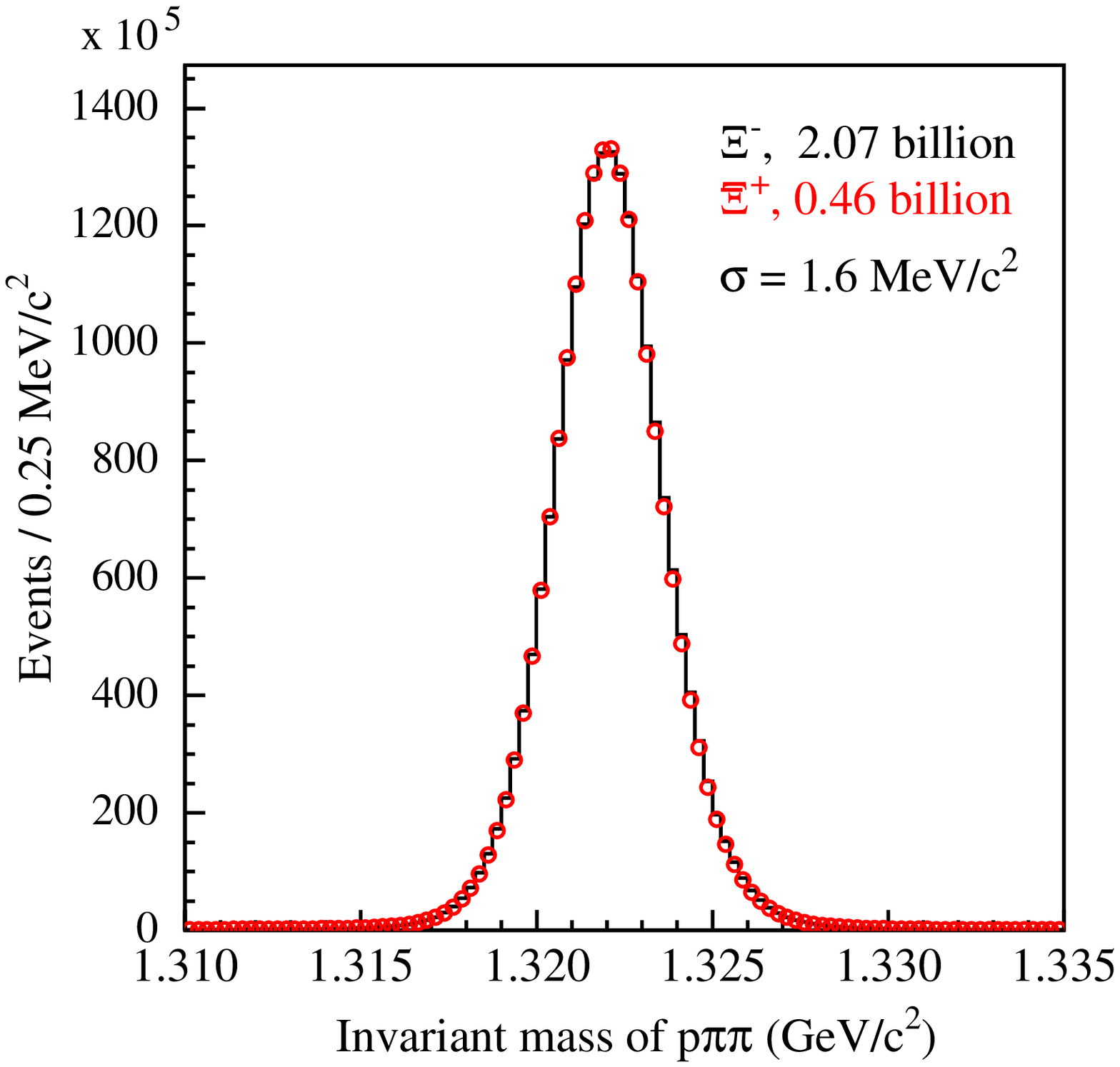}
\includegraphics[width=7.0cm]{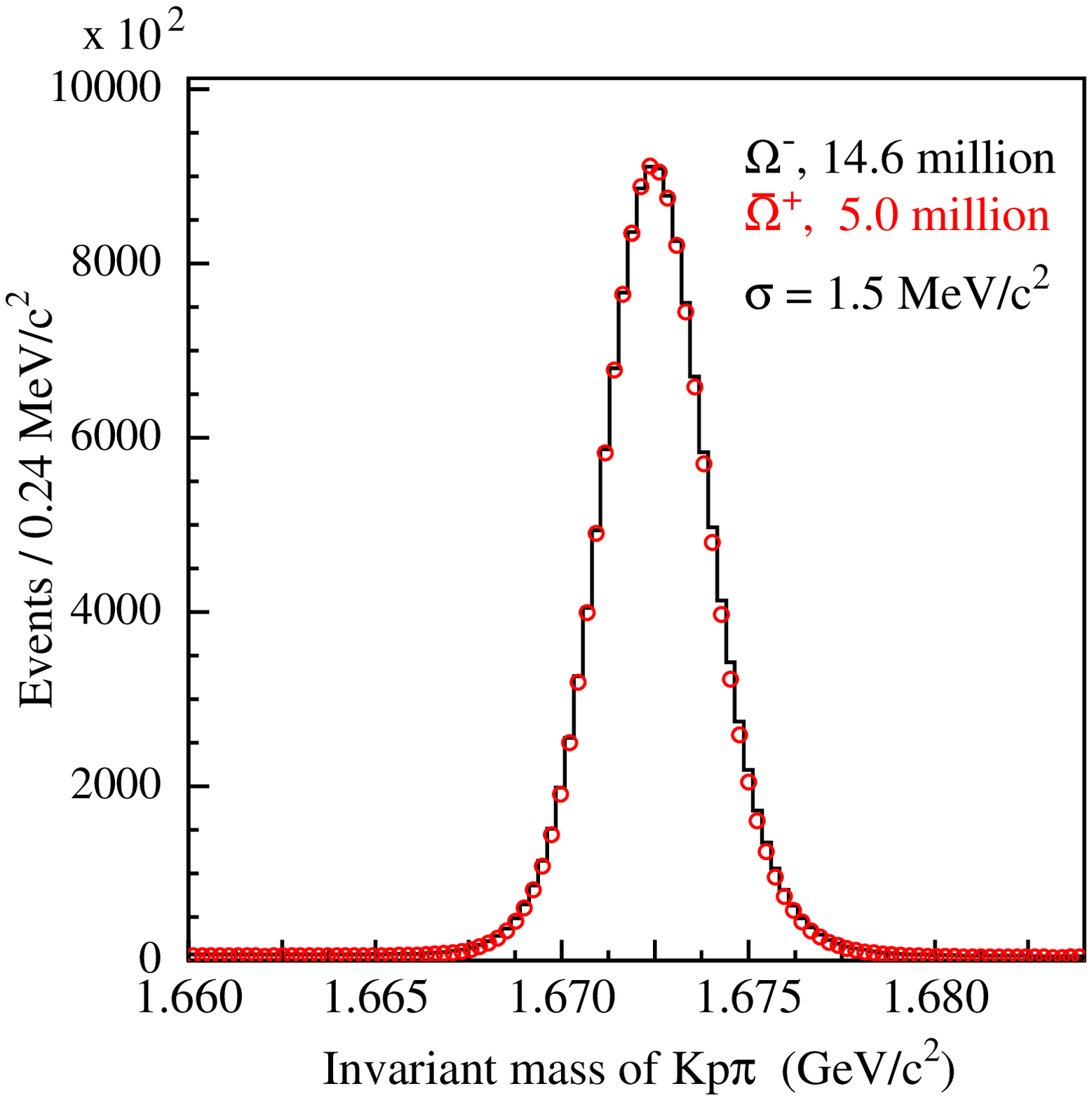}} 
\centerline{\includegraphics[width=7.0cm]{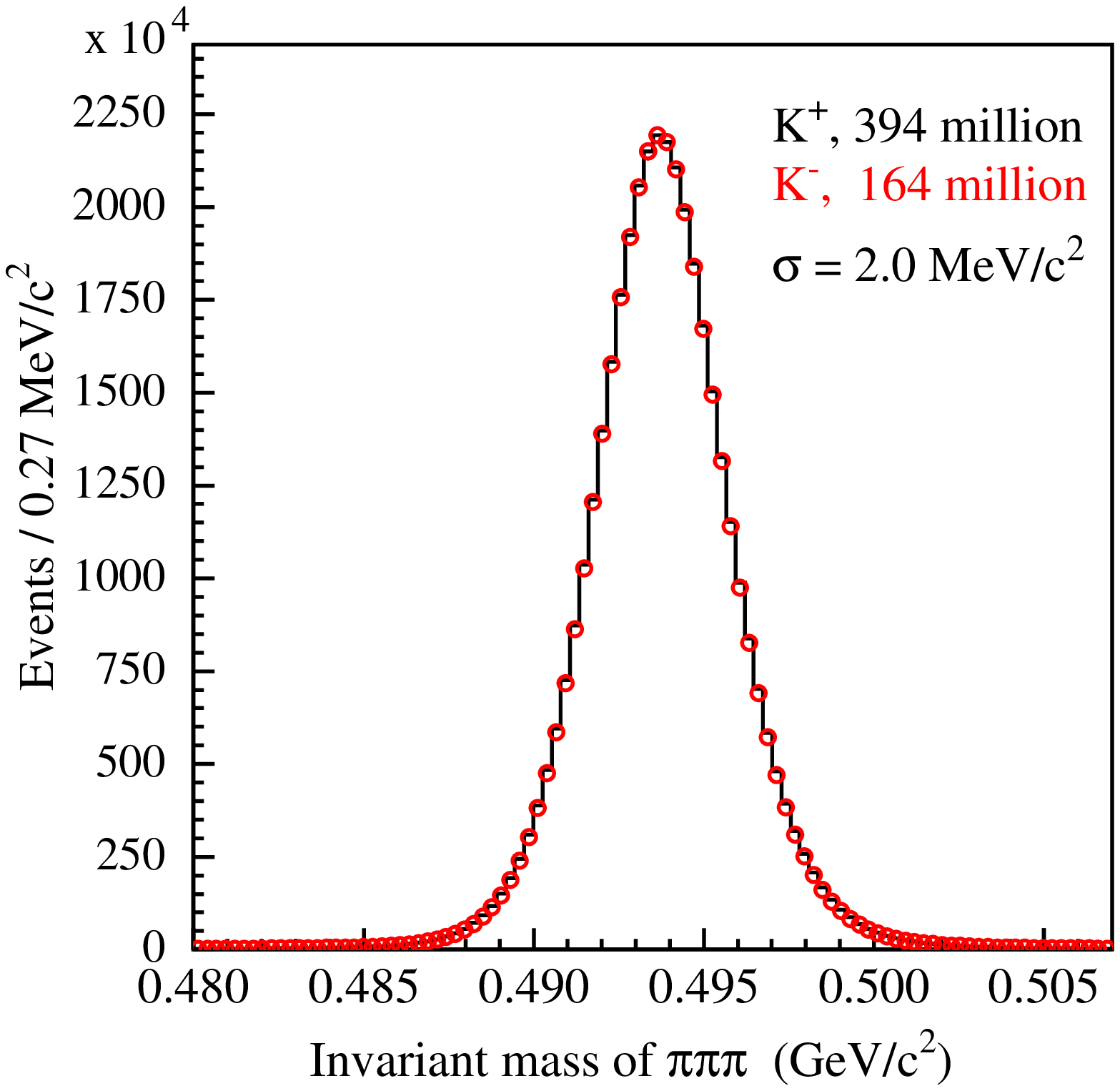}
\includegraphics[width=7.0cm]{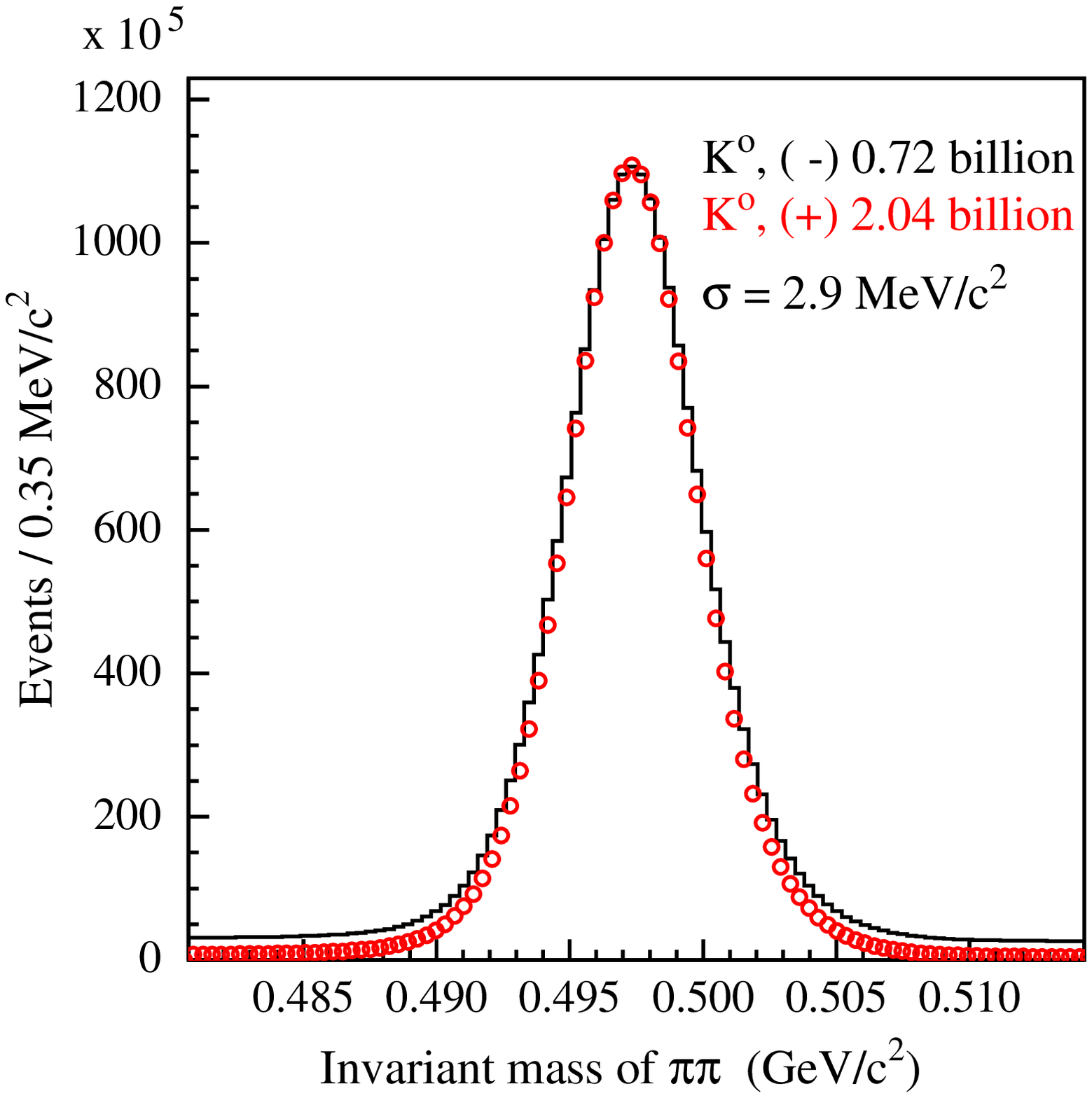}}
\mbox{}\\[0.5in]
\caption{}
\label{fig:farm_masses}
\end{figure}

Fig.~\ref{fig:farm_masses}.
The \Xipm, \Ompm, \Kpm, and \Ks\ masses from the farm analysis.
Solid lines (circles) correspond to negative (positive)
polarity running.

\clearpage

\begin{table}
\caption{The $z$ position of spectrometer elements and material,
  in fractions of interaction and radiation lengths (for those
  elements in the secondary beam).
  The position of the front is given, unless otherwise stated.  
  The helium bags have been omitted.} 
\vspace{0.2in}
\label{tab:material}
\begin{tabular}{rrrr}
\hline
& & \multicolumn{2}{c}{Fractional length} \\
    \multicolumn{1}{c}{Item}
  & \multicolumn{1}{c}{$z$ Position (m)} 
  & \multicolumn{1}{c}{$\lambda_I$} 
  & \multicolumn{1}{c}{$X_{\circ}$} \\
\hline
Target center                   & $-6.388$ 
  & \multicolumn{2}{c}{---}  \\ 
Beam pipe entrance window       & $-6.318$ 
  & $1.3{\times}10^{-4}$        & $2.7{\times}10^{-4}$ \\
Beam pipe exit window           &  0.273    
  & $1.3{\times}10^{-4}$        & $2.7{\times}10^{-4}$ \\
Vacuum Decay Region entrance    & 0.318
  & $2.8{\times}10^{-4}$        & $21.4{\times}10^{-4}$ \\
Vacuum Decay Region exit        & 13.323
  & $7.3{\times}10^{-4}$        & $15.1{\times}10^{-4}$ \\
C1                              & 13.844
  & $7.1{\times}10^{-4}$        & $24.3{\times}10^{-4}$ \\
C2                              & 15.844
  & $7.1{\times}10^{-4}$        & $24.3{\times}10^{-4}$ \\
C3                              & 17.827
  & $7.1{\times}10^{-4}$        & $24.0{\times}10^{-4}$ \\
C4                              & 19.836
  & $7.1{\times}10^{-4}$        & $24.0{\times}10^{-4}$ \\
Magnet 1 (Magnet 2) center      & 21.502 (23.860)
  & \multicolumn{2}{c}{---}  \\
C5                              & 25.651
  & $7.1{\times}10^{-4}$        & $24.3{\times}10^{-4}$  \\
C6                              & 27.670
  & $7.1{\times}10^{-4}$        & $24.3{\times}10^{-4}$  \\
C7                              & 30.674
  & $6.8{\times}10^{-4}$        & $24.6{\times}10^{-4}$ \\
C8                              & 32.676
  & $7.1{\times}10^{-4}$        & $24.7{\times}10^{-4}$ \\
SS Hodoscope                    & 41.100
  & \multicolumn{2}{c}{---}  \\
C9                              & 44.125
  & $6.7{\times}10^{-4}$        & $19.1{\times}10^{-4}$ \\
OS Hodoscope                    & 48.413
  & \multicolumn{2}{c}{---}  \\
Hadronic Calorimeter            & 54.376
  & \multicolumn{2}{c}{---}  \\
Muon Layer 1 LV (RV) tubes      & 58.272 (58.413)
  & \multicolumn{2}{c}{---}  \\
Muon Layer 1 LH (RH) tubes      & 58.310 (58.451)
  & \multicolumn{2}{c}{---}  \\
Muon hodoscope LV (RV)          & 59.396 (59.537)
  & \multicolumn{2}{c}{---}  \\
Muon Layer 2 LV (RV) tubes      & 59.472 (59.613)
  & \multicolumn{2}{c}{---}  \\
Muon Layer 2 LH (RH) tubes      & 59.510 (59.651)
  & \multicolumn{2}{c}{---}  \\
Muon Layer 3 LV (RV) tubes      & 60.697 (60.629)
  & \multicolumn{2}{c}{---}  \\
Muon Layer 3 LH (RH) tubes      & 60.735 (60.667)
  & \multicolumn{2}{c}{---}  \\
Muon hodoscope LH (RH)          & 61.362 (61.431)
  & \multicolumn{2}{c}{---}  \\
Beam Hodoscope                  & 62.975
  & $1.27{\times}10^{-2}$       & $2.35{\times}10^{-2}$ \\
\hline
\end{tabular}
\end{table}

\clearpage

\begin{table}
\caption{Geometry of the collimator. The coordinates are measured 
         from the center of the exit aperture of Segment E.} 
\vspace{0.2in}
\label{tab:channel}
\begin{tabular}{cr@{.}lr@{~~}r@{.}lc@{~~}r@{.}l}
\hline
    & \multicolumn{2}{c}{}
    & \multicolumn{5}{c}{Aperture} \\
    Segment & \multicolumn{2}{c}{Start/end}
    & \multicolumn{3}{c}{Size (mm)}
    & \multicolumn{3}{c}{Center (mm)} \\
    & \multicolumn{2}{c}{in $z$ (m)} 
    & \multicolumn{1}{c}{${\Delta}x$}
    & \multicolumn{2}{c}{${\Delta}y$}
    & \multicolumn{1}{c}{$x$}
    & \multicolumn{2}{c}{$y$} \\
\hline
        &   $-6$ & 096 & 20.3 & 25 & 4 & 0.0 & $-59$ & 46 \\
\raisebox{1.75ex}[0pt]{A}  
        &   $-3$ & 962 & 20.3 & 25 & 4 & 0.0 & $-52$ & 18 \\
        &   $-3$ & 962 & 10.0 &  5 & 0 & 0.0 & $-52$ & 18 \\
\raisebox{1.75ex}[0pt]{B}  
        &   $-3$ & 048 & 10.0 &  5 & 0 & 0.0 & $-46$ & 35 \\
        &   $-3$ & 048 & 13.0 &  7 & 6 & 0.0 & $-46$ & 35 \\
\raisebox{1.75ex}[0pt]{C}  
        &   $-2$ & 134 & 13.0 &  7 & 6 & 0.0 & $-34$ & 36 \\
        &   $-2$ & 134 & 17.0 & 20 & 0 & 0.0 & $-34$ & 36 \\
\raisebox{1.75ex}[0pt]{D}  
        &   $-0$ & 914 & 17.0 & 20 & 0 & 0.0 & $-16$ & 53 \\
        &   $-0$ & 914 & 20.0 & 10 & 0 & 0.0 & $-16$ & 53 \\
\raisebox{1.75ex}[0pt]{E}  
        & $   0$ &   0 & 20.0 & 10 & 0 & 0.0 & $  0$ & 00 \\
\hline
\end{tabular}
\end{table}

\clearpage

\begin{table}
\caption{Target positions (in mm) in the Spectrometer Coordinate System.}
\vspace{0.2in}
\label{tab:tgt_position}
\begin{tabular}{rr@{.}lr@{.}lr@{.}lr@{.}l}
\hline
    & \multicolumn{4}{c}{1997}
    & \multicolumn{4}{c}{1999} \\
  \multicolumn{1}{c}{Polarity:}
& \multicolumn{2}{c}{$+$}
& \multicolumn{2}{c}{$-$}
& \multicolumn{2}{c}{$+$}
& \multicolumn{2}{c}{$-$} \\
\hline
$x$ &   4 & 61 &  4 & 66 & $-0$ & 03 & $-0$ & 01 \\
$y$ &  62 & 83 & 62 & 86 &   65 & 00 &   64 & 96 \\
\hline
\end{tabular}
\end{table}

\clearpage

\begin{table}
\caption{MWPC parameters.}
\vspace{0.2in}
\label{tab:pwc_param}
\begin{tabular}{rcccc}
\hline
\multicolumn{1}{c}{Parameter} & C1,C2 & C3,C4 & C5,C6 & C7,C8,C9 \\
\hline
Aperture ($x{\times}y$) (mm) 
   & $457{\times}254$ 
   & $559{\times}305$
   & $1219{\times}406$
   & $2032{\times}559$ \\
Wire pitch (mm)
   & 1.016 & 1.270 & 1.501 & 2.00 \\
Wire diameter ($\mu$m)
   & 12.5
   & 12.5
   & 15.0
   & 20.0 \\
Wire tension (N)
   & 0.216
   & 0.216
   & 0.314
   & 0.588 \\
Anode-cathode gap (mm)
   & 3
   & 3
   & 3
   & 3 \\
Total wires X,X$^{\prime}$ 
     &       320 &       320 &      800 &       992 \\
Total wires U,V
     &       384 &       384 &      816 &      1008 \\   
Instrumented wires X,X$^{\prime}$
     &  320, 320 &  320, 320 & 448, 800 & 960, 960, 224 \\
Instrumented wires U,V 
     &  384, 384 &  384, 384 & 512, 816 & 1008, 1008, 288 \\
Anode plane order 
     & XUVX$^{\prime}$ & XUVX$^{\prime}$ & XVUX$^{\prime}$ & XVUX$^{\prime}$ \\
Operating voltage (V) 
     &            3000 &            2550 &            2550 & 2000, 2450, 2350 \\
\hline
\end{tabular}
\end{table}

\clearpage

\begin{table}
\caption{Properties of the MWPC preamplifier card.}
\vspace{0.2in}
\label{tab:preampspecs}
\begin{tabular}{rl} 
\hline
Channels      &    16          \\
Output        & differential   \\
Gain          & 9.5--12.5\,mV/fC \\
Rise time     & 9.0--13.0\,ns    \\  
Input noise   & $<0.65\,{\rm fC}$ @ $C_{in} = 25\,{\rm pF}$\\
Power         & +10\,V, +5\,V, $-2$\,V \\
\hline
\end{tabular}
\end{table}

\clearpage

\begin{table}
\caption{Properties of the MWPC discriminator/delay card.}
\vspace{0.2in}
\label{tab:discrspecs}
\begin{tabular}{rl} \hline
Channels      & 32          \\
Input         & differential, $112\,{\Omega}$  \\
Output        & differential ECL   \\
Threshold     & adjustable 0--50\,mV, nominal\ 25\,mV \\
Width         & adjustable, nominal\ 40--80\,ns \\  
Type          & leading edge, updating \\
Delay         & $100\pm 10$\,ns \\
Power         & +5\,V, $-5$\,V \\
\hline
\end{tabular}
\end{table}

\clearpage

\begin{table}
\caption{The HyperCP triggers. Prescale values shown are typical for the 
         1999 run.}
\vspace{0.2in}
\label{tab:triggers}
\begin{tabular}{lllr}
  & & & \multicolumn{1}{c}{Prescale} \\
\multicolumn{1}{c}{Name} &
\multicolumn{1}{c}{Symbol} &
\multicolumn{1}{c}{Components} &
\multicolumn{1}{c}{Factor} \\
  \hline
  Cascade                  &      CAS & LR{\cd}CAL(CAS)      & 1 \\
  Kaon                     &        K & LR{\cd}CAL(K)        & 2 \\
  Unlike-Sign Muon         &     MUUS & 1MUL{\cd}1MUR{\cd}LR & 1 \\
  Left 2-Muon Like-Sign    &   2MULSL & 2MUL{\cd}SS          & 1 \\
  Pulser                   &   PULSER &                      & 1 \\
  RF                       &       RF &                      & $2^{14}$ \\
  Left-Right               &       LR & SS{\cd}OS            & 100 \\
  SS and Calorimeter       &    SSCAL & SS{\cd}CAL(CAS)      & 100 \\
  OS and Calorimeter       &    OSCAL & OS{\cd}CAL(CAS)      & 100 \\
  Calorimeter (K thr.)     &   CAL(K) &                      & 1000 \\
  Calorimeter (Cas thr.)   & CAL(CAS) &                      & 1000 \\
  Single Muon Left         &   1MULT  & SS{\cd}1MUL          & 10 \\
  Single Muon Right        &   1MURT  & OS{\cd}1MUR          & 5 \\
  Advance                  &   ADVNC  & CAS 80\,ns ahead      & 100 \\
  S45                      &     S45  & S45 counter          & 1 \\
  Beam                     &    BEAM  & BHSUM{\cd}BVSUM      & 5000 \\
  \hline
\end{tabular}
\end{table}

\clearpage

\begin{table}
\caption{Physics processes considered for the trigger design.}
\vspace{0.2in}
\label{tab:trig_phys}
\begin{tabular}{l}
   \hline
    \Xim \ra \gL\pim \ra \gp\pim\pim  \\
    \Xim \ra \gp\pim\pim \\
    \Xim \ra \gp\mum\mum \\
    \Omm \ra \gL\pim \ra \gp\pim\pim \\
    \Omm \ra \gp\pim\pim \\
    \Km  \ra \pim\pip\pim \\
    \Km  \ra \pim\mup\mum \\
\hline
  \end{tabular}
\end{table}

\clearpage

\begin{table}
\caption{Hadronic Calorimeter specifications.}
\vspace{0.2in}
\label{tab:specs}
\begin{tabular}{rl}
\hline
Type:                         & Sampling (Fe:scintillator, 5:1) \\
Composition:                  & 24.1\,mm Fe, 5.0\,mm PS scintillator  \\
Layer depth:                  & 36.93\,mm  \\
Number of layers:             & 64   \\
Size ($x{\times}y{\times}z$): & $0.990{\times}0.980{\times}2.388$\,m$^3$ \\
Mass:                         & 12\,667 kg \\
Cell size ($x{\times}y{\times}z$): & 0.495\,m\,${\times}$\,0.980\,m\,${\times}$\,16 layers \\
Total cells:                  & 8 ($2x {\times}1y{\times}4z$)  \\
Fiber diameter:               & 2.0\,mm         \\
Fiber separation:             & 30.0\,mm  \\
Fibers per cell:              & 16\,${\times}$\,16 = 256  \\
Total fibers:                 & 8\,${\times}$\,256 = 2048  \\
Interaction length:           & $2.40\,\lambda_I$ per cell  \\
                              & $9.62\,\lambda_I$ total   \\
Radiation length:             & $22.1\,X_0$ per cell   \\
                              & $88.5\,X_0$ total   \\
Sampling fraction:            & 3.54\%    \\
\hline 
\end{tabular} 
\end{table}

\clearpage

\begin{table}
\caption{Data read out by the FastDA System}
\vspace{0.2in}
\label{tab:fastda_data}
\begin{tabular}{p{0.5cm}rl}
\hline
\multicolumn{3}{l}{Encoded Data} \\
  &  19\,680 & MWPC wires \\
  &      624 & Muon proportional tubes \\
  &       50 & Muon hodoscope counters \\
\multicolumn{3}{l}{Unencoded Data} \\
  &      10 & 14-bit ADCs \\
  &       4 & 8-bit TDCs \\
  &       1 & 32-bit Time stamp \\
  &       1 & 16-bit Run condition switch \\
  &       1 & 16-bit Spill number \\
  &      72 &  1-bit Hodoscope counters \\
  &      64 &  1-bit Trigger information \\
\hline
\end{tabular}
\end{table}

\clearpage

\begin{table}
\caption{Data read out by the SlowDA System}
\vspace{0.2in}
\label{tab:slowda_data}
\begin{tabular}{p{0.25cm}@{~$\bullet$~}p{6.0cm}} 
\hline
\multicolumn{2}{l}{Spectrometer Information} \\
  &  MWPC voltages and currents \\
  &  Hyperon and Analyzing Magnets Hall probes (5/spill) \\
  &  Prescaled trigger scalers (inhibited and uninhibited) \\
  &  Unprescaled trigger scalers \\
  &  Hodoscope counter scalers  \\
\multicolumn{2}{l}{Beam Line Information} \\
  &  Beamline magnet currents \\
  &  SWIC wire scans at five locations along the beamline
                 (10/spill) \\
  &  Date, time, spill number \\
\hline
\end{tabular}
\end{table}

\clearpage

\begin{table}
\caption{Sizes of data samples recorded by HyperCP.}
\vspace{0.2in}
\label{tab:data_raw}
\begin{tabular}{lrrr}
\hline
\multicolumn{4}{c}{Recorded Data} \\
     & 1997 & 1999 & Total \\ 
\hline
\multicolumn{1}{l}{Triggers ($10^9$)}     & 58    & 173     & 231 \\
\multicolumn{1}{l}{CAS Triggers ($10^9$)} & 39    &  90     & 129 \\
\multicolumn{1}{l}{Data volume (TB)}      & 38    &  82     & 120 \\
\multicolumn{1}{l}{Tapes}                 & 8980  & 20\,421 & 29\,401 \\
\hline
\end{tabular}
\end{table}

\clearpage

\begin{table}
\caption{Reconstructed event samples.} 
\vspace{0.2in}
\label{tab:data_recon}
\begin{tabular}{lrrr}
\hline
\multicolumn{4}{c}{Reconstructed Events ($10^6$)} \\
\multicolumn{1}{c}{Polarity: }     & $-$  & $+$  & Total\\ 
\hline
\gXi\ra\gL\gp\ra\gp\gpi\gpi        & 2032 &  458 & 2490 \\
\gOm\ra\gL\gK\ra\gp\gK\gpi         &   14 &    5 &   19 \\
\gK\ra\gpi\gpi\gpi                 &  164 &  391 &  555 \\
\Ks\ra\pip\pim                     &  693 & 2025 & 2718 \\
\hline
\end{tabular}
\end{table}

\clearpage

\begin{table}
\caption{Numbers of reconstructed \gXi\ events from nonzero-targeting-angle 
         runs.}
\vspace{0.2in}
\label{tab:data_polar}
\begin{tabular}{cr@{.}lr@{.}l}
\hline
  \multicolumn{1}{c}{\rule{0pt}{2.5ex}Angle} 
    & \multicolumn{2}{c}{$\aXim\ (10^6)$}
    & \multicolumn{2}{c}{$\Xim\ (10^6)$} \\
\hline
   $+3.0$ mrad  & ~~17 & 7  &  ~~89 & 4   \\
   $-3.0$ mrad  & 10 & 2  &  75 & 1   \\
\hline
   $+2.5$ mrad  &  2 & 9  &   6 & 9   \\
   $-2.5$ mrad  &  2 & 1  &   6 & 0   \\
\hline
\end{tabular}
\end{table}

\clearpage

\def\figurename{FIGURE}

\clearpage

\clearpage

\clearpage

\clearpage

\clearpage

\clearpage

\clearpage

\clearpage

\clearpage

\clearpage

\clearpage

\clearpage

\clearpage

\clearpage

\clearpage

\clearpage

\clearpage

\clearpage

\clearpage

\clearpage

\clearpage

\clearpage

\clearpage

\clearpage

\clearpage

\clearpage

\clearpage

\clearpage

\clearpage

\clearpage

\clearpage

\clearpage

\clearpage

\clearpage

\clearpage

\clearpage

\clearpage

\clearpage

\clearpage

\begin{figure}
\centerline{\includegraphics[width=10.0cm]{lambda_mass_mc_real.eps}}
\mbox{}\\[0.5in]
\caption{}
\label{fig:mass_mc_data}
\end{figure}
\clearpage

\clearpage

\clearpage

\clearpage

\clearpage

\clearpage

\end{document}